\documentclass[12pt,a4paper]{article}

\usepackage{multirow}
\usepackage{array}
\usepackage{longtable}

\usepackage{pgfplots}
\usepackage{subcaption}

\usepackage[english]{babel}
\usepackage[utf8]{inputenc}  
\usepackage{lmodern}         

\usepackage[top=3cm,bottom=3cm,left=2.5cm,right=2.5cm]{geometry}

\usepackage{natbib}          
\setcitestyle{authoryear,open={(},close={)}}

\usepackage{hyperref}
\hypersetup{
    colorlinks=false,
    pdfborder={0 0 0}
}

\usepackage{amsmath, amssymb, amsfonts}
\usepackage{amsthm}
\usepackage{mathabx}         
\usepackage{bbm}             

\usepackage{comment}

\usepackage{algorithmic}
\usepackage{float}
\algsetup{linenosize=\scriptsize,linenodelimiter=.,indent=1em}
\floatstyle{ruled}
\newfloat{algorithm}{h}{}
\floatname{algorithm}{Algorithm}

\usepackage{graphicx}

\usepackage{tikz}
\usetikzlibrary{patterns, arrows, backgrounds, calc, decorations.pathreplacing}
\colorlet{myblue}{blue!80!green}
\colorlet{mybluelight}{myblue!50}
\tikzset{
  > = latex',
  axis/.style    = {very thick},
  aborder/.style = {draw},
  acomp/.style   = {fill=black, fill opacity=0.1},
  rect/.style    = {very thick},
  form/.style    = {font=\scriptsize},
  sm/.style      = {font=\small},
  vsm/.style     = {font=\scriptsize}
}

\usepackage{booktabs}
\usepackage{dcolumn}

\usepackage{xcolor}

\hypersetup{
    colorlinks=true, 
    linkcolor=blue,  
    citecolor=blue,  
    urlcolor=teal    
}

\usepackage{parskip}

\usepackage[capitalize,nameinlink,noabbrev]{cleveref}

\usepackage[shortlabels]{enumitem}

\title{GATE: An Integrated Assessment Model for AI Automation}
\author{\small Epoch AI\thanks{See Section~\ref{sec:authors_acknowledgements} for a detailed list of contributions.}}

\date{}

\pgfplotsset{compat=1.18}
\begin{document}
\maketitle
\begin{abstract}
\fontsize{10pt}{12pt}\selectfont
Assessing the economic impacts of artificial intelligence requires integrating insights from both computer science and economics. We present the \textbf{G}rowth and \textbf{A}I \textbf{T}ransition \textbf{E}ndogenous model (GATE), a dynamic integrated assessment model that simulates the economic effects of AI automation. GATE combines three key ingredients that have not been brought together in previous work: (1) a compute-based model of AI development, (2) an AI automation framework, and (3) a semi-endogenous growth model featuring endogenous investment and adjustment costs. The integrated assessment model allows users to simulate the economic effects of the transition to advanced AI in a wide range of potential scenarios. GATE captures the intricate interactions between economic variables (investment, automation, innovation, and growth) and AI-related inputs and outputs (such as compute and algorithms). This paper explains the model's structure and functionality, with particular emphasis on elucidating the technological aspects of AI development for economists and clarifying economic concepts for the AI community. The model is implemented in an interactive sandbox, allowing users to explore the impact of advanced AI under different combinations of parameters and policy interventions. The modeling sandbox is available at \href{https://www.epoch.ai/GATE/}{\texttt{www.epoch.ai/GATE}}.

\end{abstract}

\newpage

\tableofcontents

\newpage

\section{Introduction}

Assessing the transformative potential of artificial intelligence requires integrating insights from both computer science and economics. We introduce the Growth and AI Transition Endogenous model (GATE), an integrated assessment tool that captures how AI-driven advances—in compute, algorithms, and automation—shape macroeconomic outcomes. Embedded in an interactive simulation platform available at \href{https://www.epoch.ai/GATE/}{\texttt{www.epoch.ai/GATE}}, GATE enables researchers and policymakers to explore different development paths for AI capabilities and the economy within a unified, dynamic framework.

GATE simulates the evolution of key economic variables, including ouput growth, consumption, capital investment patterns, compute allocation (training vs. inference), and AI automation rates under a wide range of assumptions about AI development, technology, preferences, and frictions. At its core, GATE features semi‐endogenous growth through hardware and software R\&D, which in turn indirectly shapes the accumulation of computing resources by improving hardware and algorithmic efficiency. In addition, the model imposes adjustment costs on both compute and non‐compute capital, capturing real‐world frictions in scaling up production capacity. Through this combination of R\&D‐driven efficiency gains and capital‐adjustment constraints, GATE links expanding AI capabilities (and the resulting labor‐task automation) to sustained output growth.

Users can modify crucial parameters of the model such as the strength of complementarity between AI and human inputs, R\&D investment returns, capital adjustment frictions, hardware efficiency limits, and requirements for full automation to explore a wide range of  scenarios of AI development and their economic implications. By providing this integrated framework, GATE aims to assist policymakers, economists, and AI researchers in scoping potential trajectories of AI advancement and its impacts on economic growth, investment patterns, and labor automation.

A key contribution of GATE is an interdisciplinary approach to modeling AI-driven economic transformation. For economists, GATE offers a detailed framework for the engineering process driving increases in AI capabilities. In particular, we explain how the accumulation of compute resources and algorithmic improvements are expected to drive AI capabilities and provide a simple and tractable mathematical formalization of this process. We also present and embed in GATE the standard framework of thinking about the process of AI automation within the AI community. We hope that this will facilitate transparent communication and debate between the economics and AI communities regarding the likely path of future AI automation. The model fully tracks the relationship between compute, training and inference processes, and the expanding set of tasks AI can perform as well as the downstream effects of automation for aggregate output, savings and investment. 

For the AI community, GATE incorporates key economic mechanisms and constraints such as endogenous responses of investment decisions to AI developments, adjustment costs affecting the accumulation processes of capital and compute and the role of economic incentives in driving the optimal allocation of compute, as well as their effects on key economic outcomes. By bridging the fields of economics and AI capabilities research, GATE provides a more comprehensive assessment of AI's likely economic impact than single-discipline approaches. We believe this integrated approach allows GATE to surpass siloed analyses by quantifying how AI’s rapid advancements feed back into economic growth, helping decision-makers anticipate and manage the profound changes likely to arise from ongoing technological innovation.

GATE consists of three autonomous modules that jointly track the causal chain from AI development to economic outcomes and the feedback loop from economic growth to AI investment. The modules are: (1) an AI development module that links investment in AI capabilities to the stock of effective compute, the key input required for training and running AI systems; (2) an AI automation module that maps the available stock of effective computing resources to AI’s capability to automate labor tasks on both the extensive and the intensive margins;\footnote{The intensive margin of automation can be thought of as increasing the number of “digital worker” equivalents available for the delivery of automated tasks. By contrast, the extensive margin of automation can be thought of as the increase in the share of tasks that are amenable to be performed by AI/ “digital workers”.} and (3) a macroeconomic module linking labor market automation to macroeconomic outcomes such as aggregate output, consumption and investment (and hence further AI developments). We solve the model from the perspective of a benevolent social planner aiming to maximize the net present value of consumption for a representative agent, subject to a set of technological constraints and frictions.

GATE also incorporates two optional add-ons that can be activated or deactivated by users as needed, enhancing its flexibility for various analyses. The first is an uncertainty add-on for the process of AI automation, which explicitly models investor beliefs about the mapping between the computational resources dedicated to AI and the capability of these systems to automate labor tasks, and allows these beliefs to update based on observed outcomes. By simulating how economic decisions and outcomes evolve as uncertainty about AI progress is resolved, this add-on provides more realistic projections of potential economic trajectories under different scenarios of investor expectations about AI development paths.

The second add-on models the implications of positive externalities generated by AI-related R\&D. To capture the fact that positive externalities are unlikely to be fully captured in investment decisions, this add-on allows for the modeling of a ``myopic" social planner that does not fully account for the broader societal benefits of R\&D. Such a social planner systematically underinvests in AI development relative to the social optimum. The R\&D externalities module enables simulation of economic trajectories by incorporating user-defined investment wedges that capture this undervaluation of R\&D returns, allowing modeling of scenarios with different degrees of ``internalization" of externalities.

While the integrated assessment model includes many of the forces that we believe will be crucial in shaping the evolution of AI and its impact on the economy, several significant limitations remain. The AI development module abstracts away from important non-compute inputs like data availability and quality, and relies on a simplified 'effective compute' framework that reduces algorithmic progress to a single dimension and fails to capture how improvements may vary across different scales of computation.

For the AI automation module, GATE posits a simple and static labor “task space” that gets gradually automated via a simple automation rule. Both of these assumptions could be challenged. For instance, the space of tasks could be significantly more complex, tasks could display complicated patterns of complementarity and substitutability in production, and new tasks could appear along the automation process. Moreover, the task automation process could conversely be made more realistic, but at the expense of model simplicity. Tasks might be heterogeneous along multiple dimensions, such as manual vs. cognitive tasks or routine vs. non-routine tasks, and thus the current framework does not capture how these different task characteristics might affect the ease or difficulty of automation.

For the macroeconomic module, GATE currently omits non-AI related TFP growth from the analysis and so it cannot model the effects of non-AI related R\&D on productivity. Furthermore, it only allows limited flexibility in modeling the complementarity and substitutability between labor tasks, capital and non-accumulable factors of production. It also features a relatively simple specification of demand, that omits, among other factors: consumption-leisure tradeoffs, income effects, preference heterogeneities or rich patterns of complementarity and substitutability among different commodities.

The current release of GATE is an initial step in modeling AI-driven economic impacts. We plan to regularly update the model and the associated sandbox, incorporating emerging research and community feedback. In the future, this framework could potentially serve as a foundation for exploring diverse AI automation scenarios through related models. We invite the research community to engage with GATE and collaborate in building a comprehensive toolkit for understanding the economic impacts of AI.

The rest of the paper is organized as follows. Section~\ref{sec:high_level} offers a high-level overview of the model and its key components. Sections~\ref{sec:compute_model_specification} to \ref{sec:further_ingredients} then provides a formal, in-depth description of all the components of GATE. Section~\ref{sec:compute_model_specification} outlines the AI development module, that links the path of investment in AI inputs with the evolution of the model's critical AI input, effective compute. This section is perhaps the most novel and significant contribution of GATE. Section~\ref{sec:compute_based_model} describes the AI automation module linking allocations of AI resources to the pattern of automation of labor tasks. Section~\ref{sec:macro_model_specification} outlines GATE's macro module that links automation to aggregate output and also details the feedback loop from output to AI automation via AI investment flows. Section~\ref{sec:further_ingredients} describes the optional model add-ons, namely the R\&D externalities add-on and the uncertainty add-on. Section~\ref{sec:model_functionality} then proceeds to describe the model’s functionality and some of its use cases, while Section~\ref{sec:limitations} discusses current limitations of the model and suggests directions for future work. Finally, Section~\ref{sec:conclusion} outlines our future directions for researching the economic implications of advanced AI systems. Technical details, including plausible ranges for parameter values, justifications for our parameter presets and a full characterization of our model solution algorithm are provided in the Appendices.

\section{The Model at a High Level}
\label{sec:high_level}

GATE represents an early attempt at constructing an integrated assessment model of the impact of AI on the global economy. The model is designed for the purpose of allowing users to run simulations of the economic implications of AI development under different scenarios and is embedded in an interactive simulation platform available at \href{https://www.epoch.ai/GATE/}{\texttt{www.epoch.ai/GATE}}. 

\begin{figure}[h]
    \centering
    \includegraphics[width=1\textwidth]{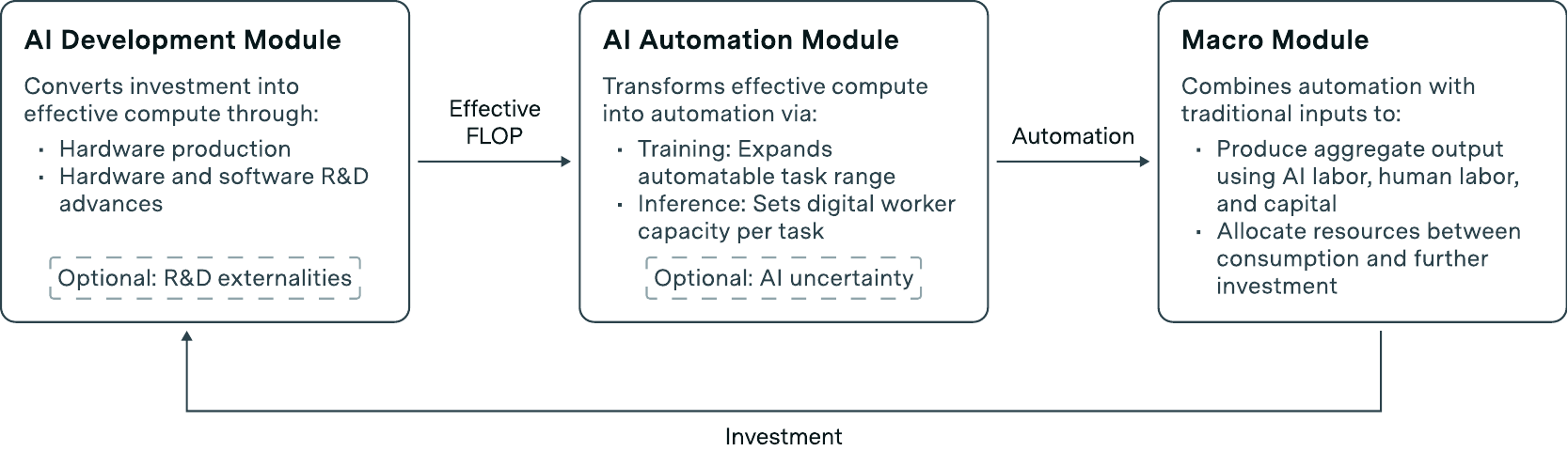}
    \caption{A high‐level schematic of GATE’s three modules and their feedback loops. The AI Development Module (left) channels investment into effective compute. The AI Automation Module (middle) applies that compute to automate tasks. The Macro Module (right) uses both AI and human labor plus capital to produce output, part of which is reinvested into AI. The dashed boxes highlight optional add‐ons—R\&D externalities and AI uncertainty—toggled on or off as needed.}
    \label{fig:GATE_model}
\end{figure}

The model follows the entire causal chain linking developments in AI to key economic outcomes, as well as the feedback loops from economic expansion to increased investment in AI development and hence increased AI capabilities. The basic set-up of the model features an economy that invests resources in order to optimize net present consumption. Crucially, this is done by investing in AI along two key dimensions: in compute-related capital to power AI training and deployment (e.g., chips, fabs, datacenters) and in software and hardware R\&D. Furthermore, increases in AI capabilities brought about by AI investment permit the gradual automation of labor tasks on both the intensive and extensive margins, which is then reflected in economic expansion. Finally, the benefits of AI automation and the additional resources made available by economic expansion motivate and facilitate further AI investment, which reinforce the two-way relationship between AI development and economic outcomes, and create potential conditions for growth accelerations resulting from AI development. Importantly, our model incorporates production bottlenecks arising from differential productivity growth, frictions affecting the investment process, and other constraints on technological progress, such as energy limitations and environmental heating effects from extensive computation, that may have a material effect on the paths of both AI development and economic growth.

The model consists of three distinct modules. The first is the AI development module, which maps investment in AI capabilities to the accumulation of effective compute, the key input required for training and developing AI systems. This module tracks how different investment profiles translate into compute resources over time. The second component is the AI automation module, which connects the available effective computing resources to the automation of labor tasks. It captures automation along both the extensive margin (which tasks can be automated) and the intensive margin (how efficiently these automated tasks are performed). The third component is the macroeconomic module, which translates labor market automation into broader economic outcomes. This final module shows how automation affects aggregate output, consumption, and investment patterns in the economy.

The AI development module is perhaps the most novel and significant contribution of GATE to the existing literature on the economic impacts of AI. We set out to model the engineering details of the AI development process in a realistic yet tractable way. We clarify why we can think of AI development as primarily driven by the gradual accumulation of a key input, which we call effective compute, and clarify how we expect the cost of this key input to evolve over time in response to investments in hardware and software R\&D. The key goal of this module is to model the relationship between any compute investment path (measured in \$) and the stock of effective compute (measured in effective FLOPs or eFLOPs) employed for the training and deployment of AI systems, taking into account the various frictions and limits affecting the AI scaling process. This module also clarifies some of the engineering tradeoffs faced in allocating effective compute between various AI-specific uses such as training larger models or running more instances of existing models (i.e., inference).

One of the key goals of the AI development module is to inform the economics community about the engineering process behind AI development and to propose a tractable modeling strategy that captures the key features of this process. The AI development module is autonomous and can easily be embedded in models that take different stances on other key economic phenomena, such as the scope of AI automation or the macroeconomic implications of any given level of automation (that we embed in the second and third modules of GATE). 

A second key component of our model is a characterization of the relationship between AI capabilities and automation, which is captured by our AI automation module. This module has two key ingredients. First, we adopt a framework that maps two inputs — the scale of training compute and the scale of inference compute—to the fraction of economic tasks that can be automated by AI systems and to the increase in the amount of effective resources (or effective ``labor") available to deliver each automated task. This approach is inspired by prior theoretical work on AI capabilities \citep{moravec1976role, cotra2020}. Intuitively, as more computing power becomes available for training and running AI systems, these systems become more capable and can perform a growing range of tasks traditionally done by humans. Moreover, as the automation process unfolds, these AI-powered `digital workers' can often operate more efficiently than humans, requiring fewer resources to complete the same tasks. 

A further ingredient of this module characterizes how human labor is reallocated as AI automation progresses. We analyze two polar opposite scenarios: perfect reallocation, where workers displaced by automation smoothly transition to remaining non-automated tasks, and complete displacement, where automated workers permanently exit the labor force. This framework enables us to study how labor mobility shapes both the automation process and broader economic outcomes. Alternative specifications of labor market dynamics can be integrated with other GATE modules in subsequent work.\footnote{Partial reallocation scenarios introduce additional complexity by requiring workers to be allocated across tasks based on both current productivity and the expected timing of future automation. We focus on these boundary cases to maintain analytical clarity while capturing key economic mechanisms.}

Finally, we embed the previous two modules into a standard macroeconomic framework (which we also call our macroeconomic module). This is a standard Ramsey-Cass-Koopmans framework with endogenous saving and a constant returns to scale aggregate production function combining a labor composite (which is gradually automated by AI), with accumulable and non-accumulable capital. The social planner is assumed to optimally choose savings, investment in AI inputs and investment in other (i.e. non-AI) capital goods in order to maximize the net present value of consumption for the representative household. The model embeds adjustment frictions governing the process of physical capital accumulation, in line with the existing literature. Moreover, given our extensive modeling of technical progress affecting AI development in the AI development module, we omit non-AI related TFP growth from our macroeconomic module. This assumption is made for simplicity as well as computational tractability, and could be relaxed in future iterations of the model.
Notably, we also allow for the consideration of a third \textit{non-accumulable} factor of production $F$\footnote{We also refer to this additional factor as non-accumulable capital}, which can be thought of as natural resources or land. This is done to allow users to specify the existence of long-run non-accumulable factors (that remain non-accumulable in the advanced stages of automation), that may yield additional bottlenecks to economic growth as AI automation advances.

Beyond the three key modules outlined above, we implement two additional add-ons to the model that can be switched on and off by users when conducting simulations. Our first add-on is the 'Positive externalities from R\&D' module. It allows us to model how society might underinvest in AI software and hardware R\&D when positive spillovers exist and are not fully internalized by the relevant economic decision makers. This add-on hence allows user to specify wedges between the social returns to R\&D and the private returns to R\&D. This allows the user to run simulations with different assumptions about how externalities might cause R\&D underinvestment, potentially resulting in socially suboptimal development of AI capabilities and automation.\footnote{As an intuition, this add-on allows users to embed market failures from R\&D externalities in a social planner setting by solving a ``myopic" social planner problem.}

The second add-on permits the modeling of uncertainty affecting the social planner's decisions concerning investments in AI inputs and AI R\&D. Intuitively, we may expect uncertainty regarding the returns to AI-related investments to depress society's incentives to invest in AI development and thus lead to more moderate paths of AI investment and automation than in our baseline perfect foresight specification.

Throughout our analysis, we focus on solving the social planner's problem: a social planner chooses the path of saving and investment (in compute, R\&D and non-compute capital) to maximize the net present value of the consumption of the representative consumer. We do this to avoid making potentially arbitrary choices about market structures, particularly concerning the markets directly involved in the development and deployment of AI. Performing simulations of the macroeconomic impact of AI development under alternative assumptions concerning market structures for key markets (e.g. the market for effective compute and ideas affecting hardware and software development) is left for subsequent work.

GATE facilitates the quantitative analysis of the macroeconomic impact of advanced AI under a wide range of scenarios and structural assumptions. The framework can be used to simulate the trajectories of key economic variables (such as output, consumption, savings and investment) under different  conditions concerning technology, preferences, frictions and uncertainty. Users can calibrate expectations about the path of investment and economic growth as well as the overall timelines of the process of AI automation. Moreover, the model can be readily extended to analyze the counterfactual impact of different policy interventions in areas such as labor market regulation, taxation and subsidies.

\section{The AI Development Module}
\label{sec:compute_model_specification}
\addtocontents{toc}{\protect\setcounter{tocdepth}{1}} 

The AI development module models how investments in AI (for example, spending on GPUs, semiconductor fabs, chip design, and improved algorithms or architectures) translate into the resources available for training and deploying AI models. At its core, this process can be viewed as a mapping from “dollars invested” to a budget of “effective compute” (measured in effective FLOP)\footnote{Floating point operations per second (FLOP) are a measure of the stock of compute available for AI training and inference. Effective compute adjusts these FLOP for improvements in software or algorithmic performance. For more details, see \cite{ho2024algorithmic}.} that powers AI systems at each point in time.

Hardware and software R\&D contribute to expanding this computational capacity in distinct ways. Software advances (e.g., novel training algorithms) enhance the efficiency of both the existing \emph{stock} of compute and all future expansions, while hardware improvements reduce the cost and increase the performance of \emph{new} compute acquired going forward. Together, these parallel R\&D activities progressively lower the price of effective compute, enabling greater AI capabilities through increased training and inference capacity. 

In addition, the AI development module specifies the technological parameters that govern how effective compute is allocated between training (to create more capable models) and inference (to deploy these models for economic tasks).

This first module showcases how increases in investment lead to increases in the computing resources available for the development of increasingly more capable AI systems. In turn, this sets the stage for the automation module, that links the development of more capable systems to the automation by AI of an increasing share of labor tasks currently performed by humans. In economic terms, the AI development module pursues two main objectives. First, it specifies the laws of motion for two key AI-related state variables—effective compute and the size of the largest training run. Second, it describes the engineering tradeoffs governing how effective compute is split between training and inference: for example, how investing more compute in training can reduce the runtime demands per task, or how adding inference compute can offset smaller training budgets.

This section presents a detailed model of the key engineering details that govern the process of AI development, aiming to make these concepts accessible to a general audience. The exposition proceeds in the following steps:
\begin{enumerate}
    \item Section~\ref{sec:effective_comp_just} explains the concept of `effective compute' - a metric combining raw computational power and algorithmic efficiency - and establishes it as the key input for AI development. Drawing on scaling laws research and empirical evidence of compute's central role in AI progress, we justify using effective compute as a sufficient statistic for tracking AI capabilities, as it captures both hardware and software improvements in a unified framework.
    \item Section~\ref{sec:training_vs_inference_tradeoff} analyzes the allocation of effective compute between training and inference phases in AI systems. Training represents a fixed cost producing non-rival model weights, while inference is a variable cost that can enhance or deploy trained models. We formalize this training-inference tradeoff and show how it affects both which tasks can be automated and how many digital workers can operate in parallel.
    \item Section~\ref{sec:cost_compute_today} examines the current empirical relationship between investment and effective compute. We establish baseline costs using individual GPU purchases, then analyze how large-scale deployments face higher costs from infrastructure requirements and supply chain frictions. This helps ground the model's treatment of the relationship between monetary investment and compute capacity.
    \item Section~\ref{sec:technological-progress} models how technological progress reduces the cost of effective compute through two channels: hardware advances (improving compute per dollar) and software advances (making computations more efficient), with an important asymmetry: software improvements enhance all compute while hardware improvements only affect new compute. 
    \item Section~\ref{sec:tech_constraints} extends our technology progress model by incorporating physical and engineering limits that bound both hardware and software efficiency gains as they approach theoretical maxima.
    \item In Section~\ref{sec:frictions}, we incorporate key constraints on expanding compute: non-linear adjustment costs from supply-chain bottlenecks, rapid hardware depreciation, and fundamental thermal limits on total usable compute.
    \item Finally, Section~\ref{sec:taking_stock_AI_dev} provides a summary of the relationship between AI related investments and the time path of effective compute endowments, as well as a summary discussion of the allocation of effective compute and its various uses.
\end{enumerate}

\subsection{Key Input of AI development: Effective Compute}
\label{sec:effective_comp_just}

The development of artificial intelligence systems relies fundamentally on computation---the processing of mathematical operations by computer hardware. Modern AI training proceeds by iteratively adjusting millions or billions of parameters in neural networks using vast amounts of computation, measured in floating point operations (FLOP). This computational process occurs in two distinct phases: training, where the AI system learns patterns from data, and inference, where the trained system is deployed to perform tasks.

A key lesson from the history of AI research, articulated by \cite{sutton2019bitterlesson}, emphasizes the importance of raw computation:
\begin{quote}
The biggest lesson that can be read from 70 years of AI research is that general methods that leverage computation are ultimately the most effective, and by a large margin... Seeking an improvement that makes a difference in the shorter term, researchers seek to leverage their human knowledge of the domain, but the only thing that matters in the long run is the leveraging of computation.
\end{quote}
This insight has been validated by recent work on scaling laws \citep{kaplan2020scaling, hoffmann2022empirical}, which shows that the performance of AI systems can be predicted remarkably well using simple functions of computational resources.

While Sutton’s “bitter lesson” underscores the primacy of \emph{raw} computational power, modern evidence shows that algorithmic innovations can also meaningfully reduce the FLOPs needed to reach a given performance level. To capture this, researchers quantify \emph{effective compute}—an adjusted measure of FLOPs that accounts for how each real FLOP grows “more potent” as algorithms improve. Concretely, studies find that the raw compute required to train neural networks for the same benchmark accuracy roughly halves every 9 to 16 months \citep{hernandez2020measuring, erdil2022algorithmic, ho2024algorithmic}. In other words, one unit of compute today can produce results that once demanded two or more units of compute a year ago. By incorporating these efficiency gains into a single metric, effective compute provides a simpler way to track overall AI progress than raw FLOP alone.

To capture both hardware and algorithmic contributions to AI progress, we adopt \emph{effective compute} as our core input variable—a notion that has gained traction among researchers and industry practitioners.\footnote{For example, see \cite{anthropic2023rsp} for Anthropic’s Responsible Scaling Policies.} Historical analyses by Moravec, Good, and Kurzweil anticipated that once raw computational power approached certain thresholds, AI capabilities would see large breakthroughs, a view that has found more rigorous support in modern scaling-law research \citep{kaplan2020scaling, hoffmann2022empirical}. Indeed, empirical work on computer vision and language models shows that increasing training compute, together with algorithmic improvements, systematically yield higher performance \citep{hernandez2020measuring,erdil2022algorithmic,ho2024algorithmic}. Unifying gains in hardware (FLOP per dollar) and algorithmic efficiency into a single measure, effective compute, furnishes a tractable way to capture the key determinants of AI capabilities. Of course, this approach does not capture every nuance (considerations such as serial vs.\ parallel constraints or data availability are omitted), but substantial evidence (see Appendix \ref{sec:compute-based}) indicates that when forecasting AI progress at scale, few factors rival compute, duly adjusted for software improvements, in predictive power.

\subsection{AI training, Inference, and the Training-Inference Trade-off}\label{sec:training_vs_inference_tradeoff}

In this section we explain how the key input described in the previous section, effective compute, is used for the development and deployment of increasingly capable AI systems.

Modern AI systems broadly operate in two phases: a training phase, where the system learns from data, and an inference phase, where the trained system is deployed to perform tasks. During training, the system adjusts its parameters to improve performance, typically through iterative optimization over large datasets. During inference, these learned parameters are used to process new inputs and generate outputs.

In practice, the distinction between training and inference is often less clear-cut. Models can be fine-tuned or continually updated after deployment, incorporating new data to improve performance~\cite{Shi2024}. Additionally, intermediate versions of models---checkpoints---may be deployed even as training continues, allowing for early utilization of the model's capabilities. Techniques like online learning blur the lines further by enabling models to adapt in real time during inference, as demonstrated by test-time training, where models update their parameters with each new data sample before making predictions~\cite{Sun2020}.

Nevertheless, there remains a fundamental economic distinction between training and inference compute. Training compute represents a fixed cost that produces non-rival goods---the model weights---which can then be deployed across any number of instances without needing to be retrained. Once a model is trained, its capabilities can be replicated across many deployments, much like how research and development costs produce knowledge that is non-rival and can subsequently be widely applied. In contrast, inference compute represents a recurring (or variable) cost: each use of the model requires computational resources. This distinction between fixed costs producing non-rival capabilities and variable costs of deployment proves crucial for modeling the economics of AI development.

\paragraph{AI training runs} 

AI training is the computationally intensive process through which models acquire their capabilities by iteratively updating parameters---often numbering in the trillions---based on training data. While training requires substantial computational resources, it produces non-rival outputs: once trained, a model's parameters can be copied and deployed across any number of instances without additional training costs, much like traditional research and development investments in standard R\&D-based growth models.

We first consider the evolution of the stock of effective compute that is devoted to AI training. If the largest training run at time \( t \) is given by \( C_{T}(t) \), then we define a simple update rule:
\begin{equation}
\dot C_{T}(t) = D(t).
\label{eq:largest_training_run}
\end{equation}
Here, \( D(t) \) is the amount of effective compute allocated to training AI at time \( t \). 

This model assumes continuous training, where compute is incrementally added to the largest training run at each timestep, rather than being allocated in discrete runs. Two main considerations justify this simplification. First, the exponential growth of training budgets means that the most recent increments dominate the total, so the difference between a smooth versus staggered allocation becomes negligible \citep{epoch2023thelimitedbenefitofrecyclingfoundationmodels}. Second, modern training pipelines are in fact highly iterative, making an essentially continuous framework a natural abstraction.\footnote{Modern AI development involves multiple continuous training phases including pre-training, synthetic data generation and training, reinforcement learning from human feedback (RLHF), specialized task-specific training, and continuous model updating. Early checkpoints can be deployed while training continues, and pre-trained models can be enhanced through additional training on new datasets or objectives. This multi-phase, continuous nature of AI development makes the continuous training assumption appropriate for modeling purposes.} The exponential growth in compute means that recent training runs tend to dominate the total compute allocation, making the precise timing of individual training runs less significant for modeling long-run trajectories.

\paragraph{AI Inference} 

Compute can be used for two primary purposes: training, which not only improves models to automate a larger set of tasks but can also reduce inference costs on tasks already automated (e.g., through ``overtraining'' or distillation), and inference, which applies models to perform economically valuable tasks. Additionally, inference can enhance the capabilities of AI models by extending runtime, allowing for deeper analysis or reasoning, and enabling parallel execution of multiple model instances \citep{openai2024openaio1, brown2024large}. For example, a chess engine can utilize more computation to search deeper, evaluating more moves before making a decision \citep{campbell2002deep}. Similarly, language models can be run in parallel, and intelligently combining their outputs often improves performance, especially when solutions can be automatically verified \citep{wang2022self}.

Training helps AI systems acquire base capabilities, while inference compute builds upon this by improving task execution. With additional inference compute, AI systems can devote more ``thought'' to solving complex problems, such as generating many candidate solutions and selecting the best, or performing intermediate reasoning steps. These processes enable AI to tackle a broader range of tasks more effectively, even when the training resources remain fixed.
\begin{figure}[h]
    \centering
    \includegraphics[width=1\textwidth]{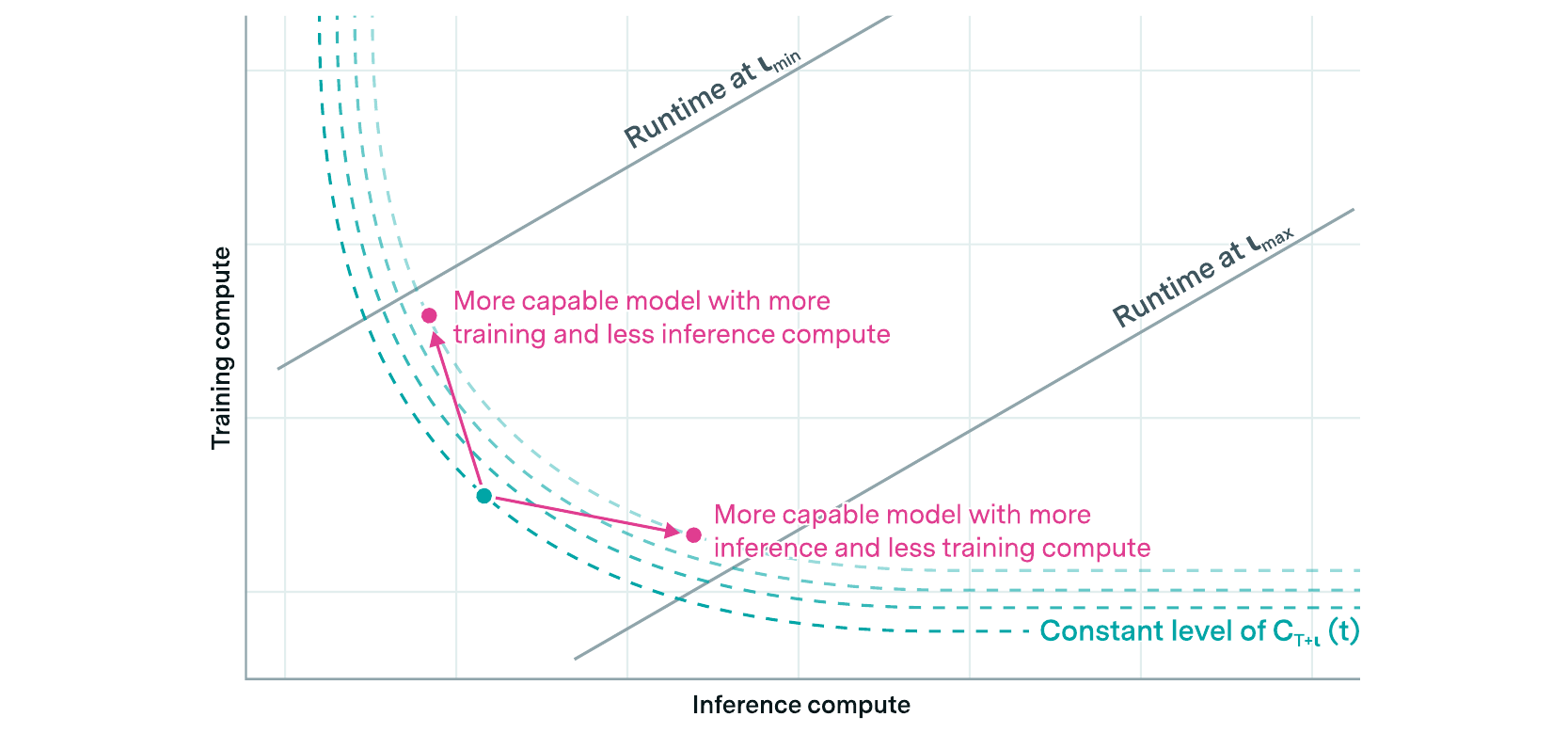}
    \caption{A schematic of the training--inference tradeoff, showing that the same overall capability (curves labeled ``constant level of $C_{T+\iota}(t)$'') can be reached by different mixes of training compute (vertical axis) versus inference compute (horizontal axis). Moving ``up'' invests more in training to reduce inference needs, while moving ``right'' adds inference to compensate for a smaller training run. The lines labeled ``Runtime at $\iota_\text{min}$'' and ``Runtime at $\iota_\text{max}$'' illustrate these feasible extremes, highlighting how a model’s performance can be maintained or enhanced by balancing the fixed (training) and variable (inference) portions of the total compute budget.}
    \label{fig:inference_training_tradeoff}
\end{figure}

We account for this by adjusting the total compute expenditure for model capability using an ``inference multiplier,'' which extends the range of tasks that can be automated beyond the effective training compute, though these additional tasks have correspondingly high inference requirements if they are actually deployed at scale. Formally, we let $\iota$ vary between some lower bound and a maximum $\iota_{\max}$. Then, to measure total system capability (training plus scaled-up inference), we write:
\begin{equation}
    C_{T+\iota}(t) \;=\; C_{T}(t)\,\times\,\iota^{\tfrac{1}{m}},
\label{eq:training_vs_inference_tradeoff}
\end{equation}
where $m$ is the slope of the training--inference tradeoff. Empirical observations suggest $m \approx 1-2$: each 1-2~OOMs of inference can substitute for about 1~OOM of training \citep{epoch2023tradingoffcomputeintrainingandinference}. 

Setting $\iota$ to $\iota_{\max}$ effectively allows very deep runtime for tasks that require it, without imposing any mandatory additional cost on simpler tasks. This is because inference usage can be chosen independently for each task. Thus, there is no downside to choosing $\iota = \iota_{\max}$; simpler tasks remain inexpensive, while more complex tasks gain the option—but not the obligation—to leverage additional inference.
\footnote{%
For instance, if the largest training run $C_{T}(t)$ involves $10^{30}$~FLOP and $\iota_{\max}=100$, then with a tradeoff slope of $m=2$ we get
\[
    C_{T+\iota}(t)\;=\;C_{T}(t)\,\times\,\iota_{\max}^{\tfrac{1}{m}}
    \;=\;
    10^{30}\,\times\,100^{1/2}
    \;=\;
    10^{31}\,\text{FLOP}.
\]
This exemplifies how leveraging additional inference capacity (up to $100\times$ in this example) can increase effective compute by one order of magnitude. In practice, the social planner (or firm) selectively allocates that extra runtime only to tasks where it is beneficial, so no global overhead arises from letting $\iota=\iota_{\max}$.}

Note that while $\iota$ sets an upper limit on how much further inference can extend a model’s base capability, it does not itself eliminate inference costs: each additional unit of runtime compute must still be purchased within the model’s resource budget. In other words, $\iota$ specifies what is technologically possible, but the economy only scales inference up to the point where the resultant gains in automated tasks justify the associated costs.

Intuitively, the total compute budget captures how a given performance level can be achieved in multiple ways: a smaller model (trained with less effective compute) but run with more inference, or a larger model (trained with more effective compute) but run with less inference. This tradeoff reflects fixed vs.\ variable costs: large models incur higher one-time (training) costs but often require fewer inference resources per task, whereas smaller models demand lower training costs but more inference effort to match the same performance. As discussed next in Section~\ref{sec:compute_based_model} (especially subsections~\ref{sec:extensive_margin_automation} and~\ref{sec:intensive_margin_automation}), these distinctions feed into how AI development affects both the \emph{extensive} (which tasks can be automated) and \emph{intensive} (how many digital workers can be deployed) margins of labor automation.

As we suggested earlier, the tradeoff outlined above is bounded. According to \citet{epoch2023tradingoffcomputeintrainingandinference}, most individual techniques (e.g., pruning, repeated sampling) can shift at most about 1--2 orders of magnitude (OOM) between training and inference while maintaining the same performance. Even when combining techniques, the total span rarely exceeds a few OOMs overall (Ibid.). Consequently, no amount of inference alone can fully replace a sufficiently large training run beyond these bounds. In particular, if a task is too complex relative to the size of the trained model, extra inference compute will not suffice, so training a larger model remains the only way to automate it.

Taking stock of the above, expanding the training compute has the primary effect of increasing automation along the extensive margin, expanding the set of tasks that AI can do. Inference compute affects automation along both the intensive margin (by powering more AI systems performing economic tasks) and the extensive margin (by effectively ``stretching'' a smaller model). More details about the mapping between compute allocations and both margins of automation are provided in Section~\ref{sec:compute_based_model}. While solving GATE, we assume that the social planner optimally allocates finite compute across training and inference to advance automation and maximize the net present value of consumption for the representative agent.

\subsection{The Cost of Compute Today}\label{sec:cost_compute_today}

To ground our model empirically we need to estimate the relationship between investment and effective compute. For small-scale purchases, this relationship is approximately linear. The NVIDIA H100, currently the leading GPU for AI training, serves as a useful benchmark: at a price point of approximately \$30,000, it can perform around $10^{15}$ FLOP/s, or roughly $3 \times 10^{22}$ FLOP/year. This translates to approximately $10^{18}$ FLOP/year per dollar when purchasing individual units or small quantities.

The economics change significantly when scaling to the levels required for large AI training runs, which require thousands of interconnected GPUs. At these scales, two factors drive up costs. First, there are substantial additional infrastructure requirements beyond the raw hardware, including cooling systems and power delivery infrastructure, high-speed interconnects for GPU communication, electricity and maintenance costs, and expenses for software development and specialized personnel \citep{cottier2024rising}.

Second, and perhaps most importantly for our analysis, at the macroeconomic level the cost per unit of compute increases super-linearly with the scale of compute investments due to adjustment frictions affecting the semiconductor supply chain (see Section~\ref{sec:frictions}). This means that while small purchases face roughly constant marginal costs, at the global level large-scale compute investments face sharply increasing marginal costs.

\subsection{Technological Progress}
\label{sec:technological-progress}

Given its key role in determining the capabilities of AI systems, tracking the path of effective compute endowments over time is at the core of GATE. In turn, the exercise of forecasting the evolution of effective compute involves two key components: tracking the build-up of the stock of compute (which we will denote $Q$, measured in FLOP/year) and predicting future advancements in the underlying technology.\footnote{In economics jargon we can think of this exercise as predicting the path of investment flows into effective compute and the path of the cost per unit of effective compute, with the latter being expected to decline over time due to hardware and software improvements.}

Technological improvements in computing occur through two channels: hardware advances that increase the computational power per dollar \citep{epoch2023trendsinmachinelearninghardware}, and software/algorithmic advances that make each computation more efficient \citep{ho2024algorithmic,erdil2022algorithmic, hernandez2020measuring}. We model these through two terms: hardware efficiency $H(t)$ and algorithmic efficiency $S(t)$.

Hardware efficiency $H(t)$, measured in FLOP/year/\$, represents how many floating point operations per year one dollar of hardware investment can perform. We let $S(t)$ denote \emph{software efficiency}, measured in eFLOP/FLOP. In other words, $S(t)$ tells us how many ``effective FLOPs'' one raw FLOP is worth. Both efficiency terms increase over time through endogenous research efforts.
The dollar cost per effective FLOP/year at time $t$ is thus given by:
\begin{equation}
    \frac{1}{H(t) \cdot S(t)}
\end{equation}
We model the evolution of hardware and software efficiency using a continuous-time analog of the R\&D-driven growth framework originally introduced in \citet{jones1995r} and subsequently adapted for various technologies \citep{bloom2020ideas}. While this framework was originally employed to model total factor productivity, it is sufficiently flexible to capture the dynamics of more specific technologies. 

In our formulation, each technology’s efficiency evolves according to a law of motion that depends on the current stock of knowledge (or efficiency) and the intensity of R\&D investment. For hardware, let $H(t)$ represent the efficiency level at time $t$, and let $I_H^{RD}(t)$ represent the R\&D investment devoted to hardware improvements. For software, let $S(t)$ denote the software efficiency level, and $I_S^{RD}(t)$ the corresponding software R\&D investment.
We assume that improvements in hardware and software efficiency occur according to the following continuous-time equations, presented side-by-side:
\begin{equation}
\frac{\dot{H}(t)}{H(t)} = \theta_H [H(t)]^{-\phi_H} [I_H^{RD}(t)]^{\lambda_H},
\qquad\qquad
\frac{\dot{S}(t)}{S(t)} = \theta_S [S(t)]^{-\phi_S} [I_S^{RD}(t)]^{\lambda_S}\label{eq:r_d_equation}
\end{equation}
Here, $\dot{H}(t)$ and $\dot{S}(t)$ represent the time derivatives of hardware and software efficiency, respectively. The parameters $\theta_H$ and $\theta_S$ determine the baseline effectiveness of R\&D investment for hardware and software. The terms $H(t)^{-\phi_H}$ and $S(t)^{-\phi_S}$ capture the intertemporal knowledge spillovers. Depending on the signs of $\phi_H$ and $\phi_S$, these can reflect a ``standing-on-shoulders'' effect (if negative) or a ``fishing-out'' effect (if positive).

By using total R\&D investment as the key input, we simplify the modeling framework and avoid the complexities of explicitly tracking intermediate inputs like human or AI labor. If we wish to remain consistent with traditional views of human-centric R\&D bottlenecks, we can adjust the elasticity parameters (the $\lambda$ terms) downward to obtain more conservative outcomes. Additionally, as AI-driven automation reduces the reliance on human labor in certain innovation processes, framing inputs in terms of overall investment rather than labor becomes increasingly relevant. This approach, while simpler, is thus flexible enough to accommodate both conventional and more automated R\&D scenarios.

When thinking through the consequences of investments in hardware and software R\&D it will be important to keep in mind a key difference between hardware and software improvements: while software improvements apply to the entire stock of existing compute, hardware improvements only apply to new flows of compute. The implications of this asymmetry for the overall law of motion for effective compute are characterized formally in Section~\ref{sec:taking_stock_AI_dev}.

\subsection{Technological Constraints}
\label{sec:tech_constraints}

The simplest version of the laws-of-motion described in the previous section have a major shortcoming: they fail to account for \textit{ceilings} and physical limits on technological progress. We know that there are theoretical limits to key engineering variables such energy efficiency, information processing and more (See Table 2 for some examples). Similarly, there are known limits to algorithmic improvements. For example, we know that a non-trivial fraction of standard algorithms have (asymptotic) runtime complexities equal to some theoretical lower bound, and therefore do not permit further improvements to runtime complexities. Hence, it is highly likely that progress in software, like hardware, will be bound from above.
\begin{table}[h]
\centering
\resizebox{\textwidth}{!}{%
\begin{tabular}{@{}lllll@{}}
\toprule
\textbf{Limits Engineering} & \textbf{Design and Validation} & \textbf{Energy, Time} & \textbf{Space, Time} & \textbf{Information, Complexity} \\
\midrule
Abbe (diffraction) & Error-corr. \& dense codes & Einstein E=mc\textsuperscript{2} & Speed of light & Shannon \\
Amdahl & Fault-tolerance thresholds & Heisenberg $\Delta E \Delta t$ & Planck scale & Holevo \\
Gustafson & & Landauer kT ln2 & Bekenstein & NC, NP, \#P \\
& & Bremermann & Fisher T(n)1/(d+1) & Turing (decidability) \\
\bottomrule
\end{tabular}%
}
\caption{\small \centering Some fundamental limits to computation, adapted from \protect\citealt{markov2014limits}.}
\label{tab:my_label}
\end{table}

To incorporate such upper bounds to technical progress into our framework, we modify our model for both hardware and software efficiency levels, \(H(t)\) and \(S(t)\). Specifically, we introduce an additional free parameter to the laws of motion for the efficiency variables that allows these efficiency terms to approach some maximum achievable level. This modification provides substantial flexibility for conducting robustness checks of the model. It also enables the incorporation of different sets of beliefs about the ultimate limits of hardware and software efficiency improvements.

To take into account technological limits, weincorporate a function that bounds potential efficiency as these terms approach their theoretical maxima. In continuous time, the laws of motion for hardware and software become:

\begin{equation}
    \frac{\dot{H}(t)}{H(t)} = g_H(t) \, \Lambda_H(H(t)),
    \qquad\qquad
    \frac{\dot{S}(t)}{S(t)} = g_S(t) \, \Lambda_S(S(t))
\end{equation}

Here, $g_H(t)$ and $g_S(t)$ capture the underlying growth potential for hardware and software at time $t$, respectively, i.e. the growth rates in hardware and software in the absence of technological constraints (i.e. the growth rates given by equation \eqref{eq:r_d_equation}). The functions $\Lambda_H(H(t))$ and $\Lambda_s(S(t))$ take values strictly between $0$ and $1$, and approach $0$ as $H(t)$ or $S(t)$ approach their terminal (maximum) values, which we denote $H^\text{max}$ and $S^\text{max}$ respectively. In particular, we define 
\begin{equation}
    \Lambda_H(H(t)) = \frac{\log H^\text{max} - \log H(t)}{\log H^\text{max} - \log H(0)}, \qquad\qquad \Lambda_S(S(t)) = \frac{\log S^\text{max} - \log S(t)}{\log S^\text{max}-\log S(0)},
\end{equation}
thus the growth rates of hardware and software efficiency diminish as they approach their respective theoretical limits.

\subsection{Frictions Affecting Investments in Compute}
\label{sec:frictions}

At the macroeconomic level, the expansion of compute capacity faces three distinct types of frictions that affect both the speed and cost of growth that we incorporate into our model:
\begin{itemize}
    \item Investment adjustment costs: Rapidly scaling compute hardware investment encounters non-linear costs due to supply chain and production bottlenecks. These costs grow super-linearly with the rate of investment.
    \item Depreciation: Compute hardware depreciates through both physical degradation and technological obsolescence at a substantially faster rate than traditional capital.
    \item Physical limits: Fundamental physical constraints, particularly physical heat dissipation limits, impose fundamental bounds on maximum usable compute, regardless of investment level.
\end{itemize}
We describe each of these frictions below.

\paragraph{Adjustment costs}

Expanding the compute stock involves more than the direct financial outlay for new hardware; it also entails overcoming production bottlenecks and supply chain limitations. We model these challenges using adjustment costs that increase non-linearly with the rate of investment in compute stock. Specifically, we assume that the cost of increasing the compute stock \( Q(t) \) by an amount \( I_q(t) \) is given by:

\begin{equation}
    I_Q(t) = \frac{Q(t)}{\chi a_Q H(t)} \left( \left( \frac{a_Q I_q(t) H(t)}{Q(t)} + 1 \right)^{\chi} - 1 \right),
    \label{eq:compute_adjustment_cost}
\end{equation}

where \( I_Q(t) \) is the total investment required, \( a_Q \) is an adjustment cost parameter, and \( \chi > 1 \) captures the increasing marginal costs associated with rapid expansion. This reflects the reality that significant increases in compute capacity cannot be achieved instantaneously or without incurring disproportionately higher costs.

The parameter \( \chi \) represents the degree of non-linearity in adjustment costs. Empirical observations suggest that expanding compute capacity faces steeper challenges compared to traditional capital investments. For instance, constructing semiconductor fabrication plants (fabs) involves substantial capital expenditure and long lead times, often taking 1.6 to 2.2 years from construction to production \citep{verwey2021permits}. Additionally, the supply chain for advanced AI hardware depends on specialized equipment and materials that are difficult to scale quickly. As a result, we consider values of \( \chi \) greater than 2, indicating that costs increase more than quadratically with the rate of investment.

\paragraph{Depreciation of compute stock}

Compute hardware depreciates more rapidly than general capital due to technological obsolescence and hardware failures. Rapid advancements in hardware efficiency render older equipment less competitive, while physical wear and tear reduces operational reliability. To model depreciation, we assume that the compute stock \( Q(t) \) decays at a constant rate \(\delta_Q\), yielding:
\begin{equation}
    \dot{Q}(t) = I_q(t) H(t) - \delta_Q Q(t),
\end{equation}
where \( I_q(t) \) is the rate of investment in new compute stock and \(\delta_Q\) is the depreciation rate. Empirical evidence suggests that compute hardware depreciates at a faster rate than standard capital, with estimates around 30\% per year \citep{ostrouchov2020gpu}. This high rate reflects both the physical degradation of components and the rapid pace of technological innovation that can render existing hardware obsolete within just a few years.

\paragraph{Physical limits and heating bottlenecks}

Physical constraints impose fundamental limits on the expansion of effective compute. One significant constraint is the issue of heat dissipation. As computational processes consume energy, they generate heat, and there is a limit to how much heat can be dissipated without causing environmental harm or damaging the hardware itself.

We incorporate this constraint by introducing a function \( g(Q) \) that adjusts the usable compute based on physical limits:
\begin{equation}
    \tilde{Q}(t) = g(Q(t)),
\end{equation}
where \( \tilde{Q}(t) \) is the \emph{usable} compute available for AI at time \( t \) after accounting for physical constraints. Specifically, it is \(\tilde{Q}(t)\)---rather than the raw stock \(Q(t)\)---that is ultimately allocated to training and inference.

The function \( g(Q) \) is chosen to smoothly cap usable compute at \( C_L \): for small \(Q\), \(g(Q)\approx Q\), but as \(Q\) grows large, \(g(Q)\) saturates near \( C_L \). This reflects the physical reality that beyond a certain heat-dissipation limit, additional hardware investment yields little extra usable compute.\footnote{We set $C_L = 2.0 \times 10^{38}$ FLOP/year based on theoretical maximum CMOS efficiency (4 $\times 10^{15}$ FLOP/J \citep{ho2023limits}) multiplied by 1\% of Earth’s annual solar energy budget, providing a physically-grounded upper bound on sustainable computation.} Specifically, we model \( g(Q) \) as:
\begin{equation}
    g(Q) = \frac{Q}{\frac{Q}{C_L} + 1},
\end{equation}
ensuring \( g(Q) \to C_L \) for large \( Q \). This captures how environmental constraints ultimately limit compute expansion. These physical limits are informed by fundamental principles of thermodynamics and environmental capacity. For example, the Earth's ability to dissipate heat without significant temperature increases is finite, and large-scale computing operations could contribute to global heating.

\subsection{Taking Stock: From Investment to the Path of Effective Compute and Beyond}
\label{sec:taking_stock_AI_dev}

The main goal of Section~\ref{sec:compute_model_specification} was to fully characterize GATE's AI development module, which maps any given path of investment in AI specific inputs (with its three components, namely hardware investment, hardware R\&D investment and software R\&D investment) into a path of effective compute endowments over time. Moreover, the AI development module also laid out the problem of allocating effective compute resources at each point in time across training and inference uses, with allocations of effective compute to training determining the path of another key state variable of the model, namely the size of the largest training run.

In other words, the key goal of the AI development module was to describe the state update rules for the two main AI-related state variables of our model: effective compute and the size of the largest training run. In what follows, we summarize the discussion provided in the rest of Section~\ref{sec:compute_model_specification}, state the main results (i.e. the two state update rules) and clarify the economic intuition.

\paragraph{Law of motion for effective compute} Effective compute is the key input for the development and deployment of AI systems in GATE. Consequently, tracking the path of effective compute across time is critical for mapping out the path of AI automation and hence the downstream impact of AI on the economy.

Effective compute was modeled as a summary statistic of the computational resources available to society at each point in time. It aims to capture a notion of ``quality adjusted" compute that takes into account the evolution of the quality of both hardware and software over time. 

The process of effective compute accumulation sketched over the previous subsections embeds three different margins of compute accumulation: accumulation of hardware, improvements in hardware efficiency (as a result of hardware R\&D investment) and improvements in software efficiency (as a result of software R\&D investment). Importantly, hardware efficiency gains and software efficiency gains display a key asymmetry: while software efficiency gains apply to the entire stock of existing compute, hardware gains only apply to new hardware/new flows of compute.

Formally we think of the model being initialized at a certain time 0, with society having a given stock of effective compute at its disposal. This is given by:
\begin{equation}
    C(0) = \tilde{Q}(0) S(0)
\end{equation}
where we typically specify that \(S(0) = 1\), i.e. we measure algorithmic efficiency relative to the start of the simulation.
Subsequently, three state variables jointly determine the accumulation of effective compute: the usable compute stock $\tilde{Q}(t)$, hardware efficiency $H(t)$, and software efficiency $S(t)$. When the physical compute stock $Q(t)$ is sufficiently small compared to the thermal limit $C_L$,\footnote{%
  In practice, we use a saturating function $g\bigl(Q\bigr) = \tfrac{Q}{\tfrac{Q}{C_L}+1}$ 
  to capture heat dissipation constraints.
  When $Q \ll C_L$, we have $g(Q) \approx Q$, 
  making the simpler, linear form a good approximation. 
  In regimes where $Q$ approaches or exceeds $C_L$, 
  the law of motion must be adjusted to incorporate $\dot{C}(t) = g\bigl(Q(t)\bigr)\times S(t)$ 
  (plus appropriate product-rule terms).
}
it is reasonable to model the evolution of effective compute using the law of motion:
\begin{equation}
    \dot C(t)=C(t)\frac{\dot S(t)}{S(t)}-\delta_Q C(t) + I_q(t)H(t)S(t)
\end{equation}

Intuitively, software efficiency improvements apply to the entire stock of effective compute such that the growth rate of software efficiency applies to the entire stock of effective compute (RHS term 1), the depreciation of compute hardware leads to the loss of the entire amount of effective compute supported by that compute (i.e. the depreciation rate of compute and effective compute are the same - RHS term 2), and investments in compute augment the stock of effective compute in direct proportion to each period's hardware and software efficiency (RHS term 3).

\paragraph{Law of motion for the size of largest training run}

At each point in time, society  faces the problem of how to allocate its computing resources to training and inference, as explained in Section~\ref{sec:training_vs_inference_tradeoff}. In the context of GATE the compute allocation decision is assumed to be taken by a social planner aiming to maximize the net present value of utility flows for the representative household (see Section~\ref{sec:macro_model_specification}). 

The social planner chooses the amount of effective compute to allocate to training each period, which we denote $D(t)$ taking into account 4 considerations:
\begin{enumerate}
    \item the training vs inference tradeoff in determining the capability and inference costs of AI systems (see equation \eqref{eq:training_vs_inference_tradeoff} in Section~\ref{sec:training_vs_inference_tradeoff})
    \item the mapping between AI system technical performance and the extensive margin of automation (see equation \eqref{eq: automation function} in Section~\ref{sec:macro_model_specification})
    \item the mapping between model size and runtime effective compute requirements for each automated task (see equation \eqref{eq: runtime compute req} in Section~\ref{sec:intensive_margin_automation}).
    \item the economic benefits from automating tasks, pinned down by the aggregate production function (see equation \eqref{eq:final_goods_production_function} in Section~\ref{sec:macro_model_specification})
\end{enumerate}
Intuitively, items 1 to 3 above pin down the relative costs of extensive vs intensive margin automation, whereas item 4 pins down the relative benefits of extensive vs intensive margin automation of labor tasks. These considerations lead to the following law of motion governing the size of the largest training run:
\begin{equation*}
\dot C_{T}(t) = D(t)
\end{equation*}
This equation captures how the social planner's optimal allocation of compute to training $D(t)$ directly determines the growth rate of the largest AI training run over time.

\section{The AI automation module: Mapping Effective Compute to Automation}
\label{sec:compute_based_model}

This section explains how we model the automation of economic tasks by AI systems, outlining the link between AI investments and the resulting automation of human labor (we also call this component of the model the ``automation module"). We employ a compute-based framework to forecast AI automation, in which the amount of computation available to AI systems at each point in time determines their capabilities. We thus connect physical compute, hardware improvements, and algorithmic improvements to the fraction of tasks AI can automate and to the increase in the effective labor force brought about by AI for each automated task. 

Intuitively, as AI systems are trained on more effective compute, they can replicate a greater share of human labor if provided sufficient inference compute. Moreover, for each automated task (i.e. task that AI systems are able to perform at each point in time), allocating more runtime compute to systems performing that task can be thought of as creating a larger number of ``digital workers" assigned to completing that task. By linking training compute to automation, the automation module formalizes the key mechanism linking advancing AI technology to economic outcomes. For an economics audience, the key technical goal of this module is to clarify the mapping between a certain allocation of effective compute and the distribution of ``digital workers" across the space of labor tasks, taking into account both the extensive and the intensive margins of the AI automation process.

To fully specify the AI automation process, the exposition in this section is organized in three parts. We first focus on the modeling of the most salient aspect of the automation process: the extensive margin of automation or the process through which increased resources dedicated to AI leads to more advanced systems that are capable of completing a wider and wider fraction of labor tasks. This is done in Section~\ref{sec:extensive_margin_automation}. We then proceed to characterizing the intensive margin of automation, or the mapping between allocations of runtime compute to automated tasks and the supply of ``digital workers" brought to bear in delivering these tasks (Section~\ref{sec:intensive_margin_automation}). Section~\ref{sec:labor_reallocation} discusses the issue of human labor reallocation in response to AI automation, though this typically has modest implications for the overall path of output growth associated with AI automation.

\subsection{Task Automation: The Extensive Margin}\label{sec:extensive_margin_automation}

Perhaps the most salient mechanism through which AI development affects macroeconomic outcomes in GATE is via the gradual automation of labor tasks. Intuitively, as the effective compute endowments of society increase, we can expect more capable AI systems to be trained. In turn, these systems
are able to complete an increasing share of tasks currently performed by humans. We call this process the extensive margin of AI automation. Moreover, at the same time, increased effective compute resources mean that more compute can be allocated to runtime compute, such that more instances of the models can be operated at the same time. We think of this as equivalent to running more ``digital workers" in parallel, and refer to this process of running a larger number of instances of AI models as the intensive margin of AI automation.

In this section, we characterize the process of AI automation on the extensive margin within GATE. Following much of the existing literature, we model the space of labor tasks as a continuum of unit measure. We then model the extensive margin of the automation process as a mapping between the compute-based capability measure $C_{T+\iota}(t)$ (i.e., the theoretical AI capability derived from the largest training run $C_T(t)$ enhanced by inference scaling) and the fraction of labor tasks that can be automated at each point in time, which we denote $f(t)$. Our approach to specifying this process is inspired by \cite{davidson2023} and is similar to \cite{korinek2024scenarios}. For a detailed justification of our compute-based approach to modeling AI capability development and task automation, we refer readers to Appendix~\ref{sec:compute-based}.

In line with previous work, we model the automation function $f$ as a smooth function of effective compute. This means that as $C_{T+\iota}(t)$ increases over time, the share of tasks in the economy that AI can perform grows gradually and continuously. This process continues until AI capabilities reach parity with human capabilities and all economically relevant tasks become automated.

Both our model and those of Davidson and Korinek and Suh share a core assumption: AI systems trained with more compute become capable of automating increasingly complex tasks. This creates a direct link between advances in computing power and the expansion of AI's economic impact through task automation.

\begin{figure}[h!]
    \centering
\includegraphics[width=0.55\linewidth]{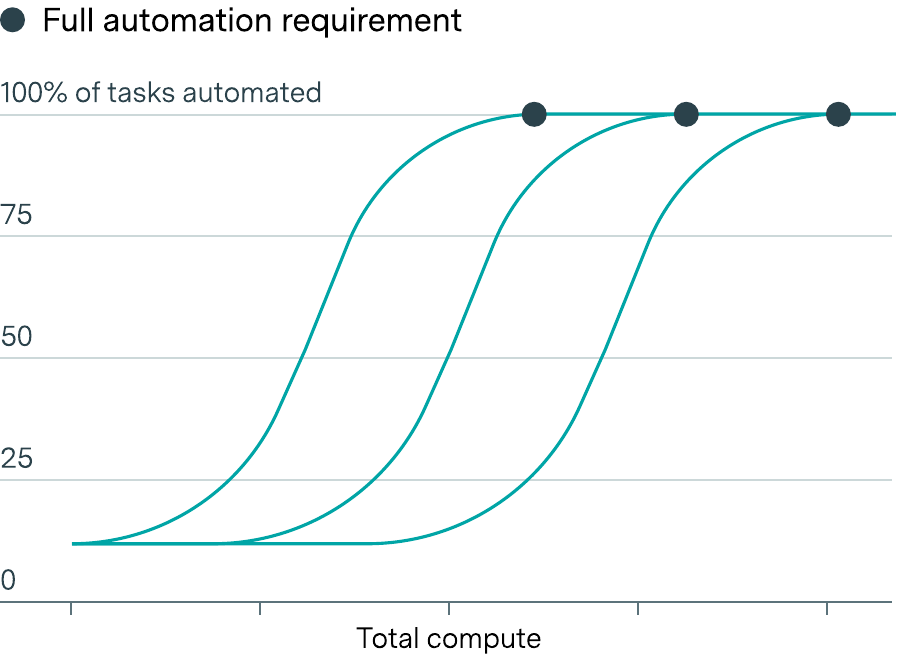}
    \caption{Relationship between total compute and task automation. Each curve represents a scenario where increasing compute gradually automates a greater fraction of tasks, with full automation achieved at varying levels of total compute. The dots indicate the points at which 100\% of tasks are automated for each scenario.}
    \label{fig:enter-label}
\end{figure}

To formalize the process of AI automation we opt for a simple piece-wise log-linear function that links effective compute to the fraction of labor tasks that can be automated. This function has three key parameters. The first parameter is the initial fraction of automated tasks ($f_\text{init}$).\footnote{We calibrate $f_\text{init}$ by matching today’s observed ratio of compute spending to total output. Specifically, we fix all other relevant parameters (e.g., elasticity of substitution, capital stock) at their baseline values, then solve for the fraction of tasks automated ($f_\text{init}$) that makes the model’s equilibrium spending on compute align with the current actual share of GDP devoted to computing resources.} This parameter initializes the share of tasks automated at the start of simulations and aims to capture the fact that the current level of automation is already non-zero. The second is the compute required for full automation ($T$). More concretely, $T$ represents the total effective compute (in FLOP) needed to train AI systems capable of performing essentially all tasks in the economy, in line with the “compute-based” perspective discussed in Appendix~\ref{sec:compute-based}. The third is the ``FLOP gap'' $(\Delta_\text{FLOP})$, which governs how much compute (relative to $T$) is needed before systems begin automating tasks above their initial level of $\,f_\text{init}$.
 
Formally, we map the relationship between effective compute $ C_{T+\iota}(t)$ and the fraction of tasks automated via the function $f : \mathbb R^{\geq 0} \to [0, 1]$ given by:
 \begin{equation}
    f(C_{T+\iota}(t)) =
    \begin{cases} 
        f_\text{init}, & \text{if } C_{T+\iota}(t) \leq \frac{T}{10^{\Delta_\text{FLOP}}} \\
        f_\text{init} + (1 - f_\text{init}) \frac{\log C_{T+\iota}(t) - \log \frac{T}{10^{\Delta_\text{FLOP}}}}{\log T - \log \frac{T}{10^{\Delta_\text{FLOP}}}}, & \text{if } \frac{T}{10^{\Delta_\text{FLOP}}} < C_{T+\iota}(t) \leq T \\
        1, & \text{if } C_{T+\iota}(t) > T
    \end{cases}\label{eq: automation function}
\end{equation}
It is important to note that while we use this simple parametric functional form for tractability, our model's framework can accommodate any arbitrary function mapping compute to automation levels.

With this type specification in place, the key feature of economic interest is the slope of the automation function $f$. This represents how much an increment of effective compute results in additional automation at each point along the automation path. To pin down this slope, we adopt the ``FLOP gap'' (denoted $\Delta_\text{FLOP}$) concept introduced by \cite{davidson2023}. This parameter quantifies the log-space difference between two critical points: the effective compute required for full task automation $(T)$, and the threshold $\tfrac{T}{10^{\Delta_\text{FLOP}}}$ at which AI systems begin to ``significantly impact''\footnote{%
By ``significantly impact,'' we mean that the fraction of tasks automated starts to rise materially above $f_\text{init}$. One could, for instance, interpret this threshold as the point at which 10\% or 20\% of tasks become automated, rather than waiting until it is strictly zero.
} the global economy. Because there is substantial uncertainty around this gap, and hence around how quickly automation picks up once we exceed $\tfrac{T}{10^{\Delta_\text{FLOP}}}$, we allow users to sample from a wide range of plausible values for $\Delta_\text{FLOP}$. In practice, a larger FLOP gap implies a more gradual onset of automation, while a smaller gap implies that once we pass the initial threshold, automation accelerates relatively quickly.

Intuitively, as more resources are allocated to AI over time, the size of the largest training run grows,\footnote{Assuming that some portion of effective compute is consistently allocated to training.} thereby expanding the set of tasks AI can automate (see Section~\ref{sec:training_vs_inference_tradeoff}). With a larger trained model, the social planner can deploy AI systems on an increasing share of tasks, provided that the benefits of automation exceed its costs. These benefits depend on factors such as the slope of the automation function, the substitutability of different tasks, and labor market frictions (detailed in Section~\ref{sec:macro_model_specification}). Consequently, we expect the fraction of automated tasks to rise gradually over time, tracing out an extensive margin of AI-driven automation.

We recognize that the specification of the ``labor task space" and of the automation process are simplified in our setting, with key aspects of heterogeneity between tasks documented in the previous literature (e.g., cognitive vs. manual tasks, routine vs. non-routine tasks) and the possibility of new labor tasks being created during the automation process both currently omitted from our analysis. We hope some of these features and extensions will be added in subsequent work.

\subsection{Task Automation: The Intensive Margin}\label{sec:intensive_margin_automation}

A second key component of the automation process is the intensive margin. Intuitively, this captures the answer to the question: ``When a task is automated, by what order of magnitude do we expect the effective supply of labor for that task to increase?''

As mentioned above, AI models have two types of computational demands. The first is training compute, which is the computation required to learn from data and update a model's weights to implement the algorithms required to perform various tasks. The second is inference or runtime demands. This is the computation required to run an AI system once it has been created. The runtime compute demands vary by task, with ``more challenging'' tasks typically requiring models that use both more training compute and runtime compute.

To model intensive margin automation, we create the notion of ``digital workers'' or ``digital worker equivalents,'' which captures the increase in the task-level labor supply associated with a certain allocation of computing resources to the delivery of each task in an intuitive way: ``how many human workers would we need to perform the same amount of each task as an AI system of a certain size and allocated a certain amount of runtime compute?''

We compute the number of ``digital workers'' allocated to each task ($N_i$) by taking the ratio of the runtime compute allocated to that task and the runtime compute requirements for the task. Formally, if $i \le f(t)$ denotes the set of automated tasks at time $t$, we write:
\begin{equation}
    N_i(t)\;=\;\frac{C_{I,i}(t)}{R_i(t)},\label{eq: number of digital workers}
\end{equation}
where $C_{I,i}(t)$ is the total runtime computation allocated to task $i$ at time $t$, and $R_i(t)$ is the corresponding runtime compute requirement. Both are measured in effective floating point operations (eFLOP) per unit time.

Recall that $f(C_{T+\iota}(t))$ is the function mapping effective compute to the fraction of tasks automated at time $t$. In other words, if $f(C_{T+\iota}(t)) = x$, then tasks in the interval $[0,x]$ can be automated. Since $i$ here refers to the task index in $[0,1]$, small values of $i$ represent relatively simpler tasks and larger values of $i$ represent more complex tasks.

We model $R_i(t)$ according to
\begin{equation}\label{eq: runtime compute req}
    R_i(t) = \underbrace{\max \left(1, \frac{f^{-1}(i)}{C_T(t)} \right)^m}_{\text{Task-specific inference multiplier}} \times \underbrace{10^{\gamma_0 + \gamma_1 i}}_{\text{Minimum runtime cost}},
\end{equation}
where \( i \) is the task index, and \( f^{-1} \) is the inverse of \( f \). Intuitively, if \( f(C) \) gives the fraction of tasks automated by compute level \( C \), then \( f^{-1}(i) \) tells us how much effective training compute is needed to automate a task indexed by \( i \) at minimal inference cost. Meanwhile, \(C_T(t)\) is the largest training run at time \(t\), \(m\) is the slope from \eqref{eq:training_vs_inference_tradeoff}, and \(\gamma_0,\gamma_1\) determine the baseline and slope of the task‐level runtime requirements in orders of magnitude.

It is important to note the distinction between the ``inference multiplier" $\iota$ (which features in equation \eqref{eq:training_vs_inference_tradeoff})  and the task-specific inference multiplier introduced in equation \eqref{eq: runtime compute req} above. While \eqref{eq:training_vs_inference_tradeoff} introduces \(\iota\) as a global scale factor that raises system capability from \(C_T(t)\) to \(C_{T+\iota}(t)\) and thus permits the automation of a higher fraction of tasks, the task specific inference multiplier captured by expression \(\max \Bigl(1, \frac{f^{-1}(i)}{C_T(t)}\Bigr)^m\) applies individually to each task, and captures the fact that tasks with compute requirements greater than $C_T(t)$ require additional inference costs to be incurred. Since tasks demanding more than $\iota_{\max}^{1/m} \times C_T(t)$ are infeasible, the maximum chosen per‐task multiplier equals $\iota_{\max}$.

This specification of runtime compute requirements, combined with our earlier definition of how the fraction of automated tasks is computed, has the following effects:\footnote{The one wrinkle in the definition above is the inverse \( f^{-1} \): it must be defined carefully because the function \( f: \mathbb R^{\geq 0} \to [0, 1] \) is not necessarily injective or surjective, though it is guaranteed to be increasing. To ensure that the inverse is a well-defined and single-valued function \( [0, 1] 
\to \mathbb R^{\geq 0} \), suppose that the infimum and supremum of the image of \( f \) are given by \( f_{\text{inf}} \) and \( f_{\text{sup}} \) respectively, and that \( f \) has a compact image so that it attains both its infimum and supremum. We make the following definition:

\begin{equation}
    f^{-1}(i) = \begin{cases}
        0 & \text{if } i < f_{\text{inf}} \\
        \inf \{ c \in \mathbb R^{\geq 0} : f(c) = i \} & \text{if } f_{\text{inf}} \leq i \leq f_{\text{sup}} \\
        \infty & \text{if } i > f_{\text{sup}}
    \end{cases}
\end{equation}

This is well-defined under the assumptions that \( f \) is increasing and has compact image, both of which are satisfied for the functions we use in this work. The rough intuition is that \( f^{-1}(i) \) tells us how much effective training compute is required to perform task \( i \) at its minimal possible inference cost.}

\begin{itemize}
    \item Tasks which require less than \( C_T(t) \) effective compute to automate are done at the minimum possible inference cost of \( 10^{\gamma_0 + \gamma_1 i} \) which increases exponentially with the task index (which reflects task complexity) \( i \).

    \item Tasks that require between \( C_T(t) \) and \( C_{T+\iota}(t) \) of effective compute to automate, i.e. tasks that can only be automated by scaling up inference, are done at an inference cost penalty scaling with the ratio of the task's training compute requirements \( f^{-1}(i) \) to the size of the largest training run \( C_T(t) \).

    \item Tasks that require more than \( C_{T+\iota}(t) \) compute to automate cannot be automated at all, as their execution is infeasible for the frontier models even with the global inference multiplier $\iota$ set at its maximum.
\end{itemize}

Equations \eqref{eq: runtime compute req} and \eqref{eq:training_vs_inference_tradeoff} capture the training vs inference tradeoff outlined earlier in the paper (see Section~\ref{sec:training_vs_inference_tradeoff}). Larger models have larger fixed costs in terms of effective compute, as more compute needs to be allocated to training them and hence cannot be used as runtime compute for ``digital workers". On the other hand, larger models will make it cheaper to create ``digital workers'' in terms of runtime compute requirements, particularly for more complex tasks. This is because a smaller model that does not meet a task’s training threshold must rely heavily on inference compute, while a model that does meet that threshold requires proportionally fewer inference resources. As a result, smaller models tend to deliver ``digital workers'' for complex tasks at higher variable cost in terms of effective compute relative to larger models.\footnote{Note that the lower bound of the inference multipliers $\iota_\text{min}$ creates the possibility that when running larger models, the planner might be forced by technological constraints to assign overcapable models to simple tasks. In other words, in the presence of lower bounds for the inference multiplier, it may sometimes be cheaper to execute simpler tasks with smaller rather than larger models. In our modeling we ignore this possibility and assume that all automated tasks are executed at each point in time using the frontier models. We expect this assumption to have very modest quantitative implications.}

All in all, the intuition behind the AI automation process described in the current and the previous section can be summarized as follows. As society's effective compute resources increase as a result of investments in hardware and AI related R\&D (both hardware and software R\&D), we expect more capable AI models to be trained (as more and more compute is allocated to training) and more digital workers to be created for each automated task, i.e.\ for each task that frontier AI models are able to perform (as more and more runtime compute is allocated to each automated task). The model features we described specify the mapping between any allocation of compute and distributions of ``digital workers'' across ``task space.'' Over time, we expect these distributions of digital workers to ``broaden'' over task space (i.e.\ have broader support over the unit interval of tasks) as well as increase in ``height'' (i.e.\ for each automated task we expect the number of ``digital workers'' made available by the automation process to increase over time).

\subsection{Labor Reallocation}\label{sec:labor_reallocation}

A third key component of the automation process, is the process of reallocation of human labor from automated to yet-to-be automated tasks. Intuitively, as AI automation proceeds, the automation process can be expected to result in a significant increase in effective labor supply for automated tasks (more on this in Section~\ref{sec:macro_model_specification}). In a setting in which labor tasks are assumed to be complements in production (as we assume in GATE, see Section~\ref{sec:macro_model_specification}) a potentially significant determinant of the macroeconomic impacts of AI automation will be given by the extent to which the economy is ``bottlenecked" by the non-automated tasks, in which effective labor supply can be expected to increase at a much slower rate than in automated tasks.

In turn, the strength of this bottlenecking phenomenon depends on two key factors. One is the elasticity of substitution between labor tasks. The lower this parameter, the less substitutability there is between tasks and the stronger the bottleneck effect exerted by non-automated tasks is. The second key determinant is the extent to which labor can quickly reallocate from automated to non-automated tasks. Intuitively, when labor reallocates quickly to non-automated tasks this pushes up effective labor supply for these tasks and weakens their bottlenecking effect on overall output. By contrast, if labor is sluggish to reallocate to non-automated tasks this tend to amplify the break on overall economic growth exerted by the remaining non-automated tasks. It is this process of labor reallocation that is the key focus of this section.

We model the process of labor reallocation in response to AI automation in a simplified way, by considering two polar extreme cases.\footnote{In subsequent work, we plan to extend the framework here to allow for more realistic retraining and reallocation of workers} In the first case, which is the default setting of our model playground, we assume human labor can be seamlessly reallocated across tasks (and hence from automated to non-automated tasks). Hence at each point in time, the social planner is free to reallocate the entirety of the human labor endowment across the (ever-decreasing) set of non-automated tasks, thus increasing effective human labor supply for each of these tasks. Formally in this specification, we only need to keep track of one state variable, the labor force, whose law of motion is given by
\begin{equation}
L(t)=L(0) e^{g_L t}
\end{equation}
The frictionless reallocation specification is one in which the strength of the bottlenecks generated by the non-automated tasks is relatively weak and could thus present a relatively aggressive estimate of the pace of the economic impact of AI. 

In the polar opposite case, we impose the assumption of no labor reallocation in response to AI automation. In effect, workers are assumed to be ``born" with human capital that is specific to a single task, and they cannot be reallocated to any other task if their task is automated. In other words the number of human workers engaged in each task is exogenous and given by the initial conditions of the model and the exogenous rate of population growth. Formally we have a law of motion for each task:
\begin{equation}
L_i(t)=L_i(0) e^{g_Lt}
\end{equation}
However, given that we assume that tasks enter symmetrically in the production function, tasks start with equal labor endowments and the labor supply for each task exogenously grows at the same rate (the overall population growth rate), in the solver keeping track of only one state variable (the total population) will be sufficient.
The no reallocation case embeds an extremely conservative assumption (i.e. the bottlenecks from non-automated tasks are set at their maximum). This should allow GATE users to explore the worst case scenario from the perspective of labor reallocation. In future releases of GATE we envision allowing for intermediate cases of labor reallocation frictions (e.g. cases with incomplete labor reallocation in response to AI automation).\footnote{Note that allowing for such intermediate cases significantly complicates the decision problem of the social planner, as the labor allocation decisions and the AI investment decisions become dynamically interdependent.}

\section{The Macroeconomics Module: Translating Automation into Economic Output}
\label{sec:macro_model_specification}

We complete the description of the model by outlining its final module: the macroeconomic module. This module specifies two key elements of the model:
\begin{enumerate}
    \item The economy's production technology, which links labor tasks, as well as other production inputs to output. This element is key in tracking the impact of AI automation on output.
    \item The decision problem of the social planner, that drives the path of investment in compute, hardware R\&D, software R\&D and investment in non-compute capital, as well as the optimal allocation of effective compute across its different uses. This element clarifies how the levels of output and the incentives faced by the representative agent feed-back into investment in AI and non-AI inputs, thus driving further advances in AI capabilities.
\end{enumerate}
Our exposition in this section is split into two parts. In the first, we outline the basic setup of GATE's macroeconomic module. In the second, on account of the complexity of the model, we provide a review of the update rules affecting all of the model's key state variables.

\subsection{Basic Setup}\label{sec:macro_model_basic_setup}

\paragraph{Social planner's problem}

For conceptual and computational simplicity, GATE is set up as a social planner's problem, which provides a benchmark of the potential evolution of the economy under minimal market frictions and distortions.\footnote{Beyond this benchmarks, more realistic simulations may be run either by imposing constraints/ introducing wedges into the social planner's problem, as we do in Section~\ref{sec:further_ingredients} or by explicitly modeling market structures in key markets and solving the market allocation problem. We plan to pursue both of these areas in subsequent work.} Perhaps the most straightforward way to think of our macroeconomic setting is as describing a social planner's problem in an otherwise standard Ramsey-Cass-Koopmans setting that incorporates AI-driven automation.
 
 The social planner chooses:
 \begin{enumerate}
     \item a sequence of allocations of output (i.e. one allocation at each point in time for an indefinite time horizon) into consumption, investment in compute hardware, investment in hardware R\&D, investment in software R\&D and investment in non-compute physical capital
     \item the fraction of tasks to be automated, and a sequence of allocations of effective compute (one allocation at each point in time for an indefinite horizon) across training compute and runtime compute used for the delivery of each automated task.
     \item a sequence of allocations of the available human labor across tasks
 \end{enumerate} 
 to maximize the net present value of intertemporal utility for a representative household given by:
\begin{equation}
    \label{eq:objective-function}
    \max_{\{\mathbf{x}_t\}}\int_{t=0}^{\infty} e^{-(\beta-g_L) t} \, \, \frac{c(t)^{1-\eta} - 1}{1-\eta} \: dt,
\end{equation}
where \( c(t) \) is consumption at time $t$, $\{\mathbf{x}_t\}$ is the vector of choice variables at the social planner's disposal, $\beta$ is a time discount rate parameter and $\eta$ is the coefficient of relative risk aversion. As usual, taking $\eta = 1$ corresponds to choosing log utility. 

The choice variable vector $\{\mathbf{x}_t\}=\{c(t),I_Q(t),I_H^{RD}(t),I_S^{RD}(t),I_K(t),D(t),C_{I,i}(t), L_i(t)\}$ contains the following variables: $c(t)$ - consumption, $I_Q(t)$ - investment in compute hardware, $I_H^{RD}(t)$ - investment in hardware R\&D, $I_S^{RD}(t)$ - investment in software R\&D, $I_K(t)$ - investment in non-compute physical capital, $D(t)$ - (effective) compute spent on training, $C_{I,i}(t)$ effective compute spent on runtime compute for automating task $i$, $L_i(t)$ - human workers allocated to task $i$ .

When solving the welfare maximization outlined problem above, the social planner faces the following set of resource constraints:
\begin{align}
    &L_i(t), \, C_{I,i}(t),\,I_Q(t), \, I_H^{RD}(t), \, I_S^{RD}(t), \, I_K(t), \, D(t)\geq 0\\
    &C_{I,i}(t) = 0 \text{ for } i > f(t)\\
    &\int_0^1 L_i(t) \, di = L(t)\label{eq: human labor mkt clearing}\\
    &D(t)+\int_0^1 C_{I,i}(t) \leq C(t) \label{eq: compute budget constraint}\\
    &L(t)c(t)+I_Q(t)+I_H^{RD}(t)+I_S^{RD}(t)+I_K(t)\leq Y(t)\label{eq: output budget constraint}
\end{align}

where \(C(t)\) is total endowment of effective compute at time $t$, $L(t)$ is the total available human labor at time $t$, $f(t)$ is the fraction of automated tasks at time $t$ and $Y(t)$ is total output. Intuitively, at each point in time, the social planner takes the state variables as given\footnote{The list of state variables and their update rules are reviewed in Section~\ref{sec:state-update-rules}}, and chooses the allocation of compute across training and runtime compute, the allocation of runtime compute to automated tasks, and the allocation of labor to tasks in order to maximize output. Conditional on period-by-period output maximization, the social planner faces an otherwise standard dynamic macro problem: allocating output between consumption and investment. The only difference between our setting and plain vanilla settings is that the planner faces a richer menu of investment options, being able to invest not only in the accumulation of physical capital but also in the accumulation of compute, hardware R\&D and software R\&D.

\paragraph{Technology}

We adopt a slightly modified version of a standard neoclassical production function to model the production side of the economy, that allows us to incorporate AI-driven automation into the analysis. Our key aim is to capture the effects of the gradual automation of labor tasks by AI systems (i.e. the gradual substitution of human labor by AI systems across an increasing range of tasks) while allowing for production bottlenecks generated by tasks that remain unautomated. Specifically, we model output the production process via the Cobb-Douglas technology:
\begin{equation}
  Y(t) = A(t) T(t)^{1 - \alpha-\mu} K(t)^{\alpha} F(t)^\mu,  \,\text{where} \,\alpha \in (0,1).
\end{equation}
where, $Y(t)$ represents total output at time $t$, $A(t)$ denotes the exogenous stock of technology or total factor productivity (TFP) at time $t$, $T(t)$ represents a composite of labor tasks (further described below), $K(t)$ represents the stock of physical capital at time $t$, while \(F(t)\) represents some non-accumulable factor that is considered to be in fixed supply (this can be thought of as land or natural resources). The parameter $\alpha$ determines the output elasticity of capital, while $\mu$ determines the output elasticity of the non-accumulable factor. The somewhat non-standard factor of production $F$ is added to our framework to allow users to specify a role for non-accumulable factors of production even in advanced stages of automation. In departure from standard macroeconomic models we do not model TFP growth (i.e. we assume $A(t)$ is constant) as the model is focused on the role effects of AI automation on the extensive and intensive margins. This issue is discussed further in Section~\ref{sec:limitations}.

The key departure from standard production models is the introduction of $T(t)$, which represents a composite of labor tasks and can be thought of as the stock of effective labor input. This term includes the capacity of the economy to perform labor tasks, that when combined with other inputs (e.g. accumulable and non-accumulable capital) result in the production of final output. Importantly, this tasks composite captures the tasks produced at any one time by both human labor and ``digital workers", i.e., AI systems that can perform tasks traditionally done by humans.

The model assumes that there is a fixed (we assume unit measure) continuum of tasks in the economy and that the effective aggregate labor input $T(t)$ is a composite measure of these labor tasks. The aggregation is performed using a constant elasticity of substitution (CES) function, which allows for some degree of substitutability between tasks in production. The CES aggregator is assumed to have constant returns to scale, meaning that doubling all inputs leads to a doubling of output:
\begin{equation}
    T(t) = \left( \int_0^1  T_i(t)^{\rho} \, di \right)^{1/\rho} , \, \text{where} \, \rho \in (-\infty, 0).
\end{equation}
where $T_i(t)$ denotes the amount of task $i$ completed at time $t$ and $\rho=\frac{\sigma-1}{\sigma}$ is a substitution parameter (note: $\sigma$ represents the elasticity of substitution between tasks). As we are focused on modeling production bottlenecks we restrict attention to values of the substitution parameter (namely  $\rho \in (-\infty, 0)$) where tasks are assumed to be complements in production, and hence limits to the scale-up of non-automated tasks place significant limits on the expansion of the economy.

In turn, the amount of each task $i$ being performed at time $t$, $T_i(t)$, will be determined by the number of human and digital workers allocated to the task. Task inputs will hence be given by:
\begin{equation}
    T_i(t) = \begin{cases}
		L_i(t) + N_i(t), & \text{if $i \leq f(t)$}\\
            L_i(t), & \text{otherwise}
   \end{cases}
\end{equation}
where 
$L_i, \, N_i$ are the human and  AI labor allocations to task $i$ ($N_i(t)$ is given by equation \eqref{eq: number of digital workers} and depends on the planner's allocation of runtime compute across tasks), and $f(t)$ is the fraction of tasks automated at time $t$. As explained at length in Section~\ref{sec:compute_based_model}, the process of automation is the result of the gradual training and deployment of better AI systems, which results in both a higher fraction of tasks being automated and a larger number of ``digital workers" being brought to bear to deliver each automated task.

Putting the prior two equations together, the process of producing final goods can be expressed as follows:
\begin{equation}
    Y(t) =  A(t) \bigg( \underbrace{\int_0^{f(t)}  (L_i(t) + N_i(t))^{\rho}\, di}_{\text{automated tasks}} +  \underbrace{\int_{f(t)}^1  L_i(t)^{\rho}\, di}_{\text{Non-automated tasks}}  \bigg)^{(1-\alpha-\mu)/\rho} K(t)^\alpha F(t)^{\mu}\label{eq:final_goods_production_function}
\end{equation}
where the overall stock of human labor is assumed to grow exogenously at rate $g_L$.\footnote{The growth rate applies to the overall labor endowment or to the task level labor endowment, depending on whether the perfect labor mobility or the no labor mobility scenarios are selected by GATE users.}

As in standard growth models, investment is a key driver of growth dynamics. However, in our setting the social planner faces a much richer investment decision than in standard settings, as it is able to invest in compute, hardware R\&D, software R\&D and non-compute physical capital. The first three investment categories yield a gradual increase in society's effective compute resources. In turn, increased effective compute resources permit the training of more capable AI systems and the operation of more instances of these systems (i.e. automate more tasks and deploy more ``digital workers" per automated task). Intuitively, this process gradually turns labor from an non-accummulable to an accumulable factor of production, which coupled with more standard physical capital accummulation dynamics leads to gradual convergence to $AK$-style dynamics in our setting. Importantly, ``labor accumulation"\footnote{We employ the labor accumulation metaphor for two reasons. Firstly for simplicity, as it is easier to analyze than keeping track of additional factors of production and variable parameters governing the substitutability of different factors of production. Secondly, to emphasize what we believe is a key distinction between AI and previous automation technologies: the ability of AI to eventually automate all labor tasks currently performed by humans, leading to AI becoming a perfect substitute for all types of human labor once full automation occurs. This approach is of course of a metaphor, as AI development depends on processes of compute accumulation that are similar in nature to standard capital accumulation. Alternative modelling strategies, for instance modelling AI development as increasing the substitutability of human and AI labor at the level of each task would produce similar predictions.} is the key mechanism through which AI development affects output in GATE, as we abstract away from any effects of AI development on TFP, which is exogenously defined and kept constant over time.\footnote{In other words, the present release of GATE does not allow for the AI automation of research or other activities that affect output primarily via their effect on TFP}

\paragraph{Investment frictions}
As with compute, we assume that the process of investing in physical capital is affected by investment frictions. In line with our treatment of compute, we assume that expanding the capital stock by some amount \( I_k(t) \) has a larger cost  $I_K(t)$ given by:
\begin{equation}
    \label{eq:capital-adj-cost}
    I_K(t) = I_k(t) + \frac{a_K I_k(t)^2}{2K(t)},
\end{equation}
where \( a_K \) is an adjustment cost parameter that has dimensions of time (roughly corresponding to over what time frame we can double the capital stock without running into serious adjustment costs), \( I_{K}(t) \) is actual investment into capital at time \( t \), and \(I_k(t)\) is the effective investment after incorporating the adjustment cost. The implied update rule for the capital stock is then given by:
\begin{equation}
    \dot K(t)=I_k(t) - \delta_K K(t)
\end{equation}
where $\delta_K$ is the depreciation rate for non-compute physical capital.

\subsection{A Review of State Update Rules}
\label{sec:state-update-rules}

In Table \ref{tab:update_rules} we offer a review of the key state variables that collectively describe the resources available to the economy at each point in time. GATE is a dynamic model where the social planner makes optimal choices under resource constraints and takes into account how choices will affect the resource endowments (and further the utility levels) of future periods.

The state variables of the model include: labor $L(t)$, the physical capital stock $K(t)$, compute hardware stock $Q(t)$, hardware efficiency $H(t)$, software/ algorithmic efficiency $S(t)$, 
the size of the largest training run $C_{T}(t)$ and the size of the overall stock of effective compute $C(t)$. The evolution of each of these variables over time is described by a law of motion or a state update rule. These state update rules have been described over the course of the previous sections. However to provide an overview of the overall decision problem faced by the social planner, we provide a summary of these state update rules in the present section. Coupled with the descriptions of the planner's problem and technology in the previous section these should provide of a synthetic description of the decision environment faced by the welfare maximizing social planner.

\begin{table}[h!]
\small
\centering
\begin{tabular}{@{}ll@{}}
\toprule
\textbf{Variable} & \textbf{Update rule} \\
\midrule
\textbf{Labor} 
& \(\dot{L}(t)= g_L\,L(t)\) \quad (Perfect reallocation) \\
& \(\dot{L}_i(t)= g_L\,L_i(t)\) \quad (No reallocation) \\

\textbf{Capital stock} 
& \(\dot{K}(t) = I_k(t)-\delta_K\,K(t)\) \\

\textbf{Compute stock} 
& \(\dot{Q}(t) = I_q(t)\,H(t)-\delta_Q\,Q(t)\) \\

\textbf{Size largest training run} 
& \(\dot{C}_{T}(t)=D(t)\) \\

\textbf{Hardware efficiency} 
& \(\frac{1}{H(t)} \frac{dH(t)}{dt} = \theta_H [H(t)]^{-\phi_H} [I_H^{RD}(t)]^{\lambda_H}\) \\

\textbf{Software efficiency} 
& \(\frac{1}{S(t)} \frac{dS(t)}{dt} = \theta_S [S(t)]^{-\phi_S} [I_S^{RD}(t)]^{\lambda_S}\) \\

\textbf{Total effective compute} 
& \(\dot{C}(t)=C(t)\,\frac{\dot{S}(t)}{S(t)}-\delta_Q\,C(t) + I_q(t)\,H(t)\,S(t)\) \\
\bottomrule
\end{tabular}
\caption{\small Summary of update rules for state variables in the dynamic model. The labor update rule depends on whether perfect reallocation (where labor moves freely across tasks) or no reallocation (where workers remain in their initial task) is assumed. Laws of motion for Total effective compute apply to the regime when heating bottlenecks are sufficiently distant; otherwise, those need to be incorporated (see \Cref{sec:tech_constraints}). Constraints on hardware and software efficiency are omitted for clarity (see \Cref{sec:tech_constraints}).}
\label{tab:update_rules}
\end{table}

\section{Model Add-ons: R\&D Externalities and Uncertainty}
\label{sec:further_ingredients}

Beyond the three core modules described in sections \ref{sec:compute_model_specification} to \ref{sec:macro_model_specification}, GATE embeds two optional add-ons that can be turned on and off as needed by users. One of these add-ons allows the model to take into account the role played by incomplete internalization of positive externalities emerging from R\&D investments in shaping the plausible paths of AI automation; while the other allows for uncertainty regarding the process of AI automation to play a role in determining investment decisions related to AI development and deployment. In what follows, we describe these two add-ons in greater detail.

\subsection{Positive Externalities in R\&D}\label{sec:RD_externalities}

As we have aimed to avoid taking firm stances on the market structures of key markets affecting AI development, GATE has been set up as a social planner problem. We believe that this set-up is useful in describing how the production possibilities of the economy might expand as a result of optimal AI development, and may be a reasonable forecast of how the future may unfold under the assumption that key markets are likely to be competitive.

However, we believe there is one key feature of GATE that may result in our social planner set-up generating a substantial upward bias in the trajectory of the economy: the key role played by R\&D investments and R\&D spillovers in the path of AI capabilities and hence the overall path of the economy. In our standard set-up outlined above, the social planner fully internalizes the positive externalities associated with R\&D investment when making investment decisions. This can in principle lead to large differentials between the levels of investment chosen by the planner and the levels of investment that we may expect to see in a decentralized market where participants capture only a fraction of the benefits of R\&D. In turn, this could create a significant gap between the path of the economy under our planner set-up and the path one might expect in a realistic market equilibrium.

To mitigate this issue, we have integrated an add-on into GATE that introduces an ``R\&D wedge" between the social returns to R\&D and the actual effectiveness of R\&D investments.\footnote{Note that this same wedge applies to both hardware and software R\&D.} In essence, this module directly reduces the productivity of R\&D investments, simulating a scenario where only a fraction of the potential benefits are realized, similar to a decentralized market setting where positive spillovers from research typically go uninternalized by individual firms.

Formally, the model add-on specifies a new parameter, $\xi$ (the ``R\&D wedge''), that scales down the perceived marginal returns to each type of R\&D by modifying the investment terms. Concretely, if $\xi > 1$, the planner sees only $I_H^{RD}(t)/\xi$ and $I_S^{RD}(t)/\xi$ of the true impact of investing in hardware or software. As a result, the marginal effects of R\&D investments om hardware and software efficiency growth rates are given by
\begin{align}
\frac{\partial g_H}{\partial I_H^{RD}}
  &= \theta_H \,\lambda_H \,[H(t)]^{-\phi_H}\,\left(\frac{I_H^{RD}(t)}{\xi}\right)^{\lambda_H - 1}, \\
\frac{\partial g_S}{\partial I_S^{RD}}
  &= \theta_S \,\lambda_S \,[S(t)]^{-\phi_S}\,\left(\frac{I_S^{RD}(t)}{\xi}\right)^{\lambda_S - 1},
\end{align}
where $g_H$ and $g_S$ denote the growth rates of hardware and software efficiency respectively. Lowering the perceived returns to R\&D in this manner induces systematically lower R\&D investments, thereby producing a downward shift in the paths of $H(t)$ and $S(t)$ in GATE simulations.\footnote{It is important to note that we assume that there is no learning regarding R\&D externalities when the externality module is activated. The planner is essentially assumed to solve the model once and for all under the incorrect beliefs regarding the returns to R\&D and then to myopically implement that plan even as the returns to R\&D are revealed to be higher than expected.}

\subsection{Uncertainty About AI Automation}\label{sec:uncertainty_add-on}

Another drawback of the baseline specification of GATE is that we have not allowed for any uncertainty affecting the mapping between AI system technical characteristics and the set of tasks that the system is able to perform. In other words, we have assumed that the social planner is omniscient about the path of AI labor automation. When this assumption is relaxed, uncertainty about the relationship between compute investment and automation capabilities can significantly alter investment decisions and economic trajectories. In such conditions, uncertainty about future automation capabilities can lead to more cautious investment decisions, thereby depressing the overall level of investment in AI.

To address this limitation, we incorporate an uncertainty add-on to the model that allows users to run simulations under a variety of uncertainty scenarios affecting the AI automation process.

When the add-on is activated, the social planner is modeled as being uncertain about the path of AI automation and having beliefs about potential future automation paths. Formally, the social planner's beliefs about the amount of effective compute required for any given level of task automation are represented by a discrete probability distribution over a (finite) family of automation functions of the type described in equation \eqref{eq: automation function}. Let \( \mathcal F = \{f_1, f_2, \ldots, f_{N_{\mathcal F}}\} \) be a set of automation functions \( \mathbb R^{\geq 0} \to [0, 1] \) mapping effective training compute (measured in units of eFLOP) to a fraction of economic tasks that can be automated using a model with that amount of training compute. The belief distribution \( G: \mathcal F \rightarrow [0,1] \) assigns a probability to each \( f_i \in \mathcal F \), such that \(\sum_{i=1}^{N_{\mathcal F}} G(f_i) = 1\).

We restrict the set of automation functions that can be in the support of the planner's beliefs to automation functions of the type:\footnote{Because we are primarily interested in the economic effects of the transition to AGI, we preserve the assumption that the “true” automation function eventually achieves full automation. Given this choice, any other candidate function that deviates too soon would be immediately ruled out by the data, so we only admit alternative functions that are observationally equivalent initially but then plateau below full automation. We also include exactly one function that allows for full automation, because additional piecewise functions also reaching 100\% automation would soon receive probability 0 if they departed from the true shape.}
 \begin{equation}
    f_i(C_{T+\iota}(t)) =
    \begin{cases} 
        f_\text{init}, & \text{if } C_{T+\iota}(t) \leq \frac{T}{10^{\Delta_\text{FLOP}}} \\
        f_\text{init} + (1 - f_\text{init}) \frac{\log C_{T+\iota}(t) - \log \frac{T}{10^{\Delta_\text{FLOP}}}}{\log T - \log \frac{T}{10^{\Delta_\text{FLOP}}}}, & \text{if } \frac{T}{10^{\Delta_\text{FLOP}}} < C_{T+\iota}(t) \leq  \bar{f}^{-1}(\zeta_i)\\
        \zeta_i \leq 1, & \text{if } C_{T+\iota}(t) > \bar{f}^{-1}(\zeta_i)
    \end{cases}\label{eq: automation function uncertainty}
\end{equation}
where $\bar{f}^{-1}$ is the inverse of the ``true" automation function that allows for full automation and is specified exactly by equation \eqref{eq: automation function}, i.e it is the same as the automation function specified in the deterministic case. Thus to make use of the uncertainty add-on of GATE users need to specify:
\begin{itemize}
\item The number of functions to allow in the support of the beliefs of the social planner, i.e. $|\mathcal{F}|$.
    \item For each automation function, the maximum automation parameter \(\zeta_i\) that determines the highest feasible fraction of tasks automatable with that function.
    \item One function that allows for full automation must be included and fully specified in \(\mathcal{F}\).
\end{itemize}

\begin{figure}[h!]
    \centering
    \includegraphics[width=1\linewidth]{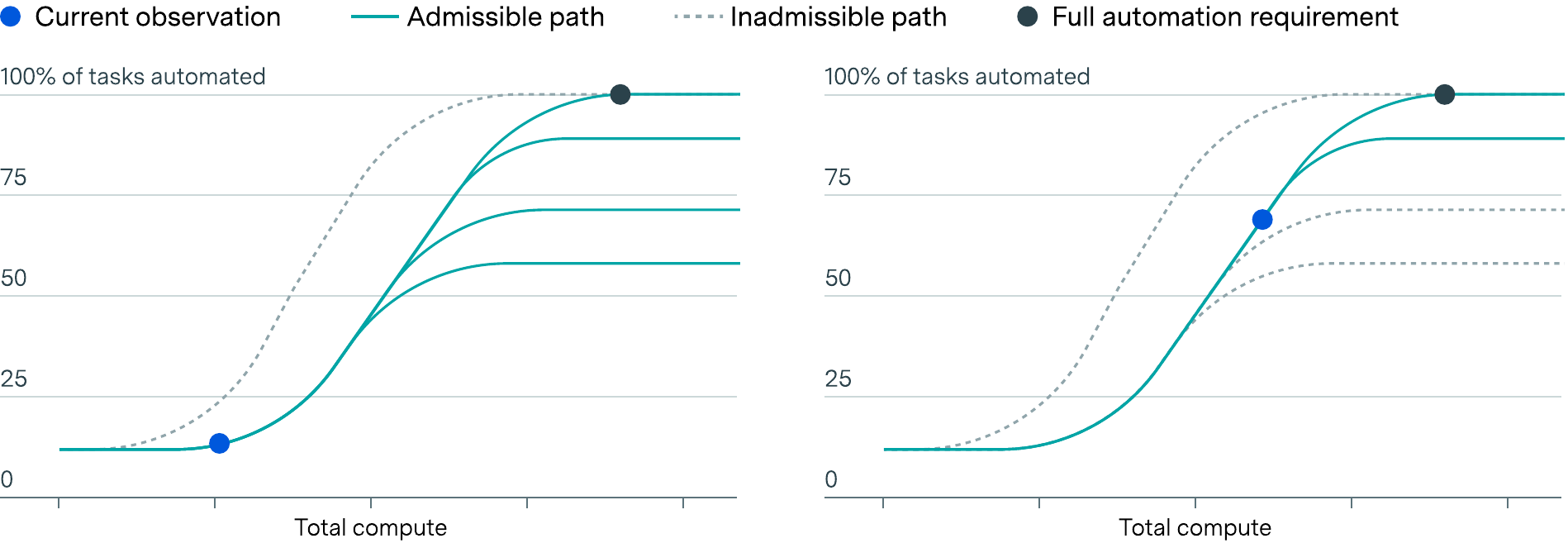}
    \caption{Belief updating after an observed compute-automation outcome. Solid lines represent admissible paths that remain consistent with the observation (black dot), while dashed lines are invalidated paths ruled out by the data. The left panel shows a larger reduction in uncertainty with more paths eliminated, while the right panel shows fewer paths ruled out, indicating a narrower range of uncertainty.}
    \label{fig:gate_automation_structure_uncertainty}
\end{figure}

Intuitively, the uncertainty add-on allows users to set social planner beliefs over a family of automation functions that are observationally equivalent up to a certain point in the unfolding of automation (all functions in the family have the same shape up to a point, and then the ``incorrect" ones taper off before full automation is achieved). As the automation process unfolds the social planner gradually learns the ``correct" automation function that allows for full automation. This is illustrated in figure \ref{fig:gate_automation_structure_uncertainty}.

Given the set of beliefs set by users as described above, at each point in time the social planner makes choices so as to maximize the expected net present value of utility flows accruing to the representative household conditional on the planner's beliefs, and taking into account the future path of the planner's beliefs under each scenario. Formally the planner solves:
\begin{equation}
    \max_{\mathbf{x}_t} \sum_{i=1}^{N_\mathcal{F}} G(f_i) \cdot U(\mathbf{x}_t, f_i)
\end{equation}

    where $\mathbf{x}_t$ represents the choice variables and \( U(\mathbf{x}_t, f_i) \) is the net present value of household utility achieved in the scenario with automation function \( f_i \) when the choices of the social planner are represented by $\mathbf{x}_t$. 
    
    As before, $\{\mathbf{x}_t\}$ is given by  $\{\mathbf{x}_t\}=\{c(t),I_Q(t),I_H^{RD}(t),I_S^{RD}(t),I_K(t),D(t),C_{I,i}(t), L_i(t)\}$

A key element of the automation uncertainty add-on is modeling the process of belief updating by the social planner along the AI automation path. These beliefs are assumed to be updated as follows:

The probability distribution over automation functions is updated based on observed compute levels and the observed fraction of tasks automated. Let \(\hat{C}\) be the observed compute level and \(\hat{f}\) be the observed fraction of tasks automated at this compute level.

The update mechanism then operates as follows: suppose that \( \bar f \in \mathcal F \) is the true function mapping compute to fraction of automated tasks in a particular run. Suppose further that at a particular time \( t \), the largest model that has been trained has a training compute of \( C_T \). In this case, at time  \( t \) the updated belief distribution \( G_t \) of the planner is supported on the set
\begin{equation}
    \mathcal F_t = \{ f \in \mathcal F: f(C_T) = \bar f(C_T) \text{ for all } C_T \leq C_{T}(t) \}
\end{equation}
and is obtained by conditioning down the original belief distribution \( G \) to this set as follows:
\begin{equation}
G_t(f_i) = 
\begin{cases} 
    \frac{G(f_i)}{\sum_{f_j \in \mathcal F_t} G(f_j)} & \text{if } f_i \in \mathcal F_t, \\[6pt]
    0 & \text{if } f_i \notin \mathcal F_t.
\end{cases} 
\end{equation}
In implementing this updating process, the model tracks and maintains distinct decision paths for each set of automation functions that remain consistent with observed data, allowing the planner to optimize decisions conditional on the current information set.

Intuitively, when the uncertainty add-on is activated, the social planner re-optimizes its vector of choice variables as more and more information is revealed about the underlying empirical automation function. Moreover, the uncertainty module allows the model to capture two drivers of AI automation missing in the baseline model. These drivers include
\begin{enumerate}
    \item risk aversion, which in the presence of uncertainty can lead to a lower/ more delayed path of investment
    \item the role of uncertainty in the decision-maker beliefs, which can push back the path of investment relative to the full information case
\end{enumerate}
By incorporating these mechanisms, the uncertainty add-on provides a more realistic framework for analyzing how imperfect information can reshape the pace, scale, and timing of AI-driven automation.

\section{Model Functionality}\label{sec:model_functionality}

The main objective of GATE is to enable users to run simulations of the global economy under a broad spectrum of potential AI automation scenarios. By offering a set of configurable parameters (detailed in Appendix~\ref{sec:model_inputs}), the model grants users substantial freedom to tailor assumptions about AI development—ranging from hardware and software R\&D trajectories to capital adjustment frictions and labor reallocation rules. This flexibility makes it possible to study plausible paths of AI-driven growth and their downstream economic effects under diverse configurations. Moreover, GATE delivers its results in real time, leveraging an efficient solution algorithm (described in Appendix~\ref{ec:solving_the_model}), and presents these forecasts in an accessible, user-friendly format.

In the remainder of this section, we first describe the model’s key output variables, showing how GATE tracks both standard macroeconomic aggregates and AI-specific development metrics. We then outline the sandbox’s principal features—such as real-time visualization and scenario comparison—and conclude with concrete examples of AI trajectories that can be explored.

\paragraph{Model Outputs}
GATE simulates the evolution of multiple endogenous variables over time, covering both macroeconomic aggregates and AI-specific development metrics. Users can visualize and export these outputs to analyze how different assumptions (e.g.\ capital adjustment frictions, R\&D returns, labor reallocation) affect the trajectory of global output, consumption, AI capabilities, and more. The key output categories include:

\begin{itemize}
    \item \textbf{Economic Indicators.}  
    These include Gross World Product \(Y(t)\) and its growth rate, consumption levels \(c(t)\), and the composition of investment (in both capital and AI-related inputs). The model also tracks the evolution of the capital stock \(K(t)\), enabling analysis of how traditional forms of capital accumulation interact with AI-driven productivity gains.

    \item \textbf{AI Development Metrics.}  
    Key outputs here comprise the size of the largest training run \(C_{T}(t)\) (i.e., how much compute is devoted to the most ambitious AI model at each point in time) and the fraction of tasks that become automated \(f(t)\).

    \item \textbf{Resource Allocation Variables.}  
    GATE simulates how effective compute \(C(t)\) is allocated between training and inference across tasks, as well as how investment flows are split among hardware, software R\&D, and non-compute capital. For instance, at each timestep users can see how much of the economy’s output is devoted to building new compute hardware \(Q(t)\), or funding R\&D aimed at pushing forward the efficiency frontiers of hardware \(H(t)\) and software \(S(t)\).

    \item \textbf{Technology Trends.}  
    Finally, the model reports the paths of hardware efficiency \(H(t)\) and software (algorithmic) efficiency \(S(t)\) over time. As these improve through R\&D, the total effective compute becomes more abundant, driving faster automation.
\end{itemize}

\begin{figure}[h!]
    \centering
    \begin{subfigure}[b]{0.31\textwidth}
        \centering
        \includegraphics[width=\textwidth]{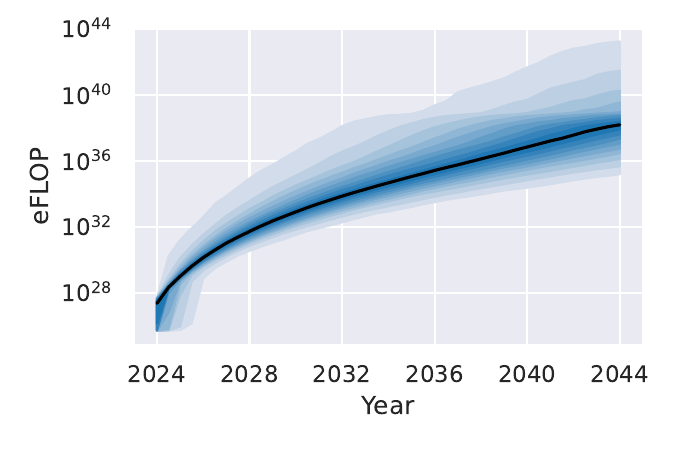}
        \caption{Projected size of largest AI training runs}
    \end{subfigure}
    \hspace{1pt}
    \begin{subfigure}[b]{0.31\textwidth}
        \centering
        \includegraphics[width=\textwidth]{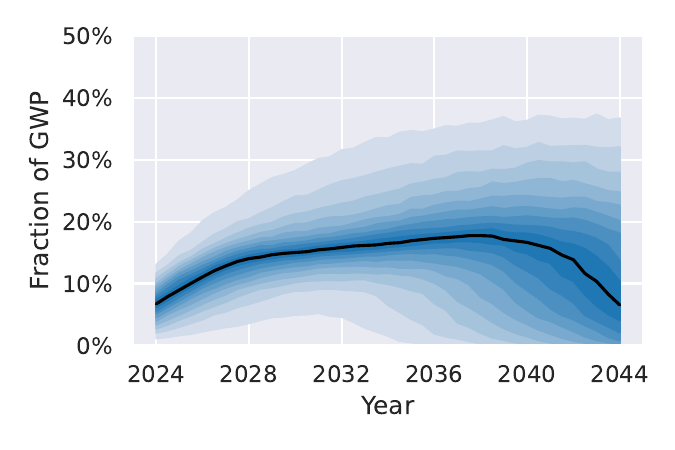}
        \caption{Predicted investment in compute-related capital}
    \end{subfigure}
    \hspace{1pt}
    \begin{subfigure}[b]{0.31\textwidth}
        \centering
        \includegraphics[width=\textwidth]{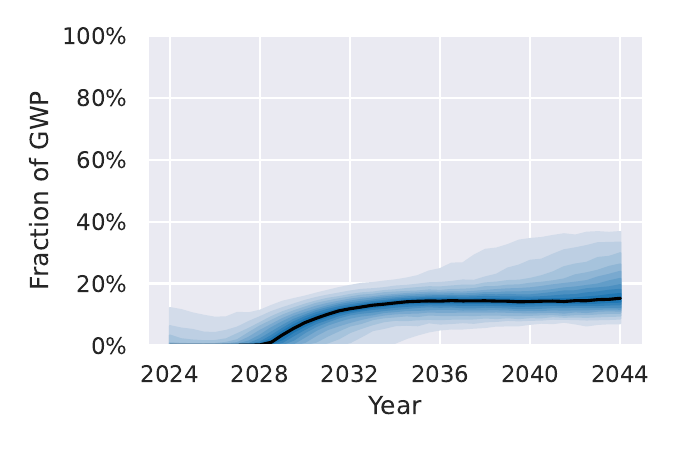}
        \caption{Projected general capital investment}
    \end{subfigure}

    \vspace{2pt}

    \begin{subfigure}[b]{0.31\textwidth}
        \centering
        \includegraphics[width=\textwidth]{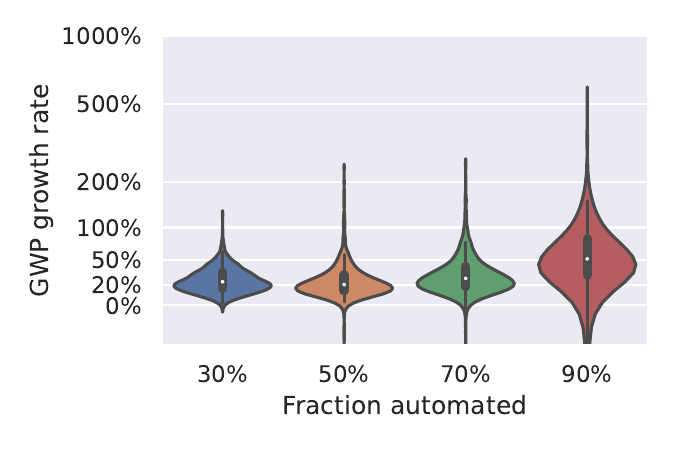}
        \caption{GWP growth vs. fraction of tasks automated}
    \end{subfigure}
    \hspace{1pt}
    \begin{subfigure}[b]{0.31\textwidth}
        \centering
        \includegraphics[width=\textwidth]{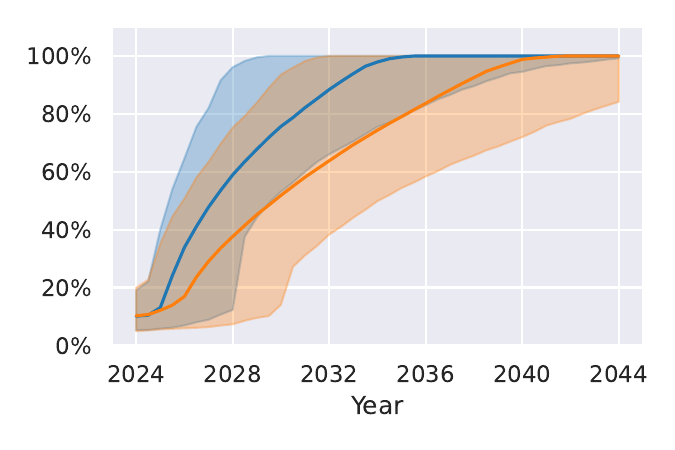}
        \caption{Automation over time}
    \end{subfigure}
    \hspace{1pt}
    \begin{subfigure}[b]{0.31\textwidth}
        \centering
        \includegraphics[width=\textwidth]{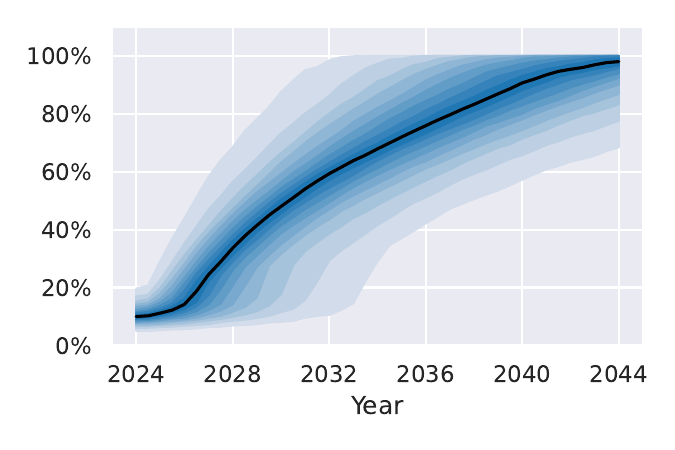}
        \caption{Predicted fraction of tasks automated over time}
    \end{subfigure}

    \vspace{2pt}

    \begin{subfigure}[b]{0.31\textwidth}
        \centering
        \includegraphics[width=\textwidth]{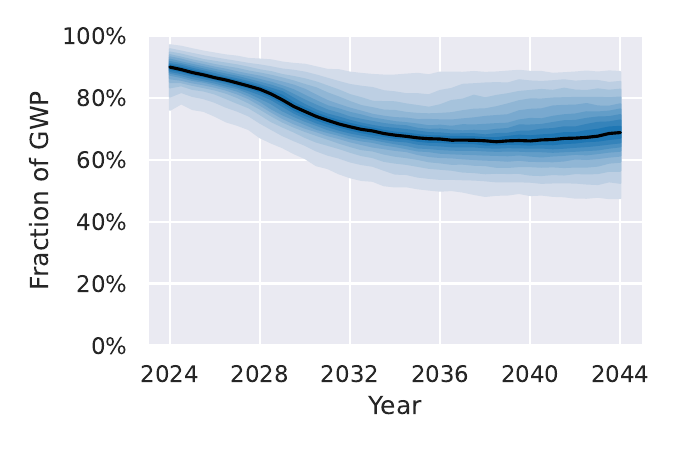}
        \caption{Projected consumption levels over time}
    \end{subfigure}
    \hspace{1pt}
    \begin{subfigure}[b]{0.31\textwidth}
        \centering
        \includegraphics[width=\textwidth]{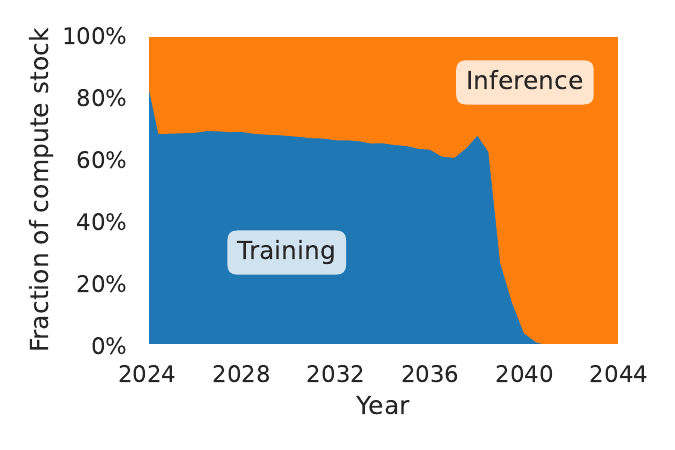}
        \caption{Distribution of compute resources over time}
    \end{subfigure}
    \hspace{1pt}
    \begin{subfigure}[b]{0.31\textwidth}
        \centering
        \includegraphics[width=\textwidth]{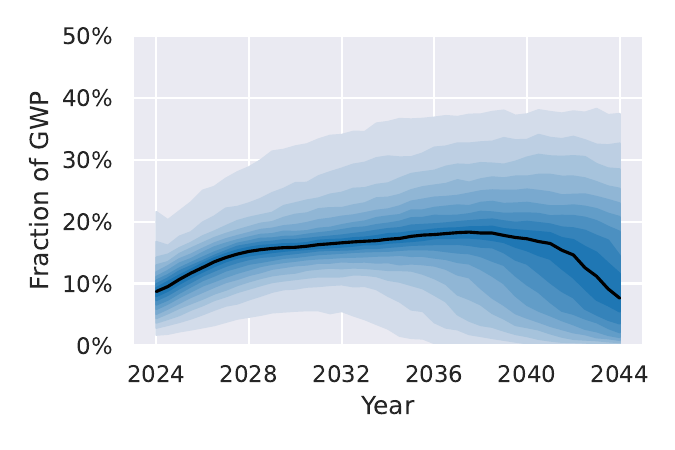}
        \caption{Investment in compute-related capital and R\&D}
    \end{subfigure}
    \caption{\small Illustrative outputs from GATE’s integrated assessment framework. Each subplot shows how key macroeconomic and AI‐specific variables evolve under different assumptions. For instance, GATE projects the scale of the largest AI training runs (a), the balance of compute‐related vs.\ general capital investment (b–c), and how rising automation (d–f) drives consumption growth (g) and reallocates compute resources (h) and R\&D (i). Together, these panels highlight how changes in AI development—e.g.\ hardware/software R\&D or labor reallocation—shape economic outcomes and further AI progress.}
\end{figure}

The model playground is initialized with a set of pre-specified default parameters (outlined and justified in Appendix \ref{sec:model_inputs}). Users are able to either manually modify parameters within their permissible ranges ( outlined in Appendix \ref{sec:model_inputs}) or select from pre-specified bundles of parameters or scenarios (e.g., median, 5th or 95th percentile values for parameters based on estimates from the existing empirical literature).

In response to user parameter choices, the model playground produces real-time visualizations of key model output variables, with graphs and charts updating dynamically to parameter changes. Moreover, to enhance functionality, the playground includes an additional feature, namely the possibility to engage in scenario comparisons. The modeling tool allows users to save and compare the outputs of multiple simulation runs, enabling side-by-side evaluations of distinct assumptions or parameter values.

\paragraph{Model functionality: Some AI development trajectories that can be explored}

With GATE’s integrated framework in place, users can investigate a broad array of possible trajectories for AI development by systematically varying the model’s economic and technological parameters. In doing so, they can glean insights about how the interplay between R\&D investments, hardware and software progress, task automation, and macroeconomic forces shapes AI’s transformative impact on global output and growth. Below are examples of particularly salient AI development paths one can explore:

\begin{itemize}
    \item \textbf{Rapid vs.\ Gradual Automation.}
    The speed of automation is primarily governed by the ``FLOP gap'' ($\Delta_{\text{FLOP}}$) parameter of the task-automation function. 
    Lowering $\Delta_{\text{FLOP}}$ (and/or adjusting the total compute $T$ required for full automation) can yield a scenario in which AI takes on a large fraction of tasks quickly once a threshold is crossed, while larger $\Delta_{\text{FLOP}}$ requires much more compute before automation truly accelerates. 
    This framing helps clarify whether advanced AI is likely to arrive abruptly or remain a gradual process over many decades.

    \item \textbf{Varying Rates of Hardware and Software Efficiency Progress.}
    GATE captures both hardware (\emph{e.g.}, GPU chips) and software (algorithmic) advances, driven by their respective R\&D processes. 
    Users can alter the returns to R\&D (parameters $\lambda_H,\lambda_S$) and the ``fishing-out'' exponents $(\phi_H,\phi_S)$, as well as imposing or relaxing upper bounds on hardware and software efficiency ($H^\text{max},S^\text{max}$). 
    This allows for simulating everything from accelerated software progress to potential stagnation in one dimension (say, hardware), revealing how bottlenecks in hardware vs.\ software shape AI’s path.

    \item \textbf{Different Levels of Investment in AI R\&D.}
    By changing macro-level parameters such as the time discount rate ($\delta$) or the representative agent’s risk aversion ($\eta$), one can shift how aggressively the planner invests in AI-specific R\&D. 
    The optional R\&D wedge $\xi$ further allows for exploring under- or over-investment in AI research when market actors do not (or do) internalize knowledge spillovers. 
    These scenarios illuminate whether society invests “too little” and thus delays breakthroughs or invests heavily up front in pursuit of accelerated automation gains.

    \item \textbf{Potential Technological Bottlenecks or Breakthroughs.}
    The model incorporates supply-chain and heat-dissipation constraints that limit the expansion of compute ($Q$). 
    Users can test cases where adjustment costs are severe (large exponent $\beta$), restricting how quickly the compute stock can grow, or conversely simulate breakthrough fabrication technologies that flatten adjustment costs. 
    In tandem, modifying $\rho$ (the substitution/complementarity of tasks) helps users gauge whether partial automation quickly boosts output or remains bottlenecked by non-automatable tasks. 
    Taken together, these experiments provide insight into how hardware constraints might bind as AI scales.

    \item \textbf{Impact of Policy Interventions.}
    GATE can incorporate a wide range of policy scenarios. 
    For instance, toggling the R\&D externalities module (adjusting $\xi$) models different degrees of subsidy or tax for AI-specific research. 
    One can also apply artificial “caps” on large training runs to mimic regulatory constraints on advanced AI or impose taxes on hardware acquisition to slow compute expansion. 
    In this way, researchers can see how strategic regulation or targeted subsidies alter the timeline and extent of automation.

    \item \textbf{Uncertainty over AI Progress.}
    Finally, the uncertainty add-on allows for multiple possible mappings between compute and the fraction of tasks automated, each with different probabilities. 
    As the planner observes real-world outcomes (e.g.\ unexpected rapid automation of certain tasks), beliefs update and investment decisions shift accordingly. 
    This framework clarifies how “surprise” breakthroughs (or disappointments) can reshape the allocation of resources to AI R\&D, thus modifying the subsequent paths of both AI capabilities and economic growth.

\end{itemize}

These potential trajectories illustrate the model’s flexibility in exploring how AI-driven automation might unfold under diverse assumptions, technological limits, and policy choices.

\section{Limitations}
\label{sec:limitations}

While GATE represents a significant step toward integrated assessment modeling of AI's economic impacts, several important limitations warrant discussion. These limitations fall into two broad categories: parametric uncertainty and structural simplifications. On the parametric side, key model inputs like training compute requirements for automation, compute stock adjustment costs, and R\&D externality wedges remain highly uncertain. These uncertainties reflect both the novelty of AI technology and limited empirical evidence about scaling dynamics in compute-intensive industries. While one could conduct sensitivity analyses across plausible parameter ranges, these analyses cannot fully address deeper structural limitations in how the model represents AI development, technological progress, and labor markets. Below, we detail the key structural limitations that may constrain GATE's ability to capture important real-world dynamics.

\paragraph{Lack of non-AI technological progress}
A significant limitation of the current specification is its handling of non-AI technological progress. Although the model does include endogenous improvements in hardware and software efficiency, the macroeconomic module keeps broader TFP exogenous and focuses exclusively on how AI affects output through labor-task automation. This approach presents two difficulties. Firstly, on a conceptual level, it precludes consideration of one of the mechanisms that has garnered most interest in the academic discourse on AI and economic growth, namely the effect of AI on general purpose R\&D and productivity.\footnote{For examples of work emphasizing these mechanisms see \citet{aghion2018artificial}, \citet{nordhaus2021singularity} and \citet{trammell2020economic}.} Secondly, on a quantitative level, the omission of this mechanism might lead to an excessively conservative estimate of the economic implications of AI.

While endogenizing TFP growth could mitigate these concerns, it would also, under many parameter values, introduce hyperbolic growth dynamics, which imply unbounded future consumption and are thus challenging to analyze within our representative-household optimization framework. Moreover, such growth scenarios would likely require incorporating additional constraints around specific planetary bounds like energy availability, heat dissipation, and raw material scarcity to remain tractable. Given these complexities, we opted for an exogenous TFP specification in this initial release, though future work might explore ways to capture AI-driven technological progress while maintaining computational feasibility.

\paragraph{Stylized model of AI development via ``effective compute''}
A further limitation arises from reducing all algorithmic progress to a single ``effective compute'' dimension, which overlooks crucial differences between cost-lowering improvements and breakthroughs that open entirely new capabilities. For instance, techniques that reduce runtime requirements may cut the compute needed for tasks already within reach, but not necessarily widen the overall range of automatable work. By contrast, new model architectures may tackle previously unattainable tasks even while yielding only modest cost efficiencies for already automated tasks.

At very large scales, the distinction between serial and parallel computing becomes especially significant: \citet{erdil2024data} show that fundamental data-movement and latency constraints can severely limit training beyond about $10^{28}$--$10^{31}$ FLOP. A single ``effective compute'' metric cannot capture the fact that some steps remain serially constrained---no amount of parallel hardware will accelerate them---whereas others scale more smoothly with increased FLOP.

As a result, key automation thresholds could arrive earlier or later than our single-index approach suggests, potentially distorting both timeline estimates and investment decisions. The relationship between efficiency improvements and capability-expanding breakthroughs remains poorly understood, making it infeasible to model them as separate dimensions of progress. Further empirical research on how scale, parallelism, and model architectures interact is needed to represent these complexities more accurately.

\paragraph{Failure to incorporate data production} 
The AI development module abstracts away from important non-compute inputs, particularly data availability, cost, and quality. While the compute-based framework captures many key dynamics of AI development, it omits potentially significant constraints around data collection, quantity, and curation requirements. This simplification could lead to overly optimistic automation timelines by not accounting for cases where data scarcity or quality issues become binding constraints, even with abundant compute resources---a concern that appears increasingly relevant given recent projections about the exhaustion of public training data \citep{villalobos2022will}. This concern may be especially salient for complex tasks that rely on real-world embodiment or physical experimentation, where data can be especially costly to gather. Future iterations of the model could benefit from explicitly incorporating data as a complementary input alongside compute, though this would inevitably introduce additional complexity into modeling the interplay between data requirements, collection costs, and AI automation capabilities.

\paragraph{Stylized labor reallocation assumptions}
An important limitation of our current specification is its stylized treatment of labor reallocation in response to automation. The model considers only two extremes---perfect reallocation where workers seamlessly transition to non-automated tasks, or complete displacement where automated workers permanently exit the labor force---while omitting crucial intermediate dynamics involving retraining periods, search frictions, and geographic mobility constraints.

Under perfect reallocation, we may underestimate short-term unemployment and skill mismatches, whereas under complete displacement, we risk overstating the severity of economic disruption and understating long-run flexibility. This simplification has material implications for the model's predictions: by omitting partial reallocation paths, GATE may overlook temporary unemployment spells, skill-upgrading costs, or the gradual movement of workers to alternative sectors. Extending the framework to incorporate realistic labor market frictions, including explicit modeling of retraining dynamics, job search processes, and wage adjustment mechanisms, would help capture more accurate transition costs and enhance the model's applicability for policy analysis.\footnote{Labor market frictions have been the object of a vast literature in economics. Two particularly relevant strands of literature are the search and matching literature (\citet{diamond1982asearch}, \citet{diamond1982bsearch}, \citet{mortensen1970asearch}, \citet{mortensen1970bsearch}, \cite{mortensenpissarides1994search}, \citet{pissarides1979search}) and the literature analyzing labor market reallocation in response to large shocks (\citet{dixcarneirokovak2017shock}, \citet{kovakmorrow2025shock}, \citet{ADH2013shock}, \citet{ADH2015shock}, \citet{AADH2014shock}). This latter literature typically finds evidence for substantial frictions affecting labor reallocation across sectors, occupations and regions.}

\paragraph{Omission of market structure and strategic behavior}
A significant limitation of GATE’s current specification is its reliance on a social planner framework that abstracts away from real-world market structures and strategic behavior among AI firms. This simplification, while preserving computational tractability and precluding the need for a detailed investigation of plausible market structures, omits crucial mechanisms including competitive investment dynamics, market power effects, and strategic timing of capability deployment.

The social planner approach may mischaracterize the pace and pattern of AI progress by assuming coordinated, welfare-maximizing investments, rather than reflecting the decentralized environment in which AI labs compete. For example, the model cannot capture how market concentration might affect AI firms' decisions to withhold or accelerate capability deployment. Moreover, our current implementation does not allow us to model quantities of interest such as prices, wages and interest rates, that could also serve efforts to empirically scrutinize the model. Future iterations of GATE could incorporate an explicit market structure layer modeling oligopolistic competition in AI development and price formation mechanisms, though this would require careful balance between added realism and computational feasibility.

\paragraph{Stylized and static labor task space} GATE employs a simple modeling of the labor task space: a unit measure of tasks that are only differentiated along one dimension, namely their complexity and hence their automatability by AI systems. This involves abstracting away from at least two layers of complexity that have been widely discussed in the existing literature. 

Firstly, we omit other dimensions of task heterogeneity that have been documented as important by previous work, such as the distinction between manual and cognitive tasks or that between routine and non-routine tasks.\footnote{See \citet{ALM2003}.} The distinction of between cognitive and manual tasks is likely to be particularly consequential given the (to date) slower progress of AI systems in the physical realm and the plausible different production function for ``digital workers" in the physical domain. Allowing for these additional dimensions of heterogeneity across labor tasks would be attractive, but would be computationally costly and highly speculative given the absence of high quality  measurements for the relative economic value of physical tasks and of the key parameters of the production technologies applicable to robotics.

Secondly, we assume that the labor task space is static and exogenous, which precludes the possibility that the space of labor itself changes along the automation process, as well as the possibility that the direction of automation might be influenced by economic considerations (such as labor market outcomes or the choices of economic decisionmakers).\footnote{Our framework allows for the timing but not the order of automation to respond to economic incentives.} Evidence in support of this dynamic joint determination of automation and labor market outcomes is provided by a recent and rapidly growing literature on directed technological change (DTC).\footnote{Some of the contributions to the literature on automation and directed technical change (DTC) include \citet{Acemoglu2002aAutomationAN}, \citet{Acemoglu2002bAutomationAN} and \citet{Acemoglu2019AutomationAN}.} Moreover, these omitted features of the space of labor tasks underpin some of the recent debates concerning the impacts of AI on labor markets, where drawing on the historical record some commentators expect the advent of novel tasks for human workers to preclude the possibilities of full automation and of large scale technological unemployment. Given the computational complexities involved in adding these features to GATE, and the model's relatively limited interest in labor markets, we have omitted these factors from the current release. We envision embedding some of these complexities in future releases of the model.

\paragraph{Other bottlenecks} The current release of GATE already includes several sources of bottlenecks that could dampen the effects of AI automation on economic growth. These include complementarities between labor tasks (and hence between automated and non-automated tasks) and allowing for the presence of long-run non-accumulable factors in the aggregate production function. However, GATE does not include the exhaustive list of potential bottlenecks limiting the growth implications of AI put forward in the literature.\footnote{For a comprehensive discussion of the potential bottlenecks to economic growth in an advanced AI regime see \citet{besiroglu_erdil2023expgrwt} and \citet{vollrath2023expgrwth}.} In particular, our modelling of the complementarities between human and AI labor is quite simplified, being captured by a single and constant over time parameter, $\rho$. By contrast, the patterns of complementarity and substitutability between AI and human labor are likely to be much richer and evolve over time.\footnote{For a more detailed discussion of this issue see \citet{idetalamas2025}.} 

While including all of the potential bottlenecks discussed in our framework would be impractical, we envision the inclusion of some of the most salient ones in future releases of GATE. These may include the role played by regulations that could restrict the automation of certain occupations and their associated labor tasks (e.g. lawyers, judges) as well as adding additional complementarities on both the production and the consumption side of the economy (e.g. complementarities between labor, capital and non-accumulable factors of production, complementarities between commodities whose production processes are easier or harder to automate etc.) that may augment the severity of Baumol effects in our setting.

\section{Concluding Remarks}
\label{sec:conclusion}

This work is an attempt to contribute to the debate surrounding the economic impact of AI by putting forward a comprehensive integrated assessment model of AI automation. We hope that this release will help guide and structure the debate about the implications of advanced AI and bring together the AI and economics communities. 

While significant effort has been invested in including some of the key drivers of AI development and AI driven automation into GATE, wide scope for further work remains. In what follows, we outline what are in our view some of the key areas for future work. We expect to continue exploring some of these research directions in the future, but we would like to encourage the research communities in economics and AI to engage and further investigate this potentially highly consequential cluster of topics at the intersection of AI and economics.

We see at least four key directions for future work. Firstly, and perhaps of most interest for the economics community, it would be interesting to analyze market equilibrium counterparts to the planner setting studied in GATE, under alternative assumptions regarding market structures. This would allow us to study the expected evolution of key market prices, particularly wages and interest rates along the path of automation, as well as test the robustness of GATE’s predictions regarding growth and investment. The ability to study prices would be particularly attractive, due to the forward-looking attributes of market prices that can allow for earlier model benchmarking and validation.

Secondly, and perhaps of most interest for the AI community, it would be helpful to explicitly model data requirements and data availability over time as an additional factor limiting the practical real-world applications of AI in the stage before capabilities reach human-level performance. This type of extension could help evaluate whether data constraints are likely to noticeably slow the pace of automation relative to current model predictions.

Third, it would be useful to incorporate endogenous idea production outside of AI-specific hardware and software technologies into the analysis. This would help bring together the ``factor accumulation” and the ``technological progress acceleration" mechanisms through which AI automation is expected to transform the economy, as well as help refine the model’s predictions about growth in the post-automation regime. In particular, this extension would allow us to assess the robustness of a frequent prediction of models focused solely on ``factor accumulation,” namely a plateau in growth in the post-full-automation regime.

Fourth, it would be valuable to treat a broader class of structural constraints — not fully captured by our existing adjustment-cost specification — that might limit the scale-up of AI compute.  Energy requirements and electricity provision are one salient example, but other factors—such as advanced chip manufacturing capacity and infrastructure bottlenecks—could equally constrain compute growth in ways not fully captured by our super-linear investment costs.  Incorporating these resource bottlenecks explicitly could substantially affect both the timing and degree of AI deployment, and remains a promising avenue for future extensions of GATE.

Finally, we encourage researchers to systematically explore GATE by adjusting parameter settings, policy interventions, and assumptions about AI development. Through scenario analyses, parameter sweeps, and sensitivity testing, one can examine how the model’s predictions shift under varied conditions—such as calibrating parameters to match current compute spending or exploring the implications of different labor‐market frictions. Incorporating new empirical insights into key parameters and refining the model’s modules can further stress‐test and enrich GATE, sharpening our ability to forecast and navigate AI’s impacts.

Beyond GATE and its natural extensions, we envision several additional areas in the Economics of AI space that we would be interested in contributing to and would encourage other interested researchers to consider as well. First of all, we observe a significant gap in the understanding of AI technology in the economics community. To bridge this gap we envision continuing our efforts to publish technical notes that describe key features of the AI engineering process in accessible terms and propose simple modeling techniques that allow these features to be incorporated into workhorse economic models.

Second, there is still very limited empirical evidence to anchor the calibration of many key parameters (e.g. R\&D parameters, automation function parameters, degree of complementarity/ substitutability between labor tasks etc) of models such as GATE or those of similar efforts to model the future path of AI and its economic impacts. To address this issue substantial additional empirical research is needed, as well as meta-studies that pool available evidence to allow for the most accurate picture of the value of these key parameters. We envision contributing to this area and encourage applied researchers to consider prioritizing this area in their own work.

Finally, as in other areas of academic research, several valuable public goods are underprovided in the Economics of AI space. These include the provision of a repository of models analyzing the potential paths of AI automation and the economy, as well as benchmarking exercises that compare the past performance and future predictions of existing models in this area. We envision contributing to mitigating this shortfall in subsequent work, but we expect substantial gaps to persist in the provision of this type of research ``infrastructure" and public goods for the foreseeable future.

\bibliography{sample}
\newpage

\begin{appendix}
\addtocontents{toc}{\protect\setcounter{tocdepth}{1}} 

\appendix
\section{Authorship, Credit Attribution, and Acknowledgments}\label{sec:authors_acknowledgements}

\textbf{Core Contributors:}  
Ege Erdil, Andrei Potlogea, Tamay Besiroglu, Anson Ho, Jaime Sevilla, and Matthew Barnett  

\textbf{Engineering and Sandbox Development:}  
Edu Roldan, Matej Vrzla, Robert Sandler  

\textbf{Roles and Contributions:}  
\begin{itemize}
    \item \textbf{Ege Erdil} initiated the project, developed the early prototype, and played a central role in advancing the key theoretical and modeling ideas.
    \item \textbf{Andrei Potlogea} contributed significantly to the technical exposition and introduced refinements to the economic model.
    \item \textbf{Tamay Besiroglu} coordinated the project, contributed to the writing, and ensured alignment across modeling, engineering, and writing efforts.
    \item \textbf{Anson Ho} provided ongoing support throughout the project, including calibration of parameter values, general model refinement, coordinating external feedback.
    \item \textbf{Jaime Sevilla} contributed to both the engineering and writing, ensuring coherence between the model’s implementation and its conceptual framework.
    \item \textbf{Matthew Barnett} contributed to the writing and parameter settings.
\end{itemize}

\textbf{Engineering and Sandbox Development:}  
\begin{itemize}
    \item \textbf{Edu Roldan} led the engineering efforts, overseeing the technical development of the model.
    \item \textbf{Matej Vrzla} contributed to the design of and implemented the interactive sandbox.
    \item \textbf{Robert Sandler} provided design support, contributing to the usability and presentation of the sandbox interface.
\end{itemize}

\textbf{Acknowledgments:}  
We are grateful to Tyler Cowen, Chad Jones, Ben Golub, Ryan Greenblatt, Kevin Kuruc, Caroline Falkman Olsson, Anton Korinek, Daniel Kokotajlo, Lev McKinney, Daan Jujin, Zachary Brown, Dan Valentine as well as seminar attendees at the 15th Oxford workshop on Global Priorities Research for their insights and feedback.

\section{A Basis for a Compute-Based Model of AI}
\label{sec:compute-based}

The idea of forecasting AI capabilities based on computational power has a rich history. Pioneers like IJ Good, Hans Moravec, and Ray Kurzweil predicted that matching human-level AI would require computational power equivalent to the human brain. This was a bold claim, as it was conceivable that a cleverly designed algorithm could have achieved general intelligence without massive computation. However, the success of modern deep learning has partially validated these compute-based predictions.\footnote{For example, \cite{moravec1976role} estimated in 1976 that computer vision systems had 1.6 million times less processing power than the human visual cortex. Given Moore's law, bridging this 6-order-of-magnitude gap was predicted to take around 40 years, putting impressive computer vision capabilities on par with humans by 2016. This prediction appears remarkably accurate today, despite the limited information available at the time.}

Compute-based models of AI forecasting were validated further by, among others, \cite{kaplan2020scaling}, when they introduced \textit{scaling laws} to the machine learning literature. A scaling law is a simple function that describes the performance of a machine learning system, in this case large language models (LLMs), using only some macroscopic properties of the system such as its parameter count and the number of examples it sees during training. Initially intended for specific neural architectures, recent scaling laws, such as \cite{hoffmann2022empirical}, seem to provide a blueprint for training models that are challenging to surpass through architectural innovations at smaller scales. This lends support to the compute-based paradigm of AI forecasting.

This perspective aligns with a broader lesson from the history of AI research, as articulated by \cite{sutton2019bitterlesson}:
\begin{quote}
The biggest lesson that can be read from 70 years of AI research is that general methods that leverage computation are ultimately the most effective, and by a large margin... Seeking an improvement that makes a difference in the shorter term, researchers seek to leverage their human knowledge of the domain, but the only thing that matters in the long run is the leveraging of computation.
\end{quote}

Still, we also have evidence that algorithmic progress is possible, despite the difficulties posed described by Sutton: every year, we are able to obtain better performance for a given budget of computation and data. There is a small but growing literature on progress in software and algorithms for common computer science problems (see \cite{ho2024algorithmic} for a review). While the exact nature of this algorithmic progress is complex, we can take it into account in an approximate way by introducing the notion of \textit{effective compute}, which is a measure of the computation done by a model adjusted for algorithmic efficiency. For example, if a Boolean satisfiability problem solver algorithms have become 300-fold better today relative to the 1990s, then for that specific problem, 1 unit of computation may be said to be effectively ``worth" 300 units of 1990s compute.\footnote{In fact, experiments by \cite{fichte2020time} suggest that modern SAT solvers achieve performance improvements equivalent to reducing compute requirements by roughly 100 to 1,000 times compared to solvers from the early 2000s, due to the developments of novel algorithms, such as Conflict-Driven Clause Learning.}

Prior work has sought to estimate the pace of algorithmic progress defined above for AI specifically. \cite{hernandez2020measuring} and \cite{erdil2022algorithmic} approach this question in the context of a famous computer vision benchmark, ImageNet-1k, by running experiments on older models and analyzing the results of old papers using statistical methods, respectively. \cite{hernandez2020measuring} estimates a doubling time of 16 months for effective compute, while \cite{erdil2022algorithmic} has a more uncertain estimate that is centered around 9 months.  Similarly, \cite{ho2024algorithmic} investigates algorithmic progress for large language models, and finds that the compute required for a certain level of performance halves roughly every 8 months. Despite variations in the data and uncertainties in individual estimates, the overall evidence indicates that the computational requirements for AI tasks tend to halve approximately every year due to algorithmic improvements.

The concept of effective compute is, of course, a simplification. It neglects many important features of algorithmic progress. To list a few:

\begin{itemize}
    \item The idealization of effective compute neglects the distinction between serial and parallel compute. When training a machine learning system, some calculations can be parallelized, meaning we can simply use more GPUs to train the model if one GPU proves insufficiently fast, while some cannot. Deep learning models themselves are usually designed such that the model can be trained in a highly parallelized way, but some non-trivial amount of serial compute is still required.\footnote{For instance, while the evaluation of gradients during gradient descent is highly parallelizable, ultimately, it's necessary to take gradient steps in a particular sequence, and this introduces an irreducible amount of serial computation to the training process.}

    \item Algorithmic progress in machine learning may be more or less pronounced at smaller or larger scales. For example, the LSTM architecture from the 1990s can perform better than the now-dominant  Transformer architecture from 2017 on some tasks \citep{droppo2021scaling}. A particular innovation might provide savings of only two times when training a model with \( 10^5 \) parameters, but ten times when training a model with \( 10^{15} \) parameters. This scale-dependent nature of progress is not captured by the simplified concept of ``effective compute", which reduces algorithmic advancements to a single dimension.

    \item Algorithmic progress in AI may advance more rapidly in reducing computational requirements for already-achieved capabilities than in expanding the frontier of AI system capabilities. For instance, advancements in Knowledge Distillation techniques (see \cite{xu2024survey}) enable the replication of top model capabilities at significantly reduced computational cost. However, these techniques contribute less to expanding the range of attainable AI capabilities. The ``effective compute" concept does not distinguish between these two types of advancement.
\end{itemize}

While we recognize these limitations, substantial evidence supports the significance of algorithmic progress in AI development. Improvements in architectures, algorithms, and application techniques have demonstrably enhanced AI capabilities. Moreover, the concept of effective compute has proven valuable in AI research and development, as evidenced by its adoption in strategic planning at major AI labs.\footnote{For instance, Anthropic incorporates the notion of effective compute in their Responsible Scaling Policies \citep{anthropic2023rsp}.} Therefore, we also think that the notion of effective compute is a sufficiently useful abstraction to warrant applying it.

\section{Solving The Model}\label{ec:solving_the_model}
In this section we present the solution algorithms employed to solve the model.

\subsection{Solution Approach: Baseline Model}\label{sec:solution_approach}

In this section we present a brief description of our solution approach for the model. How do we find the (optimal) paths for the model's key endogenous variables (i.e. output, investment, extensive and intensive margin automation, the size of AI training runs etc.) chosen by the welfare maximizing social planner? Our focus in this section is on the solution procedure for the baseline model that excludes R\&D externalities. The next section presents the solution approach taken when R\&D externalities add-on is activated.

The model is too complex to admit tractable closed-form solutions. However, since it is deterministic, the model can be solved numerically by optimizing the value function of the planner as a function of the vector of decisions that must be taken in the model. 

We proceed to solving the model via a numerical procedure called gradient descent, which can be used to find the maximum of a function, in our case the social planner's value function. Intuitively, this procedure works as follows: for a given vector of (user selected) parameters and initial values of the models key state variables, the solution algorithm begins with a guess of for the vector of endogenous variables (denoted $\{\mathbf{x}_t\}$ in Section~\ref{sec:macro_model_basic_setup}) and computes the planners value function. Then this choice is iteratively updated in the direction of increasing the value function until a stopping condition is met and the final updated guess of the vector of endogenous variables is outputted as the proposed model solution.

To make the model amenable to the above solution procedure, we discretize the task space into 20 grid points\footnote{For the purposes of allocating digital workers across tasks we discretize task space into 20 grid points. For the allocation of human workers and for solving the integral of the CES aggregator we discretize the tasks space into 100 grid points.} and also discretize time at a resolution of 1 year. We note that different time discretization schemes produce similar answers as long as the time resolution is sufficiently high. We also choose a time planning horizon $\tau_\text{plan}$ over which we aim to solve the planner's optimization problem. This is typically set at 80 periods (or 80 years in our setting). Our model has five degrees of freedom in the decisions of the social planner agent per time step. This number of degrees of freedom emerges from the fact that we have 7 endogenous variables (output shares of consumption, hardware investment, capital investment, hardware R\&D investment, software R\&D investment; as well as the compute shares of training and inference respectively) subject to two constraints (the sum of output shares needs to sum to 1 and the sum of the compute shares also needs to sum to 1).

We solve the model by performing gradient descent over the vector of parameters that define the investment, compute and labor allocation decisions. Specifically, in each time period, the social planner in the model makes the following choices:

\begin{enumerate}
    \item How to allocate total economic output between the five categories of consumption, capital investment, compute investment, software R\&D and hardware R\&D?
    \item How to allocate existing computing hardware between runtime compute and training compute?
\end{enumerate}

It is important to note that our solution algorithm does not need to specifically solve for the optimal allocations of inference compute across automated tasks or of human labor across non-automated tasks. This is because the optimal value of these variables are deterministic functions (pinned down by first order conditions) of the choice variables explicitly solved for by our solution algorithm.\footnote{It is easy to show that the optimal allocation of human labor is uniform across non-automated tasks for the case of frictionless labor allocation, and it is exogenous in the case without labor reallocation. Also, combined with the model's state variables and parameters, any choice of the share of compute allocated to runtime compute fully pins down the allocation of runtime compute across automated tasks. This is because the size of the largest training run at each point in time fully pins down the relative price (in terms of effective compute) of digital workers across tasks. This, coupled with the CES aggregator for tasks, fully specifies the efficient allocation of the total runtime compute budget across automated tasks.} 
 
 When solving the model over \( \tau_\text{plan} \) time periods (we call this the ``planning horizon"), we choose a time horizon \( \tau_\text{optim} \gg \tau_\text{plan} \) and simulate the model over this longer time horizon (we call this longer time horizon $\tau_\text{optim}$ the ``optimization horizon"). We define the planner's value function over a  choice variable vector with \( 5\tau_\text{optim} \) elements (5 degrees of freedom for each of the \(\tau_\text{optim}\) periods of the social planner optimization horizon) and apply gradient descent over this vector of choice variables. In other words, we continuously update the decision vector to increase the social planner's value function $V$. The optimization procedure terminates when the algorithms stopping conditions are met (e.g. a certain number of iterations has been exceeded, or the gradient norm has fallen below a certain tolerance threshold). At this point the program outputs the realization of the representative agent's value function as well as the optimal choices of each of the endogenous variables over the planning horizon $\tau_\text{plan}$. Formally our optimization algorithm solves for

 \[ \max V = \sum_{t=1}^{\tau_\text{optim}} e^{-(\beta-g_L) (t-1)} \frac{c_t^{1-\eta} - 1}{1- \eta} \]

   In our default implementation, one time period in the model corresponds to one year, but this can be modified by appropriately rescaling the parameters in line with the time or frequency dimensions. Moreover, picking some optimization horizon \( \tau_\text{optim} \gg \tau_\text{plan} \) is necessary to cut off the sum defining the value function at a finite threshold because \( V \) can't be computed as an infinite sum. In practice, when \( \tau_\text{plan} \) is on the order of e.g. 80 years, a value for $H$ on the order \( \tau_\text{optim} = 2\tau_\text{plan}= 160 \) years works well.\footnote{We have experimented with other optimization horizons, and we found that beyond $\tau_\text{optim}=2 \times \tau_\text{plan}$ the changes in the model solutions were negligible.}

   [If there is anything we do explicitly to avoid the problem of getting stuck at a local optimum it might be helpful to detail that here]

   [In a conversation with Ege he mentioned that the problem we have here is of a slighlty different nature than the one typically faced by economists, calling for a different model solving approach. having a bit of that discussion here might be nice, but it is not essential]

\subsection{Solution approach: R\&D externalities add-on}

Activating the R\&D externalities add-on results in some changes to the approach used to solve the model and deliver simulation output.

Intuitively, when the R\&D externalities module is activated, the social planner is operating with an incorrect model of the world: it underestimates the returns to R\&D investments by a factor $\xi$, period by period, with no updating.

To solve the model under these circumstances we employ a two step procedure. We first solve the model under the wrong specifications of the software and hardware efficiency laws of motion, and determine the values of the choice variables that would be chosen by  a social planner under this wrong model for the entire duration of the simulation. In a second step we simulate the path of the economy keeping the vector of choice variables fixed at the solution found in the optimization, but outputting the actual path of the economy given the choices of the planner and the correct laws of motion for hardware and software efficiency. This exercise can be thought of as a social planner being forced to set and commit to a plan of action ex-ante while operating under a misspecified model of the world (in this case a model of the world that is myopic of the true returns to R\&D).

An alternative solution approach would be to allow the planner to re-solve the optimization period by period. In this case, the myopic planner would be constantly surprised and wish to change its plans every period. However, this approach would be computationally significantly more expensive and has not been implemented for the current release of GATE.

\section{Model inputs}
\label{sec:model_inputs}

GATE allows users to perform simulations by selecting a wide range of key economic and technological parameters. The list of these parametes is presented in greater detail below. For each parameter we specify their ranges and their default values in our model sandbox.\footnote{In principle users can choose any value for parameters within their theoretical or plausible empirical ranges. However, for extreme parameter values our solution algorithm may not converge, in which case our model playground will typically deliver an error warning. Further details about this provided below. Moreover, for parameters for which there is very little empirical uncertainty we hard code them at their empirically documented values.}

\subsection{General economics parameters}
\begin{small}
\begin{longtable}{@{}%
>{\raggedright\arraybackslash}p{0.25\textwidth}%
>{\raggedright\arraybackslash}p{0.25\textwidth}%
>{\raggedright\arraybackslash}p{0.50\textwidth}@{}}
\caption{General Economics Parameters}\\
\toprule
\textbf{Parameter (units)} & \textbf{Range (default)} & \textbf{Explanation} \\
\midrule
\endfirsthead
\multicolumn{3}{c}{\tablename\ \thetable\ -- \textit{Continued from previous page}}\\
\toprule
\textbf{Parameter (units)} & \textbf{Range (default)} & \textbf{Explanation} \\
\midrule
\endhead
\midrule
\multicolumn{3}{r}{\textit{Continued on next page}}\\
\endfoot
\bottomrule
\endlastfoot
Initial GWP - $Y(0)$ (real USD/year)
& 105 trillion to 115 trillion (110 trillion)
& Gross world product at the start of simulations, i.e., the total economic value produced globally in 2024. \\
Initial labor force (perfect reallocation) - $L(0)$ (no. of human workers)
& 3.4 billion to 3.8 billion (3.6 billion)
& Total number of humans in the labor force at the start of 2025. \\
Population growth rate - $g_L$ (\%/year) & -1.00\% to 2\% (0.25\%)
& The annual rate of human population increase. \\
Initial capital stock - $K_0$ (real USD)
& 200 trillion to 1 quadrillion ($450$ trillion)
& Total value of physical capital in the economy at the start of 2025. \\
Output elasticity of capital - $\alpha$ (dimensionless)
& 0.2 to 0.6 (0.35)
& This measures how much GWP increases for each increase in capital, where \(\text{GWP } \propto \text{ capital}^{\alpha}\). For example, if \(\alpha = 0.5\), a 4\(\times\) increase in capital results in a 2\(\times\) increase in GWP, all else equal. \\
Initial non-accumulable capital stock - $F(0)$ (dimensionless)
& $1$ trillion
& Total value of ``nature'' endowments in the economy at the start of 2025. \\
Output elasticity of non-accumulable factor - $\mu$ (dimensionless)
& 0 to 0.3 (0)
& This measures how much GWP would increase for each unit increase in non-accumulable capital, where
$\text{GWP} \propto \text{nature}^{\mu}$. For example, if $\mu = 0.5$, a 4x increase in capital results in a 2x increase in GWP, all else equal. \\
Adjustment timescale for capital stock $\alpha_K$ (Years)
& 0.5 to 2 (1)
& The adjustment timescale captures how quickly capital investments can be converted into conventional capital. For example, if $\alpha_K$ is 1 year, investing at a rate of 100\% of the current stock per year means that around 70\% of the investment is lost to adjustment costs (assuming adjustment costs grow quadratically).  \\
Capital depreciation - $\delta_K$ (\%/year)
& 0.3\% to 20.0\% (6.5\%)
& The rate at which capital loses value over time. For example, if $\delta_K = 5\%$, this means 5\% of an existing capital stock is lost each year. \\
Substitution parameter - $\rho$ (dimensionless)
& -5 to -0.2 (-0.65)
& Captures how easily labor on different tasks substitute or complement each other. If $\rho < 0$, the tasks are primarily complementary. For instance, suppose that \( \rho = -0.5 \), and that the labor input is allocated uniformly across different tasks. If the labor input to a third of the tasks is set then to infinity, the total output would increase by 2.25\(\times\). We enforce \(\rho<0\) to ensure that tasks are gross complements.  \\
Consumption discount rate - $\beta$ (\%/year)
& 3\% to 7\% (5\%)
& The rate at which future consumption is valued less than present consumption. For example, a discount rate of 5\% means that the utility from consumption in the next year is worth 95\% of the same amount of consumption today. \\
Utility function parameter (relative risk aversion) - $\eta$ (dimensionless)
& 1 to 6 ($1.45$)
& Parameter of the isoelastic utility function $\frac{c^{1-\eta} - 1}{1-\eta}$ that determines risk preferences. \(\eta=1\) corresponds to taking log utility, and as $\eta$ increases, consumers become increasingly risk-averse. For example, \(\eta = 2\) suggests indifference between maintaining current consumption, and taking a bet with a 50\% chance of doubling consumption and a 50\% of decreasing consumption by a third.  \\
\end{longtable}
\end{small}

\begin{itemize}
    \item \textbf{Initial gross world product} \(Y(0)\): \$105T to \$115T (\$105T). \\  According to the World Bank, the GWP in 2023 was around \$106T \citep{worldbank2022gdp}. For our defaults we assume around 3\% annual GWP growth, resulting in a default of around \$110T. For the lower end of the range we use around the 2023 value, and for the upper value we add a buffer of around two more years of 3\% growth.
    \item \textbf{Initial labor force} \(L(0)\): 3.4 billion to 3.8 billion (3.6 billion). \\
    According to the World Bank, the labor force in 2023 was 3.6B \citep{worldbank2023}. This corresponds to a labor share of around 45\%, and we consider lower and upper bounds for the range by considering labor shares of around 42.5\% to 47.5\% respectively. 
    \item \textbf{Population growth rate} \(g_L\): -1.00\% to 2\% (0.25\%).\\
    We weight estimated population growth rates by GDP per capita. High-income countries grow $\sim$0.1\%year, middle income ~1\%/year, and low income $\sim$3\%/year. Since the ratio of high to lower-middle income labor is 3:2, we arrive at around $\sim$0.5\%/year \citep{worldpopulationprospects2022}. We reduce this slightly to 0.25\% given the broadly observed trend towards lower birth rates across different countries.
    \item \textbf{Initial capital stock} \(K_0\): \$200T to \$1Q (\$450T).\\
    We obtain a rough estimate of this based on standard capital depreciation rates and gross savings rates. In particular, global gross savings rates are around 27\% of GWP \citep{worldbank2022savings}, and a standard capital depreciation rate is 6\% per year \citep{Nadiri1993EstimationOT}. Given that current GWP is $\sim$ \$100 trillion \citep{worldbank2022gdp}, we estimate the worldwide capital stock as \( 0.27 \cdot \$1e14 \cdot 1/0.06 \approx \$4.5e14 \). We assume roughly 2x uncertainty in this number for our overall range. 
    \item \textbf{Output elasticity of capital $\alpha$}: 0.2 to 0.6 (0.35).\\ This is loosely based on \citet{jones2003growth}, where estimated capital shares are generally around 0.35, and between 0.2 and 0.6.
    \item \textbf{Initial non-accumulable capital stock} \(F(0)\): 1 trillion.\\ This parameter helps illustrate the possibility of a non-accumulable factor limiting the total output. The specific starting value of this parameter is not particularly important since the TFP is adjusted to make this match initial GWP. 
    \item \textbf{Output elasticity of non-accumulable factor} \(\mu\): 0 to 0.3 (0).\\ By default, this parameter is chosen to exclude the presence of the non-accumulable factor. To illustrate the significance of the non-accumulable factor, we can also choose \(\mu\) such that the production function has constant returns to scale in all inputs. 
    \item \textbf{Capital adjustment timescale $\alpha_K$}: 0.5 to 2 (1) year.\\
    \citet{Groth2005EstimatingUC} estimates capital adjustment costs using a similar functional specification on data from the UK economy between 1970-2000, and finds $\alpha_K$ to be around 1-3. Similarly, \citet{Groth2010InvestmentAC} generally report estimates of the adjustment cost parameter to roughly between 0.1 and 10, depending on the data and the precise functional form used. For our purposes we opt for 1 as a rough estimate, and assume a 2\(\times\) factor of uncertainty. 
    \item \textbf{Capital depreciation \( \delta_K \)}: 0.3\% to 20.0\% (6.5\%).\\ The capital depreciation rates are chosen to roughly correspond with the values in  \citet{Nadiri1993EstimationOT, Giandrea2022}.
    \item \textbf{Substitution parameter $\rho$}: -5 to -0.2 (-0.65).\\
     \citet{Knoblach2019TheEO} measure long-run elasticities of substitutions for the aggregate economy at in the range of 0.45-0.87. We interpolate between the corresponding substitution parameters to obtain $\frac{1}{2} \left( \frac{0.45 - 1}{0.45} + \frac{0.87 - 1}{0.87} \right) \approx -0.65$. We loosely account for a range of uncertainty, bounded above by \(-0.2 < 0\) so that labor inputs are always gross complements, and bounded below by -5, which suggests very substantial complementarities. 
    \item \textbf{Consumption discount rate} \(\beta\): 3\% to 7\% (5\%).\\
    \citet{Moore2004JustGM} suggest using a social discount rate of 3.5\%, and \citet{Zhuang2007TheoryAP} point out that developed countries tend to use consumption discount rates of 3-7\%. We opt for a round number of around 5\% as our default parameter estimate, and use 3-7\% as our range.  
    \item \textbf{Relative risk aversion in isoelastic utility} $\eta$: \(1.45\).\\
    A meta-analysis by \cite{havranek2015cross} collects 2,735 estimates of the elasticity of intertemporal substitution (EIS) in consumption. Across 1429 studies for the United States, they find a mean EIS of 0.594 ($\eta \approx 1.7 $). They find that households in countries with higher income per capita and higher stock market participation show larger values of the EIS. For example, they find an estimated posterior mean of the regression coefficient for income per capita of around 0.13, meaning that a doubling of GDP per capita results in an increase in EIS by roughly 0.13. Given the rapid economic growth in our model, we opt for a value of $\eta$ of 1.5. This also matches with the baseline calibration value of $2/3$ in the DSGE work of \cite{smets2007shocks} ($\eta =1.5)$.\footnote{\citet{elminejad2022rra_meta_analysis} perform a meta-analysis of papers that estimate the relative risk aversion $\eta$ using the consumption Euler equation. They find a mean relative risk aversion of around 1 in economics contexts (e.g. $\eta = 1.2$ under their preferred practice in their table 6), but substantially higher (lower $\eta$) in finance contexts. }
\end{itemize}

\subsection{Technological parameters}
\subsubsection{Hardware R\&D parameters}

\begin{small}
\begin{longtable}{@{}%
>{\raggedright\arraybackslash}p{0.25\textwidth}%
>{\raggedright\arraybackslash}p{0.25\textwidth}%
>{\raggedright\arraybackslash}p{0.50\textwidth}@{}}
\caption{Software R\&D Parameters} \label{tab:software-rd}\\
\toprule
\textbf{Parameter (units)} & \textbf{Range (default)} & \textbf{Explanation} \\
\midrule
\endfirsthead
\multicolumn{3}{c}{\tablename\ \thetable\ -- \textit{Continued from previous page}}\\
\toprule
\textbf{Parameter (units)} & \textbf{Range (default)} & \textbf{Explanation} \\
\midrule
\endhead
\midrule
\multicolumn{3}{r}{\textit{Continued on next page}}\\
\endfoot
\bottomrule
\endlastfoot
Initial hardware efficiency - $H(0)$ (FLOP/year/\$) 
& $1.00 \times 10^{17}$ to $1.00 \times 10^{19}$ ($1.00 \times 10^{18}$) 
& Number of computations performed per dollar per year at the start of simulations. \\

Maximum hardware efficiency - $H^{\text{max}}$ (FLOP/year/\$) 
& $1.00 \times 10^{21}$ to $1.00 \times 10^{25}$ ($1.00 \times 10^{23}$)  
& Maximum possible hardware efficiency of AI hardware, based on the current paradigm of CMOS microprocessors. See Section~\ref{sec:tech_constraints}. \\
Initial inputs to hardware R\&D - $I_{H}^{\text{RD}}(0)$ (real USD/year) 
& $1.00 \times 10^{10}$ to $1.00 \times 10^{12}$ ($1.00 \times 10^{11}$) 
& Amount spent on improving hardware efficiency per year at the start of 2025. \\
Returns to scale in hardware R\&D \(\lambda_H\) (dimensionless) & 0.0625 to 1 (0.14) & This parameter controls the extent to which investments in hardware R\&D translate into increases in hardware efficiency. If \(\lambda_H > 1\), this corresponds to a network effect where having more R\&D investment yields increasing returns. If \(\lambda_H<1\), this is a “stepping on toes” effect where increasing R\&D investment yields diminishing returns. For example, a value of \(\lambda_H = 0.5\) indicates that a doubling in investment only results in a \(\sqrt{2}\) increase in hardware efficiency growth.  \\
Returns to scale in hardware efficiency \(\phi_H\) (dimensionless) & 
0.01 to 1 (0.0769) & This parameter determines how the growth rate in hardware efficiency changes with the level of efficiency. If this is less than 0, this captures the effect where greater efficiency makes it easier to achieve further efficiency gains. If this is greater than 0, this captures the “fishing out” effect, where further efficiency gains become harder to find as further progress is made. For example, a value of \(\phi_H = 0.5\) indicates that a doubling in efficiency results in a \(\sqrt{2}\) decrease in hardware efficiency growth. The value of this parameter is often estimated based on the returns to R\&D, given by \(\lambda_H/\phi_H\).\\
Hardware R\&D productivity parameter \(\theta_H\) - 
(FLOP\(^{\phi_H}\) year\(^{-1-\phi_H-\lambda_H}\) \(\$^{-\phi_H+\lambda_H}\))
& 0.001 to 1000 (0.192) 
& A multiplicative scaling factor that determines how well hardware R\&D investments translate into hardware efficiency improvements. This is determined by rearranging equation \ref{eq:r_d_equation}, and using estimates of existing growth rates of hardware efficiency. \\

\end{longtable}
\end{small}

\begin{itemize}
    \item \textbf{Initial hardware efficiency $H$}: 1e18 FLOP/year/\$.\\
    We anchor our estimate of this parameter to the H100 GPU, which is commonly used for training state of the art AI systems \citep{lee2024meta}. It performs around 1e15 FLOP/s \citep{nvidia2024h100}, and costs around \$40k \citep{leswing2023h100}. Over a year, this corresponds to \( 1e15 \cdot 365 \cdot 24 \cdot 3600 / 40000 \approx 1e18 \) FLOP/year/\$. We allow for around 1 order of magnitude of uncertainty in this parameter. 
    \item \textbf{Maximum hardware efficiency}: 1e23 FLOP/year/\$.\\
    We estimate this parameter with the same procedure proposed in \cite{davidson2023}. In particular, we first determine an upper bound to the FLOP/\$ by decomposing it into a product of the energy efficiency in FLOP/J and the inverse of the cost per unit of energy (J/\$). The hardware efficiency in FLOP/\$/year is then determined by estimating the time range over which hardware is likely to be used, and amortizing the cost of a processor over that time period. 
    
    For the FLOP/J, \cite{ho2023limits} proposes a simple model to estimate the limits to the energy efficiency of CMOS processors, which has a geometric mean estimate of around $5 \times 10^{15}$ FLOP/J at 4-bit precision, with a log-space standard deviation of 0.7 OOMs. Assuming quadratic scaling of energy costs with number precision, this corresponds to around $3 \times 10^{14}$ FLOP/J at 16-bit precision. If we assume that energy costs can fall from \$0.1/kWh today to \$0.01/kWh, then we have \begin{equation*} (5 \times 10^{15} \: \text{FLOP/J}) \cdot (3.6 \times 10^8 \: \text{J/\$}) \approx 10^{24} \: \text{FLOP/\$}\end{equation*} If we further assume that hardware costs are amortized over 10 years, we arrive at a hardware efficiency of $10^{23}$ FLOP/year/\$. 
    
    In practice it is unclear whether the optimal substructure assumption that the maximum FLOP/\$ can be determined by individually maximizing the FLOP/J and the J/\$. It is possible for instance that other bottlenecks may bind prior to reaching limits in FLOP/J or J/\$. This introduces uncertainty that goes beyond the uncertainty inherent in the estimates of the FLOP/J, J/\$ and amortization period themselves. 
    
    We allow for a range of 2 OOMs above or below this default estimate, to account for these uncertainties. 
    \item \textbf{Initial inputs to hardware R\&D}: \$100B/year.\\
    We obtain a ballpark estimate for this parameter by considering R\&D spending from big hardware companies. In 2022, this was \$17.5B for Intel \citep{alsop2024intel}, \$5.47B for TSMC \citep{tsmc2022}, \$5B for AMD \citep{alsop2024amd}, \$5.3B for NVIDIA \citep{macrotrends2024nvidia}, and \$3.4B for ASML \citep{macrotrends2024asml}. Since the distribution of hardware R\&D spending appears very heavy tailed, we assume a Zipfian distribution with Intel as the leader, and assume there are around 100 firms doing significant hardware R\&D worldwide. This results in an estimate of  \( \$17.5B \cdot H_{100} \approx \$90B \) for global R\&D spending on hardware, where $H_{100} \approx 5.2$ is the 100th harmonic number. We round this figure up to \$100B, and allow for uncertainty around 10\(\times\) higher or lower than this default estimate. 

    \item \textbf{Returns to scale in hardware R\&D} \(\lambda_H\): 0.14. \\
    For our central estimate of the returns to labor inputs, we anchor weakly to \citet{erdil2024estimating}. They find a value of around 0.4 for human labor inputs on Stockfish, for which by far the best data was available (here we use the same value of \(\lambda\) as for software R\&D). We then multiply this by around 0.35, an elasticity parameter between monetary investment and change in the labor inputs to R\&D, which gives \(0.4 \times 0.35 = 0.14\). For our range, we consider a lower value of \(\lambda_H\) of 0.25 (with elasticity 0.25), and an upper value of 1 (with elasticity 1). 

    \item \textbf{Returns to scale in hardware efficiency} \(\phi_H\): 0.0769.\\
    The most reliable empirical estimates usually apply to the overall R\&D returns \(\lambda_H / \phi_H\), so given the value of \(\lambda_H\) and these returns we can work out the value of \(\phi_H\). \citet{davidson2023} estimates hardware R\&D returns of around 5.2 based on GPU data from 2006-2022, with a range from around 4.3 to 7.1. Given a central estimate of \(\lambda = 0.4\) before considering human R\&D input elasticities (see \(\lambda_H\)), we have \(\phi_H = 0.4 / 5.2 = 0.0769\). We consider around 1 OOM of uncertainty in this parameter on either side. 

    \item \textbf{Hardware R\&D productivity parameter} \(\theta_H\): 0.192. \\
    We work this out by rearranging the law of motion for hardware R\&D. This gives \(\delta_H = g_H H^{\phi_H} I^-{\lambda_H}\), where \(g_H(0)\) = 27.5\%/year, \(H(0)\) = 1e18 FLOP/year/\$, \(\phi_H = 0.0769\), \(I(0)\) = \$1e11/year, and \(\lambda_H = 0.14\). Hence we have \(\delta_H = 0.275 \cdot 1e18^{0.0769} \cdot 1e11^{-0.14} \approx 0.192\) (3 s.f.). We allow for three OOMs of uncertainty in this parameter, since uncertainty in the exponents can substantially change the value of this constant. 
\end{itemize}

\subsubsection{Software R\&D parameters}

\begin{small}
\begin{longtable}{@{}%
>{\raggedright\arraybackslash}p{0.25\textwidth}%
>{\raggedright\arraybackslash}p{0.25\textwidth}%
>{\raggedright\arraybackslash}p{0.50\textwidth}@{}}
\caption{Software R\&D Parameters} \label{tab:software-rd}\\
\toprule
\textbf{Parameter (units)} & \textbf{Range (default)} & \textbf{Explanation} \\
\midrule
\endfirsthead
\multicolumn{3}{c}{\tablename\ \thetable\ -- \textit{Continued from previous page}}\\
\toprule
\textbf{Parameter (units)} & \textbf{Range (default)} & \textbf{Explanation} \\
\midrule
\endhead
\midrule
\multicolumn{3}{r}{\textit{Continued on next page}}\\
\endfoot
\bottomrule
\endlastfoot
Initial software efficiency - $S(0)$ (eFLOP/FLOP)
& 1
& Software efficiency at the start of 2025. If software efficiency is 2, this is equivalent to having 2\(\times\) the budget of available physical compute. Note that this parameter is not shown in the simulation playground since it is trivially initialized to 1 by definition.  \\
Maximum software efficiency - $S_{\text{max}}$ (eFLOP/FLOP)
& 50 to $1.00 \times 10^{8}$ ($1.00 \times 10^{4}$)
& Maximum possible software efficiency of AI algorithms, based on the current paradigm of deep learning. See Section~\ref{sec:tech_constraints}.\\
Initial inputs to software R\&D - $I_{S}^{\text{RD}}(0)$ (real USD/year)
& $1.00 \times 10^{9}$ to $2.50 \times 10^{10}$ ($5.00 \times 10^{9})$
& Amount spent on improving software efficiency per year, as measured at the start of 2025. \\

Returns to scale in software R\&D \(\lambda_S\) (dimensionless) & 0.14 to 1 (0.0625) & This parameter controls the extent to which investments in software R\&D translate into increases in hardware efficiency. If \(\lambda_S > 1\), this corresponds to a network effect where having more R\&D investment yields increasing returns. If \(\lambda_S<1\), this is a “stepping on toes” effect where increasing R\&D investment yields diminishing returns. \\
Returns to scale in software efficiency \(\phi_S\) (dimensionless) & 
0.1 to 1 (0.32) & This parameter determines how the growth rate in software efficiency changes with the level of efficiency. If this is less than 0, this captures the effect where greater efficiency makes it easier to achieve further efficiency gains. If this is greater than 0, this captures the “fishing out” effect, where further efficiency gains become harder to find as further progress is made.  The value of this parameter is often estimated based on the returns to R\&D, given by \(\lambda_S/\phi_S\).\\
Software R\&D productivity parameter \(\theta_S\) - 
(year\(^{\lambda_S-1}\) (eFLOP/FLOP)\(^{\phi_S}\) \(\$^{-\lambda_S}\))
& 0.001 to 1000 (0.0307) 
& A multiplicative scaling factor that determines how well software R\&D investments translate into software efficiency improvements. This is determined by rearranging equation \ref{eq:r_d_equation}, and using estimates of existing growth rates of software efficiency. \\
\end{longtable}
\end{small}

\begin{itemize}
    \item \textbf{Iniital software efficiency} \(S(0)\): 1 eFLOP/FLOP.\\
    Since software efficiency is defined in terms of the number of ``effective FLOP" relative to each physical FLOP in 2025 (the start of simulations), this is initialized to 1 by definition. 

    \item \textbf{Maximum algorithmic efficiency}: 1e5 eFLOP/FLOP.\\
    This parameter is highly speculative, and we are forced to consider several very rough arguments to try and get a handle on its value. 
    \citet{bubeck2022universal} argue that for a Lipschitz-smooth function to fit $d$-dimensional data, at least $n \times d$ parameters are required. For a rough estimate we consider the ImageNet dataset for image classification, which contains $n \approx 1000$ classes and around $d \approx 40$ intrinsic dimensions \citep{pope2021intrinsic}. 
    
    We approximate the minimum necessary training compute as $C \approx 6 ND$, where $N=n\times d$ is the number of model parameters and $D$ is the number of datapoints. 
    In this context, we interpret each “datapoint” as a particular “patch” of an image which has been encoded as a vector. Patch dimensions of 8$\times$8 pixels are fairly common, and since the most common image size is 500$\times$500 \citep{ehrlich2021compression}, this corresponds to around 4000 tokens. If we assume that through algorithmic innovation this can in principle be reduced by an OOM, we are left with 400 tokens. 
    Combining the above results in $C \approx 6 \cdot (40 \cdot 1000) \cdot (1000 \cdot 400) \approx 1e11$ FLOP. By comparison, the current state of the art system on the ImageNet benchmark, ViT-Huge/14 \citep{dosovitskiy2021image} was trained with 4e21 FLOP \citep{epochdb}. This thus suggests that ViT-Huge/14, which achieves ~89\% top-1 accuracy on ImageNet, could have been trained with around 10 OOM fewer FLOP.
    
    Another approach, followed by \citet{davidson2023} is to anchor to the efficiency of the human brain. In particular, \citet{cotra2020} estimates that a human performs around 1e24 FLOP over the first 30 years of life, which is around 12 OOMs lower than Davidson’s median estimate for the compute needed to train AGI (at 1e36 FLOP). 
    
    As a third heuristic, \citet{ho2024algorithmic} and \citet{erdil2022algorithmic} estimate growth rates of 0.4 OOM/year for algorithmic progress in language models and computer vision respectively. This corresponds to $\sim$4 OOM of sustained algorithmic progress since 2014, and it seems broadly reasonable to expect this to persist for another 4 OOMs with greater than 50\% probability.
    
    Overall, there is a tremendous amount of uncertainty over this parameter and as such we consider a wide range of possible values, ranging from 50 to 1e8.  
    
    \item \textbf{Initial inputs to software R\&D}: \$10B/year.\\
    We estimate this based on the number of researchers and researcher salaries, since algorithmic improvements are mainly driven by researcher effort at present. \citet{benaich2019stateofai} report around 20,000 unique authors on machine learning papers in 2019, so for a ballpark figure we assume there are 50k researchers in AI today. If each researcher is paid around \$100k each, this corresponds to around \$5B per year. For lower and upper ranges we consider 5\(\times\) more or less than the default estimate.

    \item \textbf{Returns to scale in software R\&D} \(\lambda_S\): 0.14. \\
    For our central estimate of the returns to labor inputs, we anchor weakly to \citet{erdil2024estimating}. They find a value of around 0.4 for human labor inputs on Stockfish, for which by far the best data was available (here we use the same value of \(\lambda\) as for software R\&D). We then multiply this by around 0.35, an elasticity parameter between monetary investment and change in the labor inputs to R\&D, which gives \(0.4 \times 0.35 = 0.14\). For our range, we consider a lower value of \(\lambda_S\) of 0.25 (with elasticity 0.25), and an upper value of 1 (with elasticity 1). 

    \item \textbf{Returns to scale in software efficiency} \(\phi_S\): 0.32. \\
    The most reliable empirical estimates usually apply to the overall R\&D returns \(\lambda_S / \phi_S\), so given the value of \(\lambda_S\) and these returns we can work out the value of \(\phi_S\). \citet{erdil2024estimating} find returns on the order of 1.25 for software across a range of different domains (e.g. computer vision and computer chess). For our default value we therefore have 0.4/1.25 = 0.32. We account for around 1 OOM of uncertainty in this parameter. 

    \item \textbf{Software R\&D productivity parameter} \(\theta_S\): 0.0307. \\
    We work out this parameter by rearranging the law of motion for software R\&D, where we have \(\delta_S = g_S S^{\phi_S} I^{-\lambda_S}\). We also have that \(g_S(0)\) = 70\%/year, \(S(0)\) = 1 eFLOP/FLOP, \(\phi_S = 0.32\), \(I(0)\) = \$5e9/year, and \(\lambda_S = 0.14\). Hence we have \(\delta_S = 0.7 \cdot 1^{0.32} \cdot 5e9^{-0.14} \approx 0.0307\) (3 s.f.). Similar to \(\theta_H\), we allow for substantial uncertainty in this parameter. 
\end{itemize} 

\subsection{AI and compute parameters}
\subsubsection{Compute investment parameters}

\begin{small}
\begin{longtable}{@{}%
    >{\raggedright\arraybackslash}p{0.25\textwidth}%
    >{\raggedright\arraybackslash}p{0.25\textwidth}%
    >{\raggedright\arraybackslash}p{0.50\textwidth}@{}}
\caption{Compute Investment Parameters} \label{tab:compute-invest}\\
\toprule
\textbf{Parameter (units)} & \textbf{Range (default)} & \textbf{Explanation} \\
\midrule
\endfirsthead
\multicolumn{3}{c}{\tablename\ \thetable\ -- \textit{Continued from previous page}}\\
\toprule
\textbf{Parameter (units)} & \textbf{Range (default)} & \textbf{Explanation} \\
\midrule
\endhead
\midrule
\multicolumn{3}{r}{\textit{Continued on next page}}\\
\endfoot
\bottomrule
\endlastfoot

Initial compute investment - $I_{Q}(0)$ (real USD/year) 
& $5.00 \times 10^{10}$ to $8.00 \times 10^{11}$ ($2.00 \times 10^{11}$) 
& Initial dollar spending on compute production. \\
Compute adjustment cost exponent - $\chi$ (dimensionless) 
& 3 to 5 (4) 
& Determines how steep the adjustment costs are from increasing the stock of compute. An exponent of 2 means that adjustment costs grow quadratically with investment flows. \\
Compute adjustment timescale - $\alpha_Q$ (years) 
& 1 to 4 (2) 
& The adjustment timescale captures how quickly compute investments can be converted into compute stock. For example, if $\alpha_Q$ is 1 year, investing at a rate of 100\% of the current stock per year means that around 70\% of the investment is lost to adjustment costs, assuming that adjustment costs grow quadratically (this depends on the compute adjustment cost exponent).   \\
\end{longtable}
\end{small}

\begin{itemize}
    \item \textbf{Initial compute investment}: \$200B/year.\\
    Major AI firms spent on the order of \$50B per quarter in 2024 \citep{pymnts2024bigtech}, so we loosely assume around \$200B over the entire year. We account for around 4\(\times\) uncertainty in this parameter. 
    
    \item \textbf{Compute adjustment cost exponent ($\chi$)}: 4.\\
    The production of AI chips relies on an intricate supply chain that is difficult to expand due to the use of advanced technologies and materials. As a result, supply of compute-related capital is likely highly inelastic, and larger-than-usual premia may need to be paid to expand the compute-related capital stock on a timescale smaller than is typically needed to produce new fabs, datacenters, and lithography machines. Given that we choose quadratic adjustment costs for conventional capital, we assume substantially steeper costs with an exponent from 3 to 5 for compute. 
    
    \item \textbf{Compute adjustment timescale ($a_Q$)}: 2 years.\\
    This parameter is the counterpart to the capital adjustment timescale $\alpha_K$, where the reciprocal corresponds to the rate at which the compute stock can be grown before adjustment costs become significant. AI chip manufacturing depends on a complex, hard-to-scale supply chain, featuring sophisticated lithography, advanced packaging, specialized photoresists, and specialized chemicals. Using a sample of 635 fab construction projects between from 1990-2020, \cite{verwey2021permits} finds that it takes between 1.6 to 2.2 years from fab construction to production depending on the region the fab was constructed in. This suggests that under ``normal" conditions when extreme adjustment costs are avoided fab capacity can be substantially expanded just under two years. For the range, we choose values 2\(\times\) larger or smaller than this default estimate. 
    
    \item \textbf{Compute depreciation ($\delta_Q$)}: 0.3/year.\\
    \cite{ostrouchov2020gpu} analyzes the reliability and lifetimes of the 18,688 GPUs in the Titan supercomputer at Oak Ridge National Laboratory over its nearly 7 year lifespan from 2012 to 2019. On average, GPUs in the Titan supercomputer lasted around 2.8 years before experiencing failures. However, some units failed much earlier than expected, requiring replacements. GPUs installed in 2017 to replace the failing units lasted significantly longer, often beyond 3 years. The failure rate is $(\text{mean time to failure})^{-1} \approx 0.3$/year.
\end{itemize}

\subsubsection{Compute stock parameters}

\begin{small}
\begin{longtable}{@{}%
    >{\raggedright\arraybackslash}p{0.25\textwidth}%
    >{\raggedright\arraybackslash}p{0.25\textwidth}%
    >{\raggedright\arraybackslash}p{0.50\textwidth}@{}}
\caption{Compute Stock Parameters} \label{tab:compute-stock}\\
\toprule
\textbf{Parameter (units)} & \textbf{Range (default)} & \textbf{Explanation} \\
\midrule
\endfirsthead
\multicolumn{3}{c}{\tablename\ \thetable\ -- \textit{Continued from previous page}}\\
\toprule
\textbf{Parameter (units)} & \textbf{Range (default)} & \textbf{Explanation} \\
\midrule
\endhead
\midrule
\multicolumn{3}{r}{\textit{Continued on next page}}\\
\endfoot
\bottomrule
\endlastfoot

Largest training run - $C_T(0)$ (eFLOP) 
& 2e25 to 2e26 (5e25) 
& Largest training run in effective floating point operations (eFLOP) at the start of 2025. \\

Physical compute limit - $C_L$ (FLOP/year) 
& $8.60 \times 10^{34}$ to $5.00 \times 10^{41}$ ($2.00 \times 10^{38}$) 
& Bound on total FLOP that can be performed per year due to thermodynamic limits. \\
\end{longtable}
\end{small}

\begin{itemize}
    \item \textbf{Largest training run} (\(C_T(0)\): 5e25 eFLOP.\\
    At the time of writing, the largest training run performed was for Google Deepmind’s system Gemini Ultra, at roughly 5e25 FLOP \citep{epoch2023aitrends}. This estimate was obtained by combining two approaches: (1) backing out compute estimates based on known relationships between training compute and performance on certain machine learning benchmarks, and (2) using details of the hardware used during training.\footnote{The calculations can be found in \href{https://colab.research.google.com/drive/1sfG91UfiYpEYnj_xB5YRy07T5dv-9O_c?usp=sharing}{this colab notebook}.}

    \item \textbf{Physical compute limit} \(C_L\): 2e38 FLOP/year. \\
    Under the existing CMOS paradigm for microprocessors, computations are generally performed irreversibly, and this results in heat dissipation. As a result, if excessively large quantities of compute are performed in some timeframe, the resulting heat dissipation might be extremely large and have massive environmental effects. 

    As such, this parameter acts as an upper bound to the FLOP/year. We derive this by combining existing estimates of the maximum FLOP/J of CMOS microprocessors, and estimate a maximum J/year that can be devoted to computation. For the former, \citet{ho2023limits} estimate that the maximum energy efficiency of CMOS microprocessors is roughly 5e15 FLOP/J. For the latter, we assume that energy availability is capped at 1\% of the Earth’s annual energy dissipation, yielding approximately 5e22 J/year based on the Stefan-Boltzmann law.\footnotemark If we combine these two results we arrive at a limit of 2e38 FLOP/year. 
    \footnotetext{This says that $P = \sigma A T^4$, where $P$ is power, $\sigma = 5.6 \times 10^{-8} \: \text{Wm}^{-2} \text{K}^{-4}$ is the Stefan-Boltzmann constant, $A$ is the surface area of the Earth, and $T$ is the Earth's surface temperature. (Note that this setup assumes the Earth is an ideal black body with emissivity $\epsilon = 1$, which does not affect our conclusions.) Assuming the earth has a radius of $6.4 \times 10^{6}$ m \citep{Mamajek2015IAUresolution} and surface temperature of 300 K, then we arrive at a rough estimate of $5 \times 10^{24}$ J/year in annual energy dissipation. Taking 1\% of this estimate yields our result.}
    
    For the lower end of our range, we consider 2 standard deviations below the ideal CMOS efficiency, and multiply this by the energy consumption of today (i.e. 5e20 J/year \citep{OurWorldInData2024}). Thus we have 5e20 J/year * 1.6e14 FLOP/J = 8e34 FLOP/year.
    
    For the higher end of our range, we consider 2 standard deviations above the default ideal CMOS efficiency, and multiply this by 100\% of the Earth's energy budget. This is 5e24 J/year * 1e17 FLOP/year = 5e41 FLOP/year.
\end{itemize}

\subsubsection{Runtime compute parameters}
\begin{small}
\begin{longtable}{@{}%
    >{\raggedright\arraybackslash}p{0.25\textwidth}%
    >{\raggedright\arraybackslash}p{0.25\textwidth}%
    >{\raggedright\arraybackslash}p{0.50\textwidth}@{}}
\caption{Runtime Compute Parameters} \label{tab:runtime-compute}\\
\toprule
\textbf{Parameter (units)} & \textbf{Range (default)} & \textbf{Explanation} \\
\midrule
\endfirsthead
\multicolumn{3}{c}{\tablename\ \thetable\ -- \textit{Continued from previous page}}\\
\toprule
\textbf{Parameter (units)} & \textbf{Range (default)} & \textbf{Explanation} \\
\midrule
\endhead
\midrule
\multicolumn{3}{r}{\textit{Continued on next page}}\\
\endfoot
\bottomrule
\endlastfoot
Initial runtime compute $C_I(0)$ (eFLOP/year) 
& 1e27 to 1e29 (1e28)
& Total compute available for running AI systems each year, as of the start of 2025. \\

Runtime compute requirements - $10^{\gamma_0}$ (FLOP/year)
& 1e13 to 1e17 (1e15)
& Minimum runtime compute cost of substituting a worker for the easiest to automate tasks, at maximum overtraining (i.e. after maximizing training compute to minimize future inference costs). \\

Task-level runtime compute requirements \(\gamma_1\) (dimensionless) & 7 to 11 (9) & Increase in the runtime compute cost in orders of magnitude across tasks, going from automation of the first tasks to full automation. \\

Inference - training tradeoff parameter - $m$ (dimensionless)
&  1 to 4 (2)
& Slope of inference compute tradeoff in determining the effective compute size of a system. This captures how many OOMs of inference compute can be substituted for 1 OOM of training compute while achieving the same level of performance. See equation \eqref{eq:training_vs_inference_tradeoff}.\\

Max inference training compute tradeoff - $\iota_\text{max}$ (dimensionless)
&  $10^{3}$ to $10^7$ ($10^{5}$)
& Maximum value of inference multiplier. This includes effects from both decreasing overtraining and increasing inference scaling, both of which would increase the multiplier (where the "unadjusted runtime compute cost" assumes maximum overtraining).  See equation \eqref{eq:training_vs_inference_tradeoff}.\\
\end{longtable}
\end{small}

\begin{itemize}
    \item \textbf{Initial runtime compute $D(0)$}: 1e28 FLOP.\\
    While the compute devoted to training Gemini Ultra is large by historical standards, the overall compute resources that are not currently dedicated for AI training is substantially larger. For example, \citet{steinhardt2023gpt2030} estimates the total compute stock to be around 1e29 FLOP,\footnote{In particular, \citet{steinhardt2023gpt2030} estimates that GPT-4 (~1e25 FLOP) was trained on around 0.01\% of the world’s compute resources, which suggests an overall runtime compute of around 1e29 FLOP.} and Barnett 2023 similarly estimates that all the AI hardware in the world can produce at most 1e22 FLOP/s (or 3e29 FLOP/year). Both of these values are based on estimates of the quantity of sold microprocessors (e.g. H100 GPUs), and theoretical performance metrics for these processors obtained from official datasheets. However, since in practice it is unlikely that all of this theoretically-available runtime compute can be allocated to AI training, we adjust this estimate down by an OOM to obtain 1e28 FLOP. We account for around 1 OOM of uncertainty above or below this default estimate. 
    \item \textbf{Baseline runtime compute requirement} \(10^{\gamma_0}\): 1e15 FLOP/year.\\
    For this parameter, we anchor loosely to the inference compute required for one of the first automated tasks, namely optical character recognition (OCR). Each forward pass required several hundred thousand to a million mult-adds, and so we loosely anchor to around 1M FLOP per forward pass (e.g. LeNet-4 involved 260k mult-adds per forward pass \citep{bottou1994comparison}). Assuming that each forward pass can be used to analyze one character, and around 20 characters are processed per second, this corresponds to around 6e14 FLOP/year, which we round up to 1e15 FLOP/year. There is substantial uncertainty over this parameter (e.g. in terms of which task was automated first, the precise minimum runtime compute requirement, which examples to anchor to, etc.) and so we introduce a range of around 2 OOMs on either side of our central estimate. 

    \item \textbf{Task-level runtime compute requirement slope} \(\gamma_1\): 9 (OOMs). \\
    Compared to the easiest tasks, we assume that the hardest tasks to automate have a computational requirement similar to that of the human brain. To this end, we anchor to \citet{carlsmith2020computational_power}, who estimates that the human brain does the equivalent of 1e15 FLOP/s, or around 3e22 FLOP/year. Since the brain is known to be highly efficient, we shift this estimate up around 1-2 OOMs to arrive at 1e24 FLOP/year as the minimum runtime compute cost. This parameter is highly uncertain, and as such we account for 2 OOMs of uncertainty on either side of our central estimate. 

    \item \textbf{Inference-training compute tradeoff}: 2. \\
    We base our central estimate of 2 OOMs on the estimates from \citet{epoch2023tradingoffcomputeintrainingandinference}, and allow for roughly 2\(\times\) uncertainty, roughly consistent with the provided estimates. 

    \item \textbf{Max inference training compute tradeoff}: 5. \\
    \citet{epoch2023tradingoffcomputeintrainingandinference} argue that it is possible to increase inference by 2-6 OOMs in exchange for less training compute, depending on the type of task. In the opposite direction, it is possible to decrease inference compute by 1 OOM in exchange for more training compute (overtraining). For our central estimate we hence choose \(4+1 = 5\) OOMs. 
\end{itemize}

\subsubsection{Automation parameters}
\begin{small}
\begin{longtable}{@{}%
    >{\raggedright\arraybackslash}p{0.25\textwidth}%
    >{\raggedright\arraybackslash}p{0.25\textwidth}%
    >{\raggedright\arraybackslash}p{0.50\textwidth}@{}}
\caption{Automation Parameters} \label{tab:automation}\\
\toprule
\textbf{Parameter (units)} & \textbf{Range (default)} & \textbf{Explanation} \\
\midrule
\endfirsthead
\multicolumn{3}{c}{\tablename\ \thetable\ -- \textit{Continued from previous page}}\\
\toprule
\textbf{Parameter (units)} & \textbf{Range (default)} & \textbf{Explanation} \\
\midrule
\endhead
\midrule
\multicolumn{3}{r}{\textit{Continued on next page}}\\
\endfoot
\bottomrule
\endlastfoot

AGI training requirements - $T$ (eFLOP) 
& 1e33 to 1e41 (1e36.5) 
& Effective FLOP required to train AI systems capable of automating all tasks. \\

Initial fraction of automated tasks - $f_\text{init}$ (dimensionless) 
& 0.05 to 0.2 (0.1)
& Initial fraction of tasks in the economy that can be automated. Determines whether runtime compute has any economic value at the start of simulations. \\

Flop gap fraction - \(\Delta f\)
(dimensionless) 
& 0.4 to 0.8 (0.55)
& Ratio between the training compute needed to automate all tasks and the training compute needed to automate the least demanding currently-unautomated tasks. \\
\end{longtable}
\end{small}

\begin{itemize}
    \item \textbf{AGI training requirements} \(T\): 1e36.5 FLOP.\\
    We take a rough central estimate from \citet{cotra2020}, which estimates the computational requirements for training a transformative AI system using "biological anchors" (e.g. from the human brain or evolution), to get 1e36.5 FLOP. This parameter is highly uncertain, so we account for several OOMs of uncertainty ranging from 1e33 to 1e41, based on the distribution of FLOP requirements reported by Cotra. 

    \item \textbf{Initial fraction of automated tasks $f(0)$}: 0.1.\\
    This parameter describes a latent variable which loosely corresponds to the fraction of automated tasks at the start of simulation. We estimate this parameter using two main sources of evidence. First, we examine the historic share of GWP for compute, which increased from ~0\% to ~0.1\% by the 2000s.\footnote{For example, Stewart 2021 reports that global chip revenues were around 0.4\% of GWP in 2005.} Under the GATE production function, using our median runtime compute requirements, this implies an automated fraction of ~10\%. Second, we examine the estimates of \citet{Acemoglu2019AutomationAN} who investigated automation by relating the labor share, sectoral productivity changes and labor reallocation. Between 1987 and 2017 they found a net change in task content of production of approximately 10\%. As such, we choose 10\% as our median estimate for this parameter, such that runtime compute initially has some economic value. We account for around 2\(\times\) uncertainty in this parameter. 
    
    \item \textbf{FLOP gap fraction}: 0.52.\\
    This parameter describes the ratio between the effective training compute needed to automate all tasks, and the effective training compute needed to automate the least demanding currently-unautomated tasks. This is measured in OOMs, divided by the initial distance in OOMs between the largest initial training run and AGI training requirements. Thus this fraction corresponds to 
    \begin{equation}
    \frac{C(100\%) - C(20\%)}{C(100\%) - C(\text{init})},
    \end{equation}
    where $C(x\%)$ is the base 10 logarithm of the compute required for $x\%$ automation. Intuitively, this parameter captures how soon pre-AGI systems start automating the economy. If this parameter is zero, then only AGI can automate anything; if this parameter is one, then we see steady linear progress in automation at current training run sizes.
    
    Since we only have weak evidence for this parameter, we adopt a fairly broad distribution over its values. Some studies suggest that present day AI systems can already automate some tasks that older systems could not \citep{epoch2023challengesinpredictingaiautomation}, and other sources predict the potential for substantial automatability \citep{davidson2023} or exposure \citep{eloundou2023gpts} over the next decade. On current trends, a decade involves 5 OOMs compute scaling \citep{epoch2023aitrends}, i.e. 1e30.5 FLOP. As a FLOP gap fraction for our median training requirements, this results in a FLOP gap fraction of around 0.55. 

    For the lower end of our range, we note that several AI researchers anticipate rapid improvements in AI capabilities across relatively small scaling of compute. If we assume ~2 OOM of compute could raise AI systems from insignificance to automating half of the economy, then this roughly corresponds to a \(\Delta f \approx 2*2/11~=0.4\) for our range of AGI training requirements \(T\). For our upper end, we consider the situation where 2 more OOMs are needed to reach an increase in the fraction of tasks automated, which results in a fraction of \(\Delta f = (11-2)/11 \approx 0.8\). 
\end{itemize} 

\subsection{Add-ons}
\begin{small}
\begin{longtable}{@{}%
    >{\raggedright\arraybackslash}p{0.25\textwidth}%
    >{\raggedright\arraybackslash}p{0.25\textwidth}%
    >{\raggedright\arraybackslash}p{0.50\textwidth}@{}}
\caption{Add-ons} \label{tab:add-ons}\\
\toprule
\textbf{Parameter (units)} & \textbf{Range (default)} & \textbf{Explanation} \\
\midrule
\endfirsthead
\multicolumn{3}{c}{\tablename\ \thetable\ -- \textit{Continued from previous page}}\\
\toprule
\textbf{Parameter (units)} & \textbf{Range (default)} & \textbf{Explanation} \\
\midrule
\endhead
\midrule
\multicolumn{3}{r}{\textit{Continued on next page}}\\
\endfoot
\bottomrule
\endlastfoot
R\&D wedge - $\xi$ (dimensionless)
& 2 to 20 (8)
& This parameter helps account for differences between the private and social benefits of R\&D. When this parameter is 5, investments into R\&D are cut by 1/5th, to capture the effect where positive externalities decrease the private returns to investment.  \\
Number of automation functions (uncertainty add-on) - \(|\mathcal{F}|\)
(dimensionless) 
& 1 to 20 (5)
& Number of automation functions in the support of planner's beliefs when uncertainty add-on is activated. See Section~\ref{sec:uncertainty_add-on}.\\

Maximum automation parameters (uncertainty add-on) - $\{\zeta_i\}, 1 \leq i \leq |\mathcal{F}|$ (dimensionless) 
& 0 to 1 (no default)
& Parameters describing the shape of the alternative, ``untrue" automation functions that are in the support of the planners beliefs at the start of the simulation when the uncertainty add-on is activated. See Section~\ref{sec:uncertainty_add-on}.\\

Beliefs over alternative automation functions (uncertainty add-on) - $\left\{G(f_i)\right\}_i, 1 \leq i \leq |\mathcal{F}|$ (dimensionless) 
& 0 to 1 (no default)
& Parameters describing the probability of each of the possible automation functions that are in the support of the planners beliefs at the start of the simulation when the uncertainty add-on is activated. See Section~\ref{sec:uncertainty_add-on}.\\

Initial labor force task i, no reallocation - $L_{i}(0)$ (no. of human workers)
& Initial labor force per task (no reallocation) (3.6 billion)
& Total number of humans involved in task $i$ at the start of simulations. This applies to the no reallocation scenarios discussed in Section~\ref{sec:labor_reallocation}. \\
\end{longtable}
\end{small}

The model playground is initialized with a set of pre-specified default parameters (outlined in the second columns of Tables 3 to 9 above). Users are able to either manually modify parameters within their permissible ranges (also outlined in the second column of Tables 3 to 9) or select from pre-specified bundles of parameters or scenarios (e.g., median, 5th or 95th percentile values for parameters based on estimates from the existing empirical literature).

The model playground updates graphs and charts as users adjust parameters. Moreover, to enhance functionality, the playground includes an additional feature, namely the possibility to engage in scenario comparisons. The modeling tool allows users to save and compare the outputs of multiple simulation runs, enabling side-by-side evaluations of distinct assumptions or parameter values.

\paragraph{Sanity checks} To bolster confidence in GATE’s optimizer, we implement two built-in “sanity checks.” First, a simple policy check\footnote{Here, GATE re-solves the problem under a deliberately simpler, lower-dimensional policy class—e.g., using a single vector of decisions for the pre-automation phase, another for partial automation, and so on. If this simpler approach unexpectedly achieves strictly higher utility than the full solution, it signals that the main solver may have converged to a suboptimal local optimum.} ensures that the flexible, high-dimensional solution is not outperformed by a restricted baseline. Second, a spliced utility check\footnote{In this step, GATE replaces power utility with a log-based tail once consumption surpasses a certain threshold. This modified utility function avoids potential numerical flatness at extremely high consumption. If the solution found under this “spliced” utility then yields higher welfare when re-evaluated on the \emph{original} power utility, it indicates the baseline solution may have been hampered by numerical pathologies.} substitutes a log-based tail for large consumption levels and verifies that the main solution remains robust. These checks help ensure that the interactive playground does not inadvertently present suboptimal or unstable trajectories, increasing confidence in the model’s outputs.

\paragraph{Parameters requiring extra caution}
In configuring GATE, three parameters warrant particular caution: the substitution parameter $\rho$ in our CES aggregator, the coefficient of relative risk aversion $\eta$ in the utility function, and the FLOP gap $\Delta_{\mathrm{FLOP}}$ in the automation function. First, pushing $\rho$ too far below zero or extremely close to zero from below makes the tasks nearly Leontief (i.e.\ near-perfect complements), which can create numerical instability. Second, making $\eta$ overly large or extremely close to one can yield excessively steep or flat gradients in the objective function (equation~\ref{eq:objective-function}), undermining gradient descent. Finally, $\Delta_{\mathrm{FLOP}}$ governs how quickly the fraction of tasks automated, $f(t)$, transitions between its initial level and full automation (see equation~\ref{eq: automation function}). Setting $\Delta_{\mathrm{FLOP}}$ below its recommended range may produce an overly “jumpy” automation schedule, while an excessively high value can stagnate automation progress. Hence, to ensure smoothness in both production and utility—and to keep optimization tractable—users should keep $\rho,\ \eta,\ \Delta_{\mathrm{FLOP}}$ within the intervals highlighted in Sections~\ref{sec:model_inputs}.

\end{appendix}

\end{document}